\newif\ifrmp
\rmptrue

\newif\ifnotcompact
\notcompactfalse

\documentclass[a4paper,11pt]{article}
\usepackage{jcappub} % for details on the use of the package, please
\usepackage{amsmath}
\usepackage{amsfonts}
\usepackage{amssymb}
\usepackage[pass,letterpaper]{geometry}
\usepackage{aas_macros}
\usepackage{multirow}
\usepackage{graphicx}
\usepackage{hyperref}
\usepackage{bm}
\usepackage{subeqnarray}
\usepackage{epstopdf}

\newcommand{\modif}[1]{{\color{black} #1}}
\newcommand{\be}{\begin{equation}}
\newcommand{\ee}{\end{equation}}
\newcommand{\bea}{\begin{eqnarray}}
\newcommand{\eea}{\end{eqnarray}}
\newcommand{\gr}[1]{{\mathbf{#1}}}
\newcommand{\ii}{\mathrm{i}}
\newcommand{\HH}{\mathcal{H}}
\newcommand{\dd}{\mathrm{d}}

\newcommand{\dis}{{\alpha}}
\newcommand{\dist}{d}
\newcommand{\kerK}{{\cal K}}
\newcommand{\kerH}{{\cal N}}

\newcommand{\IpowJ}{{\cal J}}
\newcommand{\IpowJtilde}{{\widetilde{\cal J}}}
\newcommand{\IntI}{{\cal I}}

\newcommand{\calJ}{{\cal J}}

\newcommand{\LLL}[1]{{\mathrm{L}^2_{#1}}}

\newcommand{\Pow}{{\cal P}}

\newcommand{\MYXI}{{\Xi}}

\newcommand{\comm}[1]{}

\usepackage{accents}
\newlength{\dhatheight}

\newcommand{\Crr}{C}
\newcommand{\Ckk}{C}
\newcommand{\xidr}{\xi}
\newcommand{\xidk}{\widehat{\xi}}
\newcommand{\zetaKk}{\zeta}
\newcommand{\zetadk}{\widetilde{\zeta}}

\usepackage{moresize}
\newcommand{\FS}{{\rm pp}}
\newcommand{\Kai}{\text{\ssmall Kaiser}}

\title{Redshift-space distortions with wide angular separations}

\author[a, b]{Paulo Reimberg}
\author[a, b]{Francis Bernardeau}
\author[b]{Cyril Pitrou}
\affiliation[a]{CEA, IPhT, 91191 Gif-sur-Yvette c{\'e}dex, France}
\affiliation[b]{Institut d’Astrophysique de Paris, CNRS, UMR 7095 and Sorbonne Universit\'es, UPMC Univ Paris 6, 98 bis bd Arago, 75014 Paris, France}

\emailAdd{paulo.flose-reimberg@cea.fr}      
\emailAdd{francis.bernardeau@cea.fr}       
\emailAdd{pitrou@iap.fr}       

\abstract{
Redshift-space distortions are generally considered in the plane
parallel limit, where the angular separation between the two sources
can be neglected. Given that galaxy catalogues now cover large
fractions of the sky, it becomes necessary to consider them in a formalism which
takes into account the wide angle separations. In this article we derive an operational formula for the matter correlators in the Newtonian limit to be used in actual data sets. \modif{In order to describe the geometrical nature of the wide angle RSD effect on Fourier space, we extend the formalism developed in configuration space to Fourier space without relying on a plane-parallel approximation, but under the extra assumption of no bias evolution}. We then recover the plane-parallel limit not only in configuration space where the geometry is simpler, but also in Fourier space, and we exhibit the first corrections that should be included in large surveys as a perturbative expansion over the plane-parallel results. We finally compare our results to existing literature, and show explicitly how they are related.}

\arxivnumber{1506.06596}

\begin{document}
\maketitle
\flushbottom

%\ifrmp
%\else
%\nokeywords
%\fi

%\date{\today}

%\ifrmp
%\maketitle
%\else
%\fi

%%%%%%%%%%%%%%%%
\section{Introduction}
%%%%%%%%%%%%%%%%
Redshift-space distortion (RSD) is a general denomination for the corrections on distances attributed to objects from their redshift measurements, due to their peculiar velocities. Models for the peculiar velocities rely on the theory of structure formation, in linear or nonlinear regimes. We shall here be interested in the linear regime, and therefore we will assume that dark matter distribution is well described by linear fluid equations, and that galaxies overdensities are associated to dark matter overdensities through a linear bias.

Under these hypothesis, redshift space distortions become a geometrical effect since the corrected density field $\delta^z$ is related to the underlying linear density field $\delta$ thanks to the integro-differential operator composed of a non-local part (the inverse of the Laplacian) and the radial part of the Laplacian
\be\label{EqOperatorIntro}
\delta^z(\gr{r}) = \delta(\gr{r}) +\frac{\beta}{r^2} \partial_r [r^2 \partial_r
(\Delta^{-1} \delta (\gr{r}))]\,.
\ee
When we study correlation functions, we derive from the properties of the operator in eq.~\eqref{EqOperatorIntro} that the RSD corrections to the two-point function only depend on the shape of the triangle formed by the observer and the two objects being correlated. The limit in which the triangle is squeezed and close to isosceles corresponds to the {\it plane-parallel} approximation, or {\it flat-sky} approximation. The plane-parallel approximation is well justified if the distance between the pair of galaxies is much smaller than the distances from the observer to each of the galaxies.

In the plane-parallel limit, eq.~\eqref{EqOperatorIntro} can be simplified and leads to the classical Kaiser formulation of the problem \cite{Kaiser:1987qv, Hamilton:1997zq}. 

Current and future galaxy surveys cover sizable fractions of the sky such as BOSS~ \cite{boss} and Euclid~\cite{euclid}. For these catalogues, one cannot assume anymore that the separation angle between two galaxies is small, what makes the plane-parallel approximation unjustified and compels the incorporation of the so-called  {\it wide angle} effects. Based on the analysis of simulated samples, \cite{Raccanelli2010, Yoo2015} have shown that, although sizable, wide angle effects may be of the size of systematic errors in configuration survey analysis if suitable geometrical configurations and analysis procedure are adopted. Hence, for practical purposes, the Kaiser formula would still provide accurate prescription for the RSD corrections. One of our goals in this paper is to build a perturbative expansion around the plane-parallel limit and show that the first corrections due to wide angles are quadratic in the ratio of the distance between the galaxies to the mean distance from the observer.

The study of wide angle effects on RSD is not recent. A non-exhaustive list of works dedicated to the problem includes \cite{Zaroubi:1993qt, Heavens:1994iq, hamilton_culhane, Tegmark:1994pa, Taylor_Valentine, Szalay:1997cc, Szapudi2004}. Building on those earlier articles, \cite{Papai:2008bd} was able to calculate in full generality two-point correlation function, allowing to estimate the difference with the standard plane-parallel limit.

Using the operator in eq.~\eqref{EqOperatorIntro}, the usual treatment of wide angle effects, as in \cite{Bonvin:2011bg}, is to decompose the distorted matter density field $\delta^z(\gr{r})$ in spherical harmonics, e.g. as for the Cosmic Microwave Background analysis. For every distance, this leads to define the associated multipoles $\delta^z_{\ell m}(r)$, and since the redshift-space distortions do not break the statistical isotropy around the observer, the statistics is encoded in the diagonal part of the multipoles correlations, that is 
\be
\langle \delta^z_{\ell m}(r_1) \delta^z_{\ell' m'}(r_2)\rangle = \delta_{\ell \ell'} \delta_{m m'}C^z_\ell(r_1,r_2).
\ee

Both the expression found with the configuration space approach in \cite{Papai:2008bd} and the one derived with a multipole approach in \cite{Bonvin:2011bg} are much more complicated than the simple original Kaiser formula that holds in the plane-parallel limit.  In this article, our second goal is to relate both approaches and to link them with our perturbative expansion around the plane-parallel limit.
%%%%%%%%%%%%%%%%%%%%%%%%%%%%%%%%%%%
\section{Overview of results}\label{SecOverview}
%%%%%%%%%%%%%%%%%%%%%%%%%%%%%%%%%%%

%%%%%%%%%%%%%%%%%%%%%%%%%%%%%%%%%%%%%%%%%%%%%%%%%%%%
\subsection{Wide angle effects in configuration space}\label{SecIntroReal}
\begin{figure}[!htb]
\begin{center}
\includegraphics[scale=0.5]{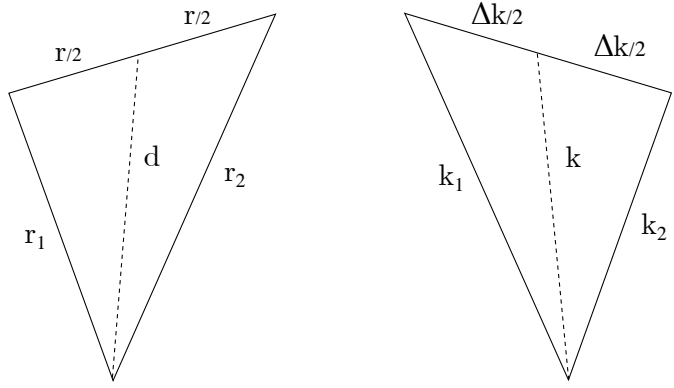}
\end{center}
\caption{Effect of Fourier transform on variables constrained to a triangular geometry. The Fourier conjugate of $\gr{r}_1$ is $\gr{k}_1$, the Fourier conjugate of $\gr{r}_2$ is $\gr{k}_2$; the Fourier conjugate of $\gr{\dist}$ is $\Delta \gr{k} = \gr{k}_1 - \gr{k}_2$, while the conjugate of $\gr{r}$ is $\gr{k}$.}
\label{conjugate}
\end{figure}

%%%%%%%%%%%%%%%%%%%%%%%%%%%%%%%%%%%%%%%%%%%%%%%%%%%%
The structure of the correlation function depends only on the shape of a triangle formed by the observer and the sources at $\gr{r}_1$ and $\gr{r}_2$. The plane-parallel limit corresponds to a squeezed configuration of this triangle where $\gr{r} = \gr{r}_2 - \gr{r}_1$ is much smaller in norm than $r_1$ and $r_2$. In order to expand the correlation functions around this plane-parallel approximation, it appears more appropriate to consider the correlation functions as depending on $\gr{r}$ and the median distance $\dist$ as illustrated in figure~\ref{conjugate}, with the definitions
\be
\xidr^z(\gr{\dist},\gr{r})\equiv \Crr^z (\gr{r}_1, \gr{r}_2) \equiv \langle  \delta^z(\gr{r}_1) \delta^z(\gr{r}_2)\rangle\,,\qquad \gr{r} \equiv \gr{r_2}-\gr{r}_1\,,\quad \gr{\dist}
  \equiv \frac{\gr{r}_1+\gr{r}_2}{2}\,\,.
\ee
In this article we give the general expression for the correlation function in configuration space, checking that we recover the results of previously existing literature, and we then expand it around its plane-parallel limit so as to grasp the structure of the wide angle corrections.
We also show that instead of using the median position $\gr{\dist}$ as an average position, it is possible to use the bisector to define another type of average position, as it leads to the same plane-parallel limit. 

If the correlation function was statistically homogeneous, then it would depend only on
$\gr{r}$, and not on $\gr{\dist}$. And if it was also statistically isotropic, it would actually depend only on one degree of freedom, $r$. This would be the case if RSD effects were ignored. However the distorted field is not homogeneous as it also depends on the velocity of the source with respect to the observer, and not just on the velocity independently. Nevertheless, the global rotational invariance around the observer removes three degrees of freedom, implying that the correlation function is only a function of three degrees of freedom  which are $r$, $\dist$ and $\mu_{\gr{\dist \, r}} \equiv \hat{\gr{r}} \cdot \hat{\gr{\dist}}$. In the plane-parallel limit, it depends on $r$ and $\mu$, but not on $\dist$. It is thus appropriate to expand the general two point
correlation function as a general angular multipole expansion
\begin{equation}\label{Teasing2}
\xidr^z (\gr{\dist},\gr{r}) = \sum_{n=0}^{\infty} \left(\frac{r}{\dist} \right)^n
\sum_{\ell=0}^{\infty} \xidr_\ell^{(n)}(r)
P_\ell(\mu_{\gr{\dist \, r}}) \, ,\qquad{\rm with}\qquad  \mu_{\gr{\dist \, r}} \equiv \hat{\gr{r}}\cdot \hat{\gr{\dist}}\,.
\end{equation}
The $\xidr_\ell^{(0)}$ are the lowest order coefficients which arise in the plane-parallel limit, and to be more precise, only $\xidr_0^{(0)}$, $\xidr_2^{(0)}$, $\xidr_4^{(0)}$ are non-vanishing. The $\xidr_\ell^{(n>0)}$ describe then corrections due to wide angle effects. The natural small parameter for this expansion is the ratio between the distance between two points being correlated, and their average distance from the observer, that is $r/\dist$. The further the two points are with respect to the observer, the smaller the corrections are. The geometrical structure at each order is described by the angular variable which depends only on the relative directions between the difference of positions $\gr{r}$ and the average direction $\gr{\dist}$, that is it depends only on $\mu$. We find that the coefficients $\xidr_\ell^{(n)}$ in the expansion~\eqref{Teasing2} are non-vanishing only if $\ell$ and $n$ are either both odd or both even. Furthermore, we show that for the median and the bisector parametrizations of the average distance, these coefficients do not vanish only if $\ell$ and $n$ are both even. As a consequence, the first order corrections $\xidr_\ell^{(1)}$ vanish, and these choices should thus be preferred to minimize the wide angle effects. In these cases, we compute explicitly the $\xidr_\ell^{(n)}$ up to second order providing the first set of corrections to the plane-parallel limit. \modif{These coefficients emerge from the analysis of eq \eqref{Fullxireal1}, explicitely written in eq \eqref{GenCorrFunc}, which constitutes one of the main results of this paper.}

%%%%%%%%%%%%%%%%%%%%%%%%%%%%%%%%%%%%%%%%%%%%%%%%%%
\subsection{Wide angle effects in Fourier space}\label{SecIntroFourier}
%%%%%%%%%%%%%%%%%%%%%%%%%%%%%%%%%%%%%%%%%%%%%%%%%%
In order to understand how such expansion arises in Fourier space, one must
first realize that when performing a double Fourier transformation on a function whose variables are related in a triangular configuration, the
Fourier conjugate to the difference of positions $\gr{r}$ is the median Fourier modes $\gr{k}$, and the
Fourier conjugate to the median distance $\gr{\dist}$ is the difference of the Fourier modes $\Delta \gr{k}$. This geometry of the Fourier space is illustrated in figure~\ref{conjugate}. This basic geometrical relation can be understood in two steps. First, for $\alpha>0$, the Fourier transform of $\Crr^z(\alpha \gr{r}_1, \gr{r}_2)$ is $\frac{1}{\alpha} \Ckk^z(\frac{\gr{k}_1}{\alpha}, \gr{k}_2)$, i.e., dilatations (resp. contractions) in configuration space lead to contractions (resp. dilatations) in Fourier space, and therefore if $r_1$ is shorter than $r_2$, $k_1$ will be longer than $k_2$. Secondly, when performing a Fourier transformation on a two point correlation function, we introduce the product
\be\label{EqCorrespkr}
\mathrm{e}^{\ii \gr{k}_1 \cdot \gr{r}_1} \mathrm{e}^{-\ii \gr{k}_2 \cdot \gr{r}_2} 
= \mathrm{e}^{\ii \gr{d} \cdot \Delta \gr{k}} \mathrm{e}^{-\ii \gr{r}
  \cdot \gr{k}} \, , \qquad{\rm with} \qquad\gr{k} \equiv (\gr{k}_1+\gr{k}_2)/2\,,\qquad \Delta\gr{k} \equiv \gr{k}_1-\gr{k}_2 \,,
\ee
from where the crossed conjugate relation among median and difference of modes can be inferred. It is thus natural to define the correlation function in Fourier space using these variables as
\be
\zetaKk^z(\Delta\gr{k},\gr{k})\equiv \Ckk^z(\gr{k}_1,\gr{k}_2)\equiv \langle \delta^z(\gr{k}_1) \delta^{z\star}(\gr{k}_2)\rangle\,.
\ee
The homogeneity in Fourier space is expressed by the fact that the correlation function depends only on the average Fourier mode, and not on the difference. The inhomogeneity introduced
by the RSD effects translates into the fact that in Fourier
space there are off-diagonal correlations. In this article, we derive the general expression for the correlation function in Fourier space and exhibit the off-diagonal contributions.
Following the correspondence~\eqref{EqCorrespkr}, the corrections introduced should be expressed as an expansion in $|\Delta\gr{k}|/k$ in the form
\begin{equation}
\label{SuperFourierExpansion}
\zetaKk^z(\Delta\gr{k}, \gr{k}) =\delta_D(\Delta\gr{k})\zetaKk_0^{(0)}(k)+\frac{1}{4\pi|\Delta\gr{k}|^3}\sum_{n=0}^{\infty} \left(\frac{|\Delta\gr{k}|}{k} \right)^n
\sum_{\substack{\ell=0\\ (\ell,n)\neq(0,0)}}^{\infty} \zetaKk_\ell^{(n)}(k) P_\ell(\mu_{\gr{k \,  \Delta}}) \, ,
\end{equation}
with $\mu_{\gr{k \, \Delta}} \equiv  \hat{\gr{k}} \cdot \widehat{\Delta \gr{k}}$.
The $\zetaKk_\ell^{(0)}(k)$ correspond to the homogeneous contribution of the plane-parallel limit, for which only $\zetaKk_0^{(0)}$, $\zetaKk_2^{(0)}$, $\zetaKk_4^{(0)}$ are non vanishing. The $\zetaKk_\ell^{(n>0)}$ are the wide angle corrections which break homogeneity. For each order, the geometrical dependence is only a function of the angle between the average Fourier modes and the difference of the Fourier modes, and this is understood from the correspondence~\eqref{EqCorrespkr}. We do not perform such expansion explicitly except for the lowest order corresponding to the plane-parallel limit.
%%%%%%%%%%%%%%%%%%%%%%%%%%%%%%%%%%%%%%%%%%%%%%%%%%%%%%%%%%%%%%%%
\subsection{Wide-angle effects in mixed configuration/Fourier space}\label{SecIntroMixed}
%%%%%%%%%%%%%%%%%%%%%%%%%%%%%%%%%%%%%%%%%%%%%%%%%%%%%%%%%%%%%%%%
In fact the RSD effects can also be apprehended using a mixed space, where the median distance is looked at in configuration space, but the dependence in the separation of the sources is considered in Fourier space.  This can be obtained either by Fourier transforming the $\gr{r}$ dependence, that is considering $\xidk^z(\gr{\dist},\gr{k})$ instead of in $\xidr^z(\gr{\dist},\gr{r})$, or by inverse Fourier transforming the $\Delta\gr{k}$ dependence, that is by considering $\zetadk^z(\gr{\dist},\gr{k})$ instead of $\zetaKk^z(\Delta\gr{k,\gr{k}})$. We check that both approaches lead to the same result as they ought to. In this mixed space the natural expansion is
\be
\xidk^z(\gr{\dist},\gr{k}) = \zetadk^z(\gr{\dist},\gr{k})  = \sum_{n=0}^{\infty} \left(\frac{1}{k \dist} \right)^n
\sum_{\ell=0}^{\infty} \Pow_\ell^{(n)}(k)
P_\ell(\mu_{\gr{k \, d}}) \,,\qquad \mu_{\gr{k \, d}} \equiv \hat{\gr{k}}\cdot\hat{\gr{d}}\,.
\ee
The functions $\Pow_\ell^{(n)}(k)$ are related to the configuration space coefficients $\xidr_\ell^{(n)}(r)$ through an integral on spherical Bessel functions as
\be
\Pow_\ell^{(n)}(k)= \sqrt{\frac{2}{\pi}}(-\ii)^\ell \int \dd r r^2 (kr)^n j_\ell(kr) \xidr_\ell^{(n)}(r)\,,\label{PlnOFxiln}
\ee
implying that they do not vanish only if $\ell$ and $n$ are either both odd or both even. Furthermore, they are directly related to the $\zetaKk_\ell^{(n)}(k)$ by numerical factors. Defining $s$ by $\ell= 2+n+2s$, these relations are simply

\be
\Pow_\ell^{(n)}(k)=\frac{\ii^\ell}{(2\pi)^{3/2}}\left[\frac{2^{n+s}(n+s)!}{(2s+3)!!}\right]\zetaKk_\ell^{(n)}(k)\, \quad{\rm if}\quad (\ell,n)\neq(0,0)\,,\quad \Pow_0^{(0)}(k)=\frac{1}{(2\pi)^{3/2}}\zetaKk_0^{(0)}(k)\,.\label{PlnOFzetaln}
\ee

In the plane-parallel limit $n=0$, and only $\Pow_0^{(0)}(k)$, $\Pow_2^{(0)}(k)$, $\Pow_4^{(0)}(k)$ are non-vanishing, as they correspond to the {\it standard Kaiser formula} which we show to be only meaningfully defined in this mixed space. Using eq.~\eqref{PlnOFxiln} we are able to deduce the coefficients $\Pow_\ell^{(n)}(k)$ of the mixed space expansion from the configuration space correlation function expansion, that is from the $\xidr_\ell^{(n)}(r)$. There is then no need to compute the coefficients $\zetaKk_\ell^{(n)}(k)$ of the full Fourier space expansion since we find that they can be easily deduced from the mixed space expansion thanks to eq.~\eqref{PlnOFzetaln}. With this mixed space approach, one is led to consider the various power spectra at each given distance $\dist$ so as to be able to take into account wide angle effects, and the $\Pow_\ell^{(n>0)}(k)$ encode the angular dependence of the corrections at each order. In practice, the inclusion of the wide angle effects in data analysis implies that we must build a spectrum for each subset of pairs having a common average distance $\dist$.
%%%%%%%%%%%%%%%%%%%%%%%%
\subsection{Outline of the article}
%%%%%%%%%%%%%%%%%%%%%%%%
The paper is structured as follows: we start by deriving eq.~\eqref{EqOperatorIntro}, and use it to obtain the general form of the two-point correlation function. We show how to obtain the plane-parallel limit of the general expression and proceed with the analysis of its counterpart in Fourier space. The mixed configuration/Fourier spaces emerge naturally from the structure of the problem and we show that the plane-parallel limit is well defined in the mixed space, from where the Kaiser formula is consistently derived. We also investigate the decomposition of the distorted density fields in spherical harmonics and the $C^z_{\ell}$s both in configuration and Fourier spaces. This constitute a first part of the paper, where we show different ways to approach the problem, and the consistency among them. In the second part of the paper we construct a perturbative treatment of the wide angle effects over the plane-parallel results. We address the problem first in configuration space using the two-point correlation function and we exhibit the wide angle corrections using different parametrizations for the triangular made by the observer and the pair of galaxies. These results are then translated into mixed configuration/Fourier space, where we introduce the idea of {\it spectrum at a given distance}. We advocate that  it is the most meaningful way to extend the Kaiser formula so as to incorporate wide angle effects. We also show  that the general correlators in full Fourier space can be well defined. 

\modif{As we shall see, although the wide angle problem can be treated in full generality in configuration space, the complexity introduced by bias evolution clouds the description in Fourier space. We shall, therefore, only study Fourier and mixed descriptions under the simplifying hypothesis of no bias evolution, what limits the practical applicability of our work. It allows, nevertheless, the investigation of the intricate geometrical nature of RSD in Fourier space, which is one of our goals in this paper.}

%%%%%%%%%%%%%%%%%%%%%%%%%%%%%%%%%%%%%%%%%%%%%%%%%%%%%%%
\section{General formalism of redshift-space distortions}\label{SecGenForm}
%%%%%%%%%%%%%%%%%%%%%%%%%%%%%%%%%%%%%%%%%%%%%%%%%%%%%%%

%%%%%%%%%%%%%%%%%%%%%%%%%%%%%%
\subsection{Physical origin}\label{basics}
%%%%%%%%%%%%%%%%%%%%%%%%%%%%%%

We consider a perturbed Friedmann-Lema\^itre (FL) universe with only scalar
perturbations $\Phi$ which is identified with the gravitational
potential. The corresponding metric is of the form
\be
\dd s^2 = - \dd t^2(1+2 \Phi) + a(t)^2 (1-2\Phi)\delta_{ij} \dd x^i \dd x^j\,,
\ee
where $a(t)$ is the scale factor. We then define the Hubble expansion rate as $H\equiv \dot a /a$, and we
also introduce the conformal time defined by $a\dd \eta \equiv  \dd
t$. The conformal Hubble rate is then simply given by $\HH = aH$.  
In a simple Newtonian description we ignore the effect of the
gravitational potential $\Phi$ on the propagation of light and the redshifting of energies, but we consider
its effect on a source velocity and the corresponding associated
redshift. If the velocity of an emitting source is decomposed as
\be
u^\mu = (1, v^i)\,\qquad v^i = \frac{\dd x^i}{\dd \eta}
\ee
then the redshift at which this source appears today ($\eta_0$) is given by
\be
\frac{1}{1+z} \simeq  \frac{a(\eta_e)}{a(\eta_0)}(1+ v_i \hat x^i)\,, \qquad
\hat x^i \equiv \frac{x^i}{x} \qquad x\equiv \sqrt{x_i x^i}
\ee
where $a(\eta_e)$ is the scale factor at emission. Given that 
\be
a(\eta_e)(1+ v_i \hat x^i)\simeq a \left[ \eta_e + \frac{v_i \hat x^i}{\HH(\eta_e)} \right]\,,
\ee 
ignoring the effect of the peculiar velocity amounts to an error in
the time of emission. This translates into an error on the conformal distance estimated from the background geometry through $| \gr{x}_0-\gr{x}_e| = \eta_0-\eta_e$.
Noting $\gr{x}=(x^1,x^2,x^3)$ the true coordinates of an object at
emission, and $\gr{s}=(s^1,s^2,s^3)$ its inferred position when the
peculiar velocity is ignored, these positions are related at first
order in the velocity by
\be
\label{stox}
s^i = x^i + \dis^i\,,\quad \dis^i \equiv \dis_r \hat x^i \,, \qquad  \dis_r \equiv \frac{1}{\HH_e}(v_j.\hat{x}^j)=\frac{v_r}{\HH_e}\,.
\ee
If we were to include all the gravitational effects, then this relation would involve more terms, and a detailed analysis can be found in e.g. \cite{Bonvin:2011bg, Challinor:2011bk, Jeong:2011as, Yoo:2009au, Bertacca:2012tp}. 

Since we consider only scalar perturbations, the velocity divergence
$\theta$ is characterized fully by the velocity field $V$ through
\be
v^i = \partial^i V,\qquad \theta \equiv \frac{\partial_i v^i}{\HH_e} = \frac{\Delta V}{\HH_e} \,.
\ee
The relation between $V$ and the matter density is obtained from
the resolution of Euler, continuity and the Poisson equations \cite{Peebles_80}.
We then find that the velocity divergence $\theta$ is directly related to the comoving matter density
contrast $\delta = \rho/\bar \rho -1$, where $\bar \rho$ and
$\rho$ are respectively the background and the full energy density
of matter. Indeed, for each Fourier mode we can relate them by a
relation of the form
\be
\theta(k,t)= -\beta(t) \delta(k,t) \qquad \Rightarrow \qquad V(k,t)=
\frac{\beta(t) \HH_e(t)}{k^2} \delta(k,t)\,.
\ee
The matter density $\rho^z$ inferred from the redshifted coordinates
$\gr{s}$, is different from the underlying matter density $\rho$, and
we thus want to relate the associated density contrast $\delta^s
\equiv (\rho^z-\bar\rho/)\bar\rho$ to the true density contrast
$\delta$. The number of galaxies cannot depend on our coordinates system and the
two matter densities are thus related by
\be\label{MatterConservation}
\rho^z(\gr{s}) \dd^3 \gr{s} = \rho(\gr{x}) \dd^3 \gr{x} 
\ee
Using the Jacobian of the transformation~\eqref{stox} which is $\dd^3 \gr{s} \simeq (1+\partial_i
\dis^i) \dd^3 \gr{x}$, we obtain 
\be
\label{rhoztorho}
\rho^z(\gr{s})\simeq \rho[\gr{x}(\gr{s})] [1- \partial_i
\dis^i] \simeq
\rho(\gr{s}) - \bar \rho \partial_i
\dis^i (\gr{s}) \,\quad \Rightarrow \quad \delta^z = \delta - \partial_i\dis^i
\ee
since $\rho[\gr{x}(\gr{s})]-\rho(\gr{s})$ is a second order quantity
as the background matter density is homogeneous. This means that at
lowest order, the volume are affected by the divergence of the
displacement $\dis^i$, and this translates into a modification of the
density with an opposite sign.

Since the displacement field $\dis^i$ is only in the radial
direction, it is useful to recast its divergence in spherical
coordinates to get 
\be\label{thetaspherique}
\partial_i \dis^i = \frac{1}{r^2}\partial_r(r^2
\dis_r)= \frac{1}{\HH_e r^2}\partial_r(r^2
\partial_r V)=-\frac{\beta}{r^2} \partial_r [r^2 \partial_r
(\Delta^{-1} \delta )]
\ee 
where we use the usual notation $r \equiv |\gr{r}|$, and therefore
\be
\label{delta_real_rad}
\delta^z(\gr{r}) = \delta(\gr{r}) +\frac{\beta}{r^2} \partial_r [r^2 \partial_r
(\Delta^{-1} \delta (\gr{r}))]\,.
\ee
In order to express the radial derivatives in this expression, we
decompose the Laplacian in three dimensions into a radial operator and
the Laplacian on the sphere according to

\be\label{DfOL}
\Delta = {\cal O}_r + \frac{\LLL{\hat{\gr{r}}}}{r^2} \,,\qquad\,{\cal O}_r
\equiv \frac{1}{r^2}\partial_r r^2 \partial_r\,,\qquad
\LLL{\hat{\gr{r}}} \equiv \LLL{\nu} +
\frac{1}{1-\nu^2}\partial_\varphi \partial_\varphi\,,\qquad \LLL{\nu}
\equiv \partial_\nu (1-\nu^2)\partial_\nu 
\ee
where $\nu$ is the cosine of the polar angle and $\varphi$ the
azimuthal angle of $\hat{\gr{r}}$ in a spherical coordinates system. 
When it brings no ambiguity, we will omit the index in
$\LLL{\hat{\gr{r}}}$. We then get from eq.~(\ref{delta_real_rad}) an equivalent
relation between the two types of densities which is
\be
\label{delta_real_ang}
\delta^z(\gr{r}) = (1+\beta)\delta(\gr{r}) -\frac{\beta}{r^2} \LLL{\hat{\gr{r}}} \,\Delta^{-1} \delta(\gr{r})\,.\qquad 
\ee
\modif{The time dependence of $\beta$ can be written as a dependence
  on $r$ thanks to the background geodesic relation $r(t)$ which can
  be inverted into a relation $t(r)$. Hence, in eq.~\eqref{delta_real_ang} $\beta$ stands for $\beta(r)\equiv \beta(t(r))$.}
\modif{The corrections considered in eq. \eqref{delta_real_ang}
  constitute a subset of all the relativistic effects in first order
  of perturbation theory. As observed by \cite{Montanari:2012} --
  based on the analysis presented in \cite{Bonvin:2011bg} --, only
  density fluctuation, redshift space distortion and lensing can
  contribute significantly to corrections of number counts in the
  context that we are considering. Lensing, however, has a distinct
  geometrical nature and we shall not be discussed here.} For
simplicity we also assume a constant selection function.
%%%%%%%%%%%%%%%%%%%%%%%%%%%%%%%%%%%%%%%%%%%%%
\subsection{Redshift-space distortions in configuration space}
%%%%%%%%%%%%%%%%%%%%%%%%%%%%%%%%%%%%%%%%%%%%%

%%%%%%%%%%%%%%%%%%%%%%%%%%%%%%%%%%%%%%%%%%%%%
\subsubsection{General expression of the correlation function}
%%%%%%%%%%%%%%%%%%%%%%%%%%%%%%%%%%%%%%%%%%%%%
We first start by deriving the two-point correlation function of distorted
densities in configuration space, using the previous relation~\eqref{delta_real_rad}. We do not assume that the angle separation is small to begin with, so as to obtain the most general
expression. In the next section, we then detail how we can recover the
usual plane-parallel limit, known as the Kaiser formula. The statistics of the fundamental field $\delta$ is known in
Fourier space, as it is given by the matter power spectrum defined by
\be\label{DefPk}
\Ckk(\gr{k}_1,\gr{k}_2)\equiv \langle \delta(\gr{k}_1) \delta^{\star}(\gr{k}_2) \rangle = \Pow(k) \delta_D(\gr{k}_1 - \gr{k}_2)\,,
\ee
\modif{where the $\delta(\gr{k})$ are the Fourier transform of $\delta$
  today.} \modif{They are related to the Fourier modes at an earlier time by
a transfer function as
\be
\delta(\gr{k},t)=G(t)\delta(\gr{k})\,,\qquad G(t_0)=1\,,
\ee
since when radiation can be neglected the modes evolve at the same
pace~\cite{Montanari:2012}.} We first need to obtain the distorted field $\delta^z$ in configuration space as a
function of the underlying density field $\delta$ in Fourier space. Given that the Fourier transform of $\Delta^{-1} \delta$ is
$-\delta(\gr{k})/k^2$, we get immediately from eq.~\eqref{delta_real_rad} that this relation takes the form
\be
\label{Basisdeltaz}
\delta^z(\gr{r}) = G(r)\int \frac{\dd^3 \gr{k}}{(2 \pi)^{3/2}}\delta(\gr{k})\left[1 -
\frac{\beta(r)}{k^2}{\cal O}_r\right]{\rm e}^{\ii \gr{k}\cdot\gr{r}} \, ,
\ee
where \modif{$G(r)\equiv G(t(r))$}. Correlating the distorted field in two different points, we use
eq.~\eqref{DefPk} to remove one of the two Fourier integrals, and the
other one is performed easily in spherical coordinates once the
exponential is expanded in spherical harmonics with the Rayleigh
expansion~\eqref{rayleigh}. We finally obtain
\be
\label{EqBasicO1}
\Crr^z(\gr{r}_1, \gr{r}_2) 
= G_1 G_2 \int \frac{k^2 \dd k}{2 \pi^{2}}\Pow(k)\left[1 - \frac{\beta_1}{k^2}{\cal O}_{r_1}\right]\left[1 -\frac{\beta_2}{k^2}{\cal O}_{r_2}\right]j_0(kr) \,,
\ee
where we used the notation $\beta_1\equiv \beta(r_1)$, $\beta_2\equiv
\beta(r_2)$ and $G_1 \equiv G(r_1)$, $G_2 \equiv G(r_2)$, and we have defined the difference of positions and the associated
norm as
\be
\gr{r} \equiv \gr{r}_2 -\gr{r}_1\,,\qquad r^2 = r_1^2 + r_2^2 - 2 r_1
r_2 \nu_{12} \,,\qquad \nu_{12}\equiv \hat{\gr{r}}_1 \cdot \hat{\gr{r}}_2 \,.
\ee 
In order to express the radial operators ${\cal O}_{r_1}$ and ${\cal O}_{r_2}$, we first use that
\be\label{Or1r2}
\Delta = {\cal O}_{r_1} + \frac{\LLL{\hat{\gr{r}}_1}}{r_1^2} = {\cal O}_{r_2} + \frac{\LLL{\hat{\gr{r}}_2}}{r_2^2} \,.
\ee
We then note that the radial operators are only applied on a function of $r$ in
eq.~\eqref{EqBasicO1}, where $r$, $r_1$ and $r_2$ are in a triangular
configuration. For any function $f(r)$ 
\be 
\label{TriangleOperators}
{\cal O}_{r_1}f(r) = \Delta f(r) - \frac{1}{r_1^2}
\LLL{\hat{\gr{r}}_1} f(r) = {\cal O}_r f(r) - \frac{\LLL{\nu_{12}}}{r_1^2}
f(r)\,, \qquad {\cal O}_{r_2}f(r) =  {\cal O}_r f(r) - \frac{\LLL{\nu_{12}}}{r_2^2} f(r)
\ee
where we have used {\it i)} that for a function of $r$ only,
the Laplacian reduces to ${\cal O}_r$, the partial derivative $\partial_r$ becoming
in that case total derivatives, and {\it ii)} that the angular operators
$\LLL{\hat{\gr{r}}_1}$ and $\LLL{\hat{\gr{r}}_2}$ do not depend on the
azimuthal angle, as they depend only on the polar angle, and thus both reduce to $\LLL{\nu_{12}}$. 

With these relations and eq.~\eqref{sph_bessel_diff_eq} we can recast \eqref{EqBasicO1} as
\bea
\label{Fullxireal1}
&&\Crr^z(\gr{r}_1, \gr{r}_2)= G_1 G_2\int
\frac{k^2 \dd k}{2 \pi^{2}} \Pow(k) I\left[j_0(kr)\right]
\,,\nonumber\\
&& I\left[j_0(kr)\right] \equiv \left[1+\beta_1+\frac{\beta_1 \LLL{\nu_{12}}}{(k r_1)^2}\right]\left[1+\beta_2+\frac{\beta_2 \LLL{\nu_{12}}}{(k r_2)^2}\right]j_0(kr) \, .
\eea
Note that this result could also have been obtained by using the
relation~\eqref{delta_real_ang} between the distorted and the
underlying matter densities. This expression for the correlation function in configuration space is the
first major result of this article. Indeed, it is from this expression
that we will extract the plane-parallel approximation and its corrections due to large angle effects.

\subsubsection{Plane-parallel limit}\label{small_angle_real}
%%%%%%%%%%%%%%%%%%%%%%%%%%%%%%%%%%%%%%%%%%%

In order to have a better insight on the general expression~\eqref{Fullxireal1}, we check in this section that we can obtain from it the standard result in the plane-parallel limit.

\begin{figure}[!htb]
\begin{center}
\includegraphics[scale=0.6]{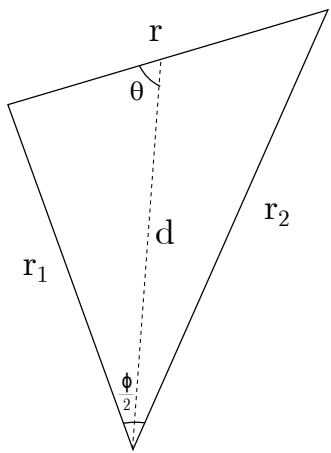}
\end{center}
\caption{Representation of the geometry of the problem. $\gr{r}_1$, $\gr{r}_2$ and ${\bm r}$ form a triangle. $r$ can also be expressed in terms of $r_1$, $r_2$ and the angle $\phi$. $\theta$ is the angle determined by the bisector and $\gr{r}$.}
\label{small}
\end{figure}

Let us first detail the geometry of the problem. The two-point correlation function is defined as a function of the two correlated positions $\gr{r}_1$ and
$\gr{r}_2$, but on its expression~\eqref{Fullxireal1} we realize that
it can also be expressed as $\Crr^z (r_1, r_2, r)$. Indeed, it is only a function of the shape
of the triangle defined by the two sources and the observer, and this
is fully characterized by the length of its three sides. This is because the global rotational invariance of the correlation function has absorbed three out of the six degrees of freedom.
Furthermore, one can equivalently describe such triangle by two sides and an angle,
see figure~\ref{small}. Defining $\phi$ as the angle between the two directions ($\cos \phi
\equiv \nu$), the correlation function can also be expressed in the
form $\Crr^z (r_1, r_2,\phi)=\Crr^z (r_1, r_2, r(r_1, r_2, \phi))$. The
third side length $r$, and the length of the bisector of sources
directions $\dist$ (see figure~\ref{small}) are obtained as
\begin{equation}
r^2(r_1, r_2, \phi) \equiv | \gr{r}_2 - \gr{r}_1 |^2 = (r_1+r_2)^2
\sin^2(\phi/2) + (r_2-r_1)^2 \cos^2(\phi/2) \, .
\end{equation}
In the plane-parallel limit, that is for $\phi \ll 1$, then $\sin^2(\phi/2) \approx \phi^2/4$, and $\cos^2 (\phi/2) \approx 1 - \phi^2/8$. Therefore,
\be
\label{r_phi}
r^2 \approx (r_2 -r_1)^2 + \frac{1}{8} ( r_1^2 + 6 r_1 r_2 + r_2^2) \phi^2 \, .
\ee
We also assume that the two sources are far away compared to their
separation, that is if $r\ll \dist$, then $\dist \approx r_1 \approx
r_2$. \modif{We then approximate $\beta_1 \simeq \beta_2  \simeq
  \beta(\dist)$ and $G_1 \simeq G_2  \simeq G(\dist)$ that we simply note
  $\beta$ and $G$.} From eqs.~\eqref{r_phi} we get immediately $\dd r/\dd \phi \approx ( r_1^2 + 6 r_1 r_2 + r_2^2)\phi/(8r)\approx \dist^2 \phi/r$, from which we deduce
\begin{equation}
\LLL{\nu_{12}} \approx \frac{1}{\phi} \partial_{\phi} \phi \partial_{\phi} \, .
\end{equation}
Instead of using the set of parameters $(r,d,\phi)$ to express the
correlation function in the plane-parallel limit, it proves useful to
keep $r$ and $d$, but to use $\theta$ (or $\mu_{\gr{\dist} \,\gr{r}}\equiv  \cos \theta$) defined as the  angle between $\gr{r}$ and the bisector (see figure~\ref{small}). In the plane-parallel limit, it is
related to $\phi$ by $\dist \phi / r \approx \sin \theta$ and $ \cos
\theta \approx |r_2 - r_1|/r$.

We are now ready to examine the correlation function given by eq.~\eqref{Fullxireal1} in the plane-parallel limit. The term linear in the differential operator $\LLL{\nu}$ is expressed in terms of $j_0''(kr)$ and $j_0'(kr)/(kr)$. Using \eqref{j0_prime_x} this can be simplified as
\be
\label{second_3}
\frac{2}{\dist^2}  \left( \frac{1}{\phi} \partial_{\phi} \phi \partial_{\phi} \right) j_0(kr(\phi))  =
- \frac{4}{3} k^2  \left[ j_0(kr) P_0(\mu_{\gr{\dist} \,\gr{r}}) + j_2(kr) P_2(\mu_{\gr{\dist} \,\gr{r}}) \right]
\ee
where the $P_\ell$ are Legendre polynomials whose explicit forms are
given in eqs.~\eqref{p024}. For the term quadratic  $\LLL{\nu}$ we obtain contributions from $j_0^{(4)}$, $j_0^{(3)}$, $j_0''$, and $j_0'$, where $j_0^{(n)}(kr)$ means the $n$-th derivative of the spherical Bessel function. Using the relations \eqref{j0_prime_x} and \eqref{j3_x}, this contribution can be cast as
\be
\label{third_3}
\frac{1}{\dist^4} \left( \frac{1}{\phi} \partial_{\phi} \phi \partial_{\phi} \right)^2 j_0(kr(\phi))  =  k^4 \left[ \frac{8}{15} j_0(kr) P_0(\mu_{\gr{\dist} \,\gr{r}}) - \frac{16}{21} j_2(kr) P_2(\mu_{\gr{\dist} \,\gr{r}}) + \frac{8}{35} j_4(kr) P_4(\mu_{\gr{\dist} \,\gr{r}})  \right] \, .
\ee
Inserting eq.~\eqref{second_3} and eq.~\eqref{third_3} into
eq.~\eqref{Fullxireal1}, we finally obtain that the plane-parallel limit
of the two-point correlation function in configuration space depends only on the difference of positions between the sources $\gr{r}$. More precisely, there is an axisymmetry around the common line of sight of the sources $\hat{\gr{\dist}}$ (that can be taken as the $z$-axis), so it depends on $r$, the distance between the two sources, and on its orientation with respect to the common line of sight which is given by $\mu$, but not on an azimuthal angle. Indeed, it takes the general form
\be
\label{xi_sa}
\xidr^z_\FS(\gr{\dist},\gr{r}) = \xidr^z_\FS(\gr{r}) \equiv \Crr^z_\FS(\gr{r}_1,\gr{r}_2) = \sum_{\ell=0, 2, 4} \xidr_\ell^{(0)} (r) P_\ell(\mu_{\gr{\dist}\,\gr{r}}) \,,
\ee
where the only non-vanishing coefficients $\xidr^0_\ell (r)$ of this
expansion are directly obtained from eqs.~\eqref{Fullxireal1},~\eqref{second_3}, and~\eqref{third_3} as
\be\label{xi_sa_all}
\xidr_0^{(0)} (r)   =    \left( 1 + \frac{2}{3}  \beta +  \frac{1}{5} \beta^2 \right) \MYXI_0^{0}(r)\,,\quad
\xidr_2^{(0)} (r)   =  -   \left(\frac{4}{3} \beta + \frac{4}{7} \beta^2 \right) \MYXI_2^{0}(r)\,,\quad
\xidr_4^{(0)} (r)   =    \frac{8}{35}  \beta^2 \MYXI_4^{0}(r) \,,
\ee
and with the convenient definition
\be\label{DefXI}
\MYXI_\ell^{m}(r)\equiv G^2\int \frac{k^2\dd k}{2 \pi^2} (kr)^{-m} j_\ell(k r)\Pow(k)\,.
\ee
Eq.~\eqref{xi_sa} is the Kaiser formula in configuration space~\cite{Hamilton:1992ApJ}.
%%%%%%%%%%%%%%%%%%%%%%%%%%%%%%%%%%%%%%%%%%%%%%%%%%%%%
\subsection{Redshift-space distortions in Fourier space}\label{SecFourier}
%%%%%%%%%%%%%%%%%%%%%%%%%%%%%%%%%%%%%%%%%%%%%%%%%%%%%

%%%%%%%%%%%%%%%%%%%%%%%%%%
\subsubsection{General expression}
%%%%%%%%%%%%%%%%%%%%%%%%%%
In this section, we investigate the Fourier conjugate of the objects
given in \eqref{delta_real_ang} and \eqref{EqBasicO1}. We first note
that for a function of $\gr{k}\cdot \gr{r}$ we can relate operators acting on the $\gr{r}$
dependence to operators acting on the $\gr{k}$ dependence. 
Indeed for scalar functions of $\gr{k}\cdot \gr{r}$,
\be
\label{PropertiesCommutekr1}
\frac{{\cal O}_r}{k^2}f(\gr{k}\cdot \gr{r}) =\frac{{\cal
    O}_k}{r^2}f(\gr{k}\cdot \gr{r})\,,\qquad
\frac{\Delta_{\gr{r}}}{k^2}f(\gr{k}\cdot \gr{r}) =
\frac{\Delta_{\gr{k}}}{r^2}f(\gr{k}\cdot \gr{r})\qquad \Rightarrow
\qquad \LLL{\hat{\gr{r}}} f(\gr{k}\cdot \gr{r}) =\LLL{\hat{\gr{k}}} f(\gr{k}\cdot \gr{r})\,.
\ee
Using these results, we find an equivalent expression for eq.~\eqref{delta_real_ang} which is
\bea
\label{fourier_step_1}
\delta^z(\gr{r}) & = & (1+\beta(r)) \delta(\gr{r})+\beta(r) G(r) \int \frac{\dd^3 \gr{k}}{(2 \pi)^{3/2}} \frac{\delta(\gr{k})}{k^2} \frac{\LLL{\hat{\gr{r}}}}{r^2}\left( {\rm e}^{\ii \gr{k}\cdot\gr{r}} \right) \nonumber\\ & = & (1+\beta(r)) \delta(\gr{r})+\beta(r)G(r) \int \frac{\dd^3 \gr{k}}{(2  \pi)^{3/2}} \frac{\delta(\gr{k})}{(kr)^2} \LLL{\hat{\gr{k}}} \left({\rm e}^{\ii  \gr{k}\cdot\gr{r}}\right)\,.
\eea
Once this step has been taken, it is then straightforward to take the
Fourier transform to find a relation between the Fourier components of
$\delta^z$ and $\delta$. \modif{In principle we must only use the property
that the Fourier transform of a product is the convolution of the
Fourier transform. However, our goal here is to understand the
geometrical structures induced by RSD effects in Fourier space, and it will be
much more transparent to consider the unrealistic case where $\beta(r)$ and $G(r)$
can be treated as constants. In that case we simply replace $\beta(r)$
by $\beta$ and $G(r)$ by $1$.} We obtain, that
\be\label{Fourierdz}
\delta^z(\gr{k}) = (1+\beta)\delta(\gr{k})+\beta \int \frac{\dd^3 \gr{p}}{p^2} \delta(\gr{p})\LLL{\hat{\gr{p}}}\left(\int \frac{\dd^3 \gr{r}}{(2\pi)^3}\frac{{\rm e}^{\ii (\gr{p}-\gr{k})\cdot\gr{r}}}{r^2}\right) \, .
\ee
Furthermore, recalling that
\be
\int \frac{\dd^3 \gr{r}}{(2\pi)^3}\frac{{\rm e}^{\ii (\gr{p}-\gr{k})\cdot\gr{r}}}{r^2}=\frac{1}{4\pi|\gr{k}-\gr{p}|} \,\, ,
\ee
we can rewrite this result in a more compact form with an integral on a kernel as
\be
\label{delta_z_k}
\delta^z(\gr{k}) = (1+\beta)\delta(\gr{k})+\beta \int \frac{\dd^3 \gr{p}}{k^2} \delta(\gr{p})\kerK(\gr{k},\gr{p}) \, ,\quad \kerK(\gr{k}, \gr{p}) \equiv  \LLL{\nu_{kp}} \left(\frac{1}{4\pi|\gr{k}-\gr{p}|}
\right)\,,\quad \nu_{kp} \equiv \hat{\gr{k}}\cdot \hat{\gr{p}}\,.
\ee
%\modif{As in the configuration space, we could keed the $k$ dependence of $\hat{\beta}$, but it would not alter the geometrical properties of the problem. In order to simplify the notation we will concentrate on the case $\beta = \mathrm{const}$, but the general case can be directly obtained from our results by replacing $\beta \to \hat{\beta}\star$, and labeling the variables accordingly to the other $k$-dependencies. We will not investigate the effects of the convolution in this paper since our main goal is to explicit the geometrical structure of the RSD problem.}

Correlating a pair of fields using eq. \eqref{delta_z_k} and the statistical properties~\eqref{DefPk} of the $\delta(\gr{k})$, we get
\bea
\label{corr_func_fourier}
\Ckk^z(\gr{k}_1,\gr{k}_2)&=&
(1+\beta)^2 \Pow(k_1)\delta^3_D(\gr{k}_1-\gr{k}_2)+\beta(1+\beta)\left[\frac{\Pow(k_1)}{k_1^2}\kerK(\gr{k}_1,\gr{k}_2)+\frac{\Pow(k_2)}{k_2^2}\kerK(\gr{k}_2,\gr{k}_1)\right]\nonumber\\
&&+\beta^2 \int \dd^3 \gr{p} \frac{\Pow(p)}{p^4} \kerK(\gr{k}_1,\gr{p}) \kerK(\gr{p},\gr{k}_2) \, .
\eea
The kernel $\kerK$ deserves a closer analysis.  An explicit form can be
obtained using the expansion
\be
\label{generating_legendre}
\frac{1}{|\gr{k} - \gr{k}'|} = \sum_\ell \frac{k_<^\ell}{k_>^{\ell+1}} P_\ell(\gr{\hat{k}} \cdot \gr{\hat{k}'} ) \, ,
\ee
where $k_< \equiv \mathrm{min} \{ k, k'\}$ and $k_> \equiv \mathrm{max}
\{ k, k' \}$. We then find
\be
\label{gen_func_der}
\kerK(\gr{k}, \gr{k'}) = \LLL{\nu_{k k'}} \left(\frac{1}{4 \pi |\gr{k} - \gr{k'}|} \right)= \frac{1}{4 \pi} \sum_\ell \frac{k_<^\ell}{k_>^{\ell+1}} \LLL{\nu_{k k'}}  P_\ell(\gr{\hat{k}} \cdot \gr{\hat{k}'} ) = - \frac{1}{4 \pi} \sum_\ell \ell(\ell+1) \frac{k_<^\ell}{k_>^{\ell+1}} P_\ell(\gr{\hat{k}} \cdot \gr{\hat{k}'} ) \, 
\ee
because of the property \eqref{legendre_eq}. The coefficients of the expansion of the kernel $\kerK$ in spherical harmonics are then easily found to be
\be
\label{kerk_l}
\kerK_\ell(k,k') \equiv - \frac{\ell(\ell+1)}{2\ell+1}
\frac{k_<^\ell}{k_>^{\ell+1}} \, ,\qquad \Rightarrow \qquad
\kerK(\gr{k}, \gr{k'}) = \sum_{\ell m}\kerK_\ell(k,k') \mathrm{Y}_{\ell m}(\hat{\gr{k}}) \mathrm{Y}^\star_{\ell m}(\hat{\gr{k}}')\,.
\ee
The most important identities satisfied by $\kerK_\ell$ are presented in appendix~\ref{prop_kernel}. This same kernel was already introduced by \cite{Taylor_Valentine}.

The two point correlation of the density fields in Fourier space is also the Fourier transform of the two point correlation function in configuration space. This means that eq.~\eqref{corr_func_fourier} should be related to eq.~\eqref{Fullxireal1} by a Fourier transformation as
\be
\Crr^z(\gr{r}_1,\gr{r}_2) = \frac{1}{(2 \pi)^3}\int \dd^3 \gr{k}_1 \dd^3 \gr{k}_2 \Ckk^z(\gr{k}_1,\gr{k_2}) {\rm e}^{\ii \gr{k}_1 \cdot \gr{r}_1-\ii \gr{k}_2 \cdot \gr{r}_2}\,.
\ee
In order to check this explicitly, we must proceed as follows. First, from \eqref{corr_func_fourier} we use \eqref{gen_func_der} to decompose the angular dependence in spherical harmonics. After taking the inverse Fourier transform, we need to also decompose the exponentials in spherical waves using the expansion~\eqref{rayleigh}. The angular integrals can be performed
easily using the orthonormality of spherical harmonics. The final angular dependence can be simplified using the addition theorem for Legendre polynomials \eqref{addition_thm}. As for the remaining radial integral, it can be simplified using the relation \eqref{int_kernel_j}. The result is finally recast in the form \eqref{Fullxireal1} if we use the addition property \eqref{j0_gegenbauer} is employed to contract  the spherical Bessel functions into $j_0(kr)$.
%%%%%%%%%%%%%%%%%%%%%%%%%%%%%%%%%%%%%%%%%%%%%%%
\subsubsection{Alternative expression of the correlation function}
%%%%%%%%%%%%%%%%%%%%%%%%%%%%%%%%%%%%%%%%%%%%%%%
It is convenient to define another kernel to recast
eq.~\eqref{corr_func_fourier} in a simpler form. Given that $-\frac{1}{4 \pi |\gr{r} - \gr{r'}|}$ is the Green's function for the Laplace equation in three dimensions, i.e., 
\be
\label{green_gen}
\Delta \left(\frac{-1}{4 \pi | \gr{r} - \gr{r'}|}\right) = \delta_D (\gr{r} - \gr{r'}) \, ,
\ee
and the decomposition~\eqref{DfOL} of the Laplacian, we define a kernel with the radial operator in Fourier space as
\be
\label{kernel_N}
\kerH(\gr{k}_1,\gr{k}_2) \equiv -k_1^2 {\cal O}_{k_1} \left(
\frac{1}{4\pi|\gr{k}_1-\gr{k}_2|} \right)= \kerK(\gr{k}_1,\gr{k}_2) +k_1^2\delta_D(\gr{k_1}-\gr{k}_2) \, .
\ee
With this kernel, the correlation in Fourier space reads simply
\bea
\label{corr_func_fourier_N}
\Ckk^z(\gr{k}_1,\gr{k_2}) & = &
P(k_1)\delta^3_D(\gr{k}_1-\gr{k}_2)+\beta\left[\frac{P(k_1)}{k_1^2}{\kerH}(\gr{k}_1,\gr{k}_2)+\frac{P(k_2)}{k_2^2}{\kerH}(\gr{k}_2,\gr{k}_1)\right] \nonumber\\ & & +\beta^2 \int \dd^3 \gr{p} \frac{P(p)}{p^4} \kerH(\gr{k}_1,\gr{p}) \kerH(\gr{p},\gr{k}_2) \, .
\eea
This could have been obtained directly if we had started with
eq.~\eqref{delta_real_rad} and not from eq.~\eqref{delta_real_ang} as chosen in the previous section.

%%%%%%%%%%%%%%%%%%%%%%%%%%%%%%%%%%%%%
\subsubsection{Plane-parallel limit in Fourier space}
%%%%%%%%%%%%%%%%%%%%%%%%%%%%%%%%%%%%%
\label{small_angle_fourier}

As emphasized in section~\ref{SecIntroFourier}, the Fourier mode associated with the difference of positions is the median Fourier mode, and the Fourier mode associated with the median position is the difference of the Fourier modes, as can be seen on eq.~\eqref{EqCorrespkr}. This is paramount to understand the structure of the Fourier space. It is thus natural to use the variables
\be
\label{p_mean_def}
\gr{k} \equiv \frac{(\gr{k}_1+\gr{k}_2)}{2}\,,\qquad \Delta\gr{k} \equiv \gr{k}_1-\gr{k}_2 \,,
\ee
to parametrize the correlation functions in Fourier space. 
Defining $\mu_{\gr{k \Delta}} \equiv  \hat{\gr{k}} \cdot \widehat{\Delta \gr{k}}$ and $\nu_{12} \equiv \hat{\gr{k}}_1 \cdot \hat{\gr{k}}_2$, they are related through
\begin{subeqnarray}\label{geom_median_pall}
k^2_1 &=& k^2+(\Delta k/2)^2 +k \, (\Delta k) \mu_{\gr{k\Delta}} \slabel{geom_medianp1} \\
k^2_2 &=& k^2+(\Delta k/2)^2 -k \, (\Delta k) \mu_{\gr{k \Delta}} \slabel{geom_medianp2} \\
k_1 k_2 \, \nu_{12} &=& k^2-(\Delta k/2)^2 \slabel{geom_medianp3}\\
k^2_1 k^2_2 (1-\nu_{12}^2) &=& k^2 (\Delta k)^2 (1-\mu^2_{\gr{k\Delta}}) \, . \slabel{geom_medianp4}
\end{subeqnarray}

Since the plane-parallel limit in configuration space is the limit where $r \ll \dist$, then in Fourier space it corresponds to $|\Delta \gr{k}| \gg k$. Following this logic, we need to find the behaviour of eq.~\eqref{corr_func_fourier_N} under this limit. We find in this section that it is of the form
\be\label{FSFourierShouldBe}
\zetaKk^z_\FS(\Delta\gr{k},\gr{k}) \equiv  \Ckk^z_\FS(\gr{k}_1,\gr{k}_2) = \delta_D(\Delta \gr{k}) \zetaKk_\ell^{(0)}(k) + \frac{1}{4\pi|\Delta \gr{k}|^3}\sum_{\ell=2,4} \zetaKk_\ell^{(0)}(k) P_\ell(\mu_{\gr{k \Delta}})\,.
\ee

The kernel $\kerH(\gr{k},\gr{k}')$ appearing in eq.~\eqref{corr_func_fourier_N} can be computed explicitly. However this requires to deal with the singular point $\gr{k}=\gr{k}'$. We thus need to use the relation (see eq. (3.46) of \cite{Blanchet:2003gy})
\be
\label{gelfand_k}
{\cal O}_k \frac{1}{4 \pi | \gr{k} - \gr{k'}|} =  {\cal O}_k^{\rm PV} \frac{1}{4 \pi | \gr{k} - \gr{k'}|} - \frac{1}{3} \delta (\gr{k} - \gr{k'}) \,,
\ee
where the superscript ${\rm PV}$ refers to the principal value. Using the geometrical relations \eqref{geom_median_pall} yields
\be
\label{CleverKernel1}
4 \pi \kerH(\gr{k}_1,\gr{k}_2) = -\frac{2 |\gr{k}|^2}{|\Delta \gr{k}|^3}
P_2(\mu_{\gr{k\Delta}}) +\frac{1}{2 |\Delta \gr{k}|} + 
4 \pi \frac{|\gr{k}|^2}{3} \delta_D(\Delta \gr{k})\,.
\ee
The first two terms are the regular part, and the last term is the distributional component of the kernel. This latter term contributes necessarily to the lowest order, that is to the plane-parallel limit, since it is necessarily diagonal in Fourier space. By dimensional analysis, the plane-parallel limit of the first two terms is necessarily proportional to $1/|\Delta \gr{k}|^3$, and this is another way to realize that the expansion must take the form~\eqref{FSFourierShouldBe}. The plane-parallel limit of the kernel $\overline{\kerH}_{\gr{k}}(\Delta \gr{k}) $ is then found to be
\be\label{ApproxKernel}
4\pi \overline \kerH(\gr{k}_1,\gr{k}_2) = -\frac{2 k^2}{|\Delta \gr{k}|^3}P_2(\mu_{\gr{k\,\Delta}})+\frac{4 \pi k^2}{3} \delta_D(\Delta \gr{k}) \,.
\ee
Using this limit, the plane-parallel limit of the Fourier space correlation~\eqref{corr_func_fourier_N} is
\bea
\label{corr_n_bar}
\zetaKk^z_\FS(\Delta \gr{k},\gr{k})  & = & 
\Pow(k)\delta_D(\Delta \gr{k})+2 \beta\frac{\Pow(k)}{k^2}\overline{\kerH}_{\gr{k}}(\Delta \gr{k}) \nonumber\\ & & + (2 \pi)^{3/2} \beta^2\frac{\Pow(k)}{k^4} \int \frac{\dd^3 \Delta \gr{p}}{(2 \pi)^{3/2}}  \overline{\kerH}_{\gr{k}}(\Delta \gr{p}) \overline{\kerH}_{\gr{k}}(\Delta \gr{k}-\Delta \gr{p})\, .
\eea
It is not expressed in the desired form~\eqref{FSFourierShouldBe} yet, due to the last integral. However if we notice that this is a convolution (defined with the $(2 \pi)^{3/2}$ factor) of a function with itself, then we can directly say that its Fourier transform is the product of the individual Fourier transforms. The inverse Fourier transform of the plane-parallel kernel is just
\be
\label{TFApproxKernel}
\widetilde{\overline{\kerH}}_{\gr{k}}(\gr{\dist}) \equiv \int \frac{ \dd^3 \Delta \gr{p}}{(2 \pi)^{3/2}} \bar{\kerH}_{\gr{k}}(\Delta \gr{p}) {\rm e}^{\ii \Delta \gr{p}\cdot \gr{\dist}}   = \frac{1}{(2 \pi)^{3/2}} \left( \frac{2}{3} k^2 P_2(\mu_{\gr{k}\,\gr{\dist}})+\frac{1}{3}k^2 \right) = \frac{k^2 \,\mu_{\gr{k}\,\gr{\dist}}^2}{(2 \pi)^{3/2}} \,,
\ee
where we used the Rayleigh formula to expand the exponential, and Weber integrals (see appendix~\ref{AppWeber}) for the radial integration on spherical Bessel functions. Finally, the coefficients of the plane-parallel limit of the Fourier space correlation function, as defined in eq.~\eqref{FSFourierShouldBe}, are

\be\label{zetaFS}
\zetaKk_0^{(0)}(k) = \left(1+\frac{2}{3}\beta+\frac{1}{5}\beta^2\right)\Pow(k)\,,\quad \zetaKk_2^{(0)}(k) = \left(-4\beta-\frac{12}{7}\beta^2\right)\Pow(k)\,,\quad \zetaKk_4^{(0)}(k) = \frac{12}{7}\beta^2\Pow(k)\,.
\ee
This result {\it is not the Kaiser formula} in Fourier space as originally derived in~\cite{Kaiser:1987qv}. In fact in the plane-parallel limit of the correlation function in Fourier space given by eq.~\eqref{FSFourierShouldBe} whose coefficients are given in eq.~\eqref{zetaFS}, there are still mode couplings. However, the Kaiser equation holds for independent Fourier modes, and is diagonal. It is thus clear that a further approximation needs to be performed to obtain an equation involving uncoupled modes. In fact, in order to find the plane-parallel result~\eqref{zetaFS}, one had to consider the configuration space part of the plane-parallel kernel in eq.~\eqref{TFApproxKernel}, using the median distance $\gr{d}$ rather than the difference of Fourier modes $\Delta \gr{k}$. This means that in order to find the plane-parallel expansion in Fourier space, we had to go through a mixed configuration/Fourier space temporarily. We shall find that the Kaiser formula arises naturally in this mixed space. And as mentioned in section~\ref{SecIntroMixed}, once the coefficients of the double expansion in this mixed space are known, the coefficients of the double expansion $\zetaKk_\ell^n$ in the Fourier space can be deduced extremely easily, since they are related by simple numerical factors. The next section is devoted to the computation of the plane-parallel approximation in this mixed spaced, allowing then to explain how the Kaiser formula arises in the subsequent section.

%%%%%%%%%%%%%%%%%%%%%%%%%%%%%%%%%%%%%%%%%%%%%%%%%%%%%%
\subsubsection{Plane-parallel limit in mixed configuration/Fourier space}\label{small_angle_mixed}
%%%%%%%%%%%%%%%%%%%%%%%%%%%%%%%%%%%%%%%%%%%%%%%%%%%%%%

The plane-parallel limit of the correlation function in the mixed space is obtained by inverse transforming the Fourier space correlation~\eqref{corr_n_bar}. With the help of the Fourier transformed plane-parallel Kernel~\eqref{TFApproxKernel}, we get
\bea
\zetadk^z_\FS(\gr{\dist},\gr{k}) &\equiv& \int \frac{ \dd^3 \Delta \gr{k}}{(2 \pi)^{3/2}} \zetaKk^z_\FS(\Delta \gr{k},\gr{k}) {\rm e}^{\ii \Delta \gr{k}\cdot \gr{\dist}} \\
&=&\frac{\Pow(k)}{(2 \pi)^{3/2}} (1 + \beta\mu_{\gr{k}\,\gr{\dist}}^2 )^2 \, . \label{almost_kaiser}
\eea
The angular dependence of this expression can be decomposed onto Legendre polynomials in the form
\be
\label{zpk_sum}
\zetadk^z_\FS(\gr{\dist},\gr{k}) = \sum_{\ell=0, 2, 4} \Pow_\ell^{(0)}(k) P_\ell(\mu_{\gr{k} \, \gr{\dist}})\,,
\ee
with coefficients

\begin{displaymath}
\Pow_0^{(0)} (k) =  \left(  1 + \frac{2}{3}  \beta +  \frac{1}{5} \beta^2 \right) \frac{\Pow(k)}{(2 \pi)^{3/2}} \,, \quad
\Pow_2^{(0)}(k) =    \left(\frac{4}{3} \beta + \frac{4}{7} \beta^2 \right) \frac{\Pow(k)}{(2 \pi)^{3/2}} \,, 
\end{displaymath}
\begin{equation}
\label{zpk_coeff}
\Pow_4^{(0)}(k) =   \frac{8}{35} \beta^2 \frac{\Pow(k)}{(2 \pi)^{3/2}} \, .
\end{equation}

Alternatively, one could also obtain the plane-parallel limit in the mixed space by taking the Fourier transform of the configuration space correlation function~\eqref{xi_sa} in that same limit, that is with
\be
\xidk^z_\FS(\gr{\dist},\gr{k}) \equiv \int \frac{\dd^3 \gr{r}}{(2\pi)^{3/2}} \xidr^z_\FS(\gr{\dist},\gr{r}){\rm e}^{-\ii \gr{k}\cdot \gr{r}} \,.
\ee
Expanding the exponential in Legendre polynomials with the Rayleigh expansion~\eqref{rayleigh}, and then performing the integral on the polar angle using the orthogonality relations of Legendre polynomials~\eqref{OrthoPl}, we get
\begin{equation}
\label{fourier_hankel}
\xidk^z_\FS(\gr{\dist},\gr{k}) =  \sum_{\ell=0, 2, 4} \left[ \sqrt{\frac{2}{\pi}} (-i)^\ell \int \dd r r^2 j_\ell(kr) \xidr_\ell^{(0)} (r) 
\right] P_\ell(\mu_{\gr{k} \, \gr{d}}) \, .
\end{equation}
The object inside squared brackets in this expression is the \emph{Hankel transform} of the $\xi_\ell^0(r)$.  Inserting eqs.~\eqref{xi_sa_all} in \eqref{fourier_hankel}, and using the orthogonality of spherical Bessel functions~\eqref{orth_j}, we can check that
\be
\xidk^z_\FS(\gr{\dist},\gr{k}) = \zetadk^z_\FS(\gr{\dist},\gr{k})\,,
\ee 
meaning that we can obtain correlation functions in this mixed configuration/Fourier space in two different, but equivalent ways. Either we Fourier transform $\xi^z(\gr{\dist}, \gr{r})$ on the second variable obtaining $\xidk^z(\gr{\dist},\gr{k})$, or we take and inverse Fourier transform on $\zeta^z(\Delta \gr{k}, \gr{k})$ on the first variable obtaining $\zetadk^z(\gr{\dist},\gr{k})$. These two objects are intermediate steps between full configuration or full Fourier space quantities, and we have explicitly verified here their equivalence in the plane-parallel limit. 

\begin{figure}[!htb]
\begin{center}
\includegraphics[scale=0.5]{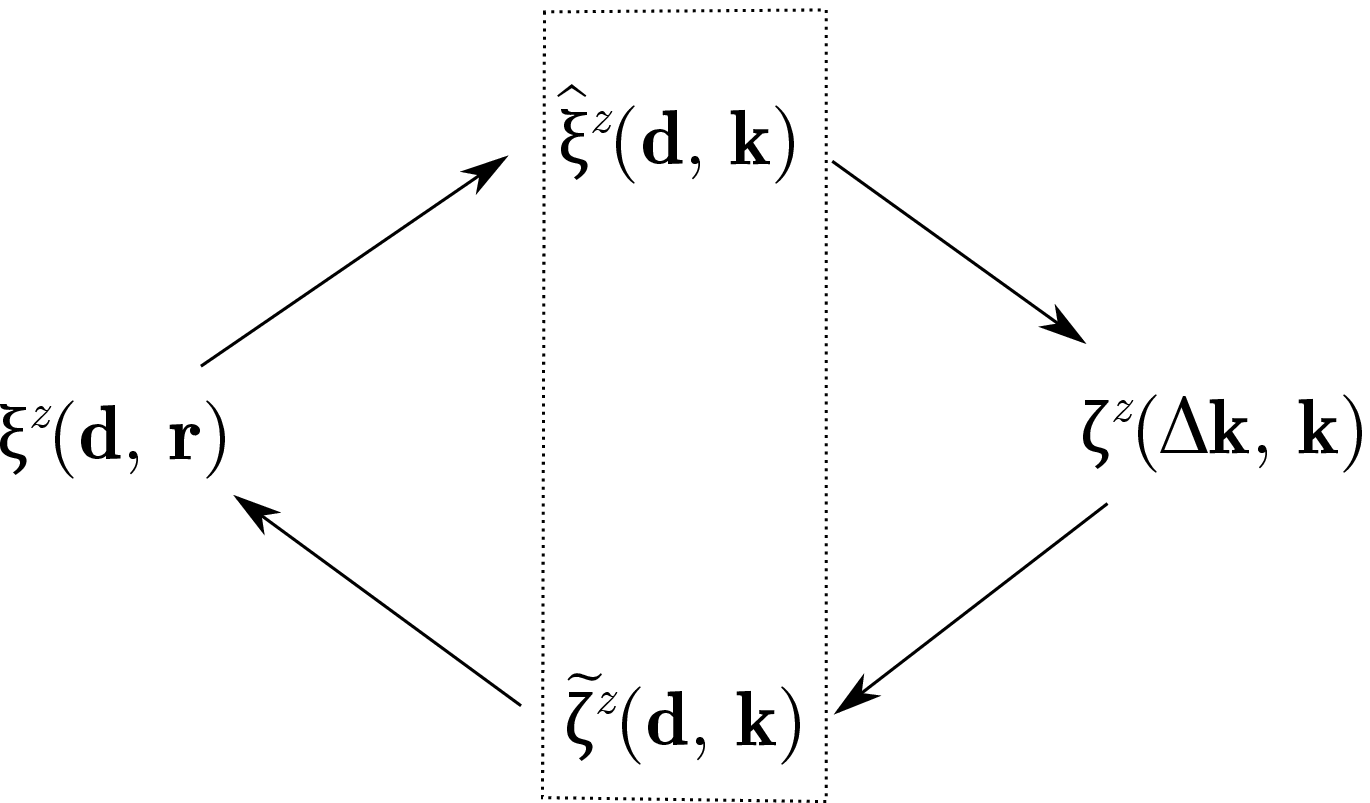}
\end{center}
\caption{Relations between $\xi^z(\gr{\dist}, \gr{r})$, $\zeta^z(\Delta \gr{k}, \gr{k})$, $\xidk^z(\gr{\dist},\gr{k})$, and $\zetadk^z(\gr{\dist},\gr{k})$. The hat denotes Fourier transform on the second variable, a tilde denotes (inverse) Fourier transform on the first variable. Arrows pointing to the right indicate Fourier transforms while arrows pointing to the left indicate inverse Fourier transforms. The objects inside the box are equivalent.}
\label{box2}
\end{figure}

%%%%%%%%%%%%%%%%%%%%%%%%%%%%%%%%%%%%%%%%%%%%%%%%%%%%%%
\subsubsection{Recovering the original Kaiser formula}\label{SecKaiserOriginal}
%%%%%%%%%%%%%%%%%%%%%%%%%%%%%%%%%%%%%%%%%%%%%%%%%%%%%%

The first method to obtain the Kaiser formula consists in making the substitution $\mu_{\gr{k}\,\gr{\dist}} \to \mu_{\gr{k}\,\gr{z}}$ in the plane-parallel limit~\eqref{almost_kaiser} of the correlation function expressed in the mixed configuration/Fourier space, i.e. we consider that the geometrical dependence is a function of the angle determined by $\gr{k}$ and the $z$ axis. This particular $\gr{z}$ direction is chosen as being some average direction of the survey considered. With this extra approximation, we can take the inverse Fourier transform to obtain
\be
\label{kaiser}
\zetaKk^z_\FS(\Delta\gr{k}, \gr{k})  \approx \zetaKk^z_\Kai(\Delta\gr{k}, \gr{k})   = \delta_D(\Delta\gr{k})\Pow(k) (1 + \beta \mu_{\gr{k}\,\gr{z}}^2 )^2  = \delta_D(\Delta\gr{k})  \sum_{\ell=0, 2, 4} \Pow_{\ell}^{(0)} (k) P_\ell(\mu_{\gr{k} \, \gr{z}})
\ee
which is now the Kaiser limit as originally derived~\cite{Kaiser:1987qv}. The main consequence of the Kaiser approximation, is that the correlation function is now perfectly diagonal, as it would be for a homogeneous distribution, but it still keeps a directional dependence.

However, it is not a well defined object in Fourier space, since the angle $\mu_{\gr{k} \, \gr{z}}$ makes reference to the direction of the $z$ axis, defined in configuration space, and for which there is no unambiguous definition. The expression~\eqref{almost_kaiser} now gives a clear meaning to eq.~\eqref{kaiser}. Indeed, only the power-spectrum at a given distance $\dist$ is well defined, and it has by definition a mixed dependence on variables in configuration and Fourier spaces. The crossed conjugation of variables in configuration and Fourier spaces allows to use geometrical arguments to define a plane-parallel limit inside the framework of Fourier space correlations, but still referring to the position space.

Since we have also shown in the previous section that the mixed space expression can be obtained from the configuration space correlation function, the second method to obtain the Kaiser formula consists in performing the replacement $\mu_{\gr{r}\,\gr{\dist}} \to \mu_{\gr{r}\,\gr{z}}$ in the expression~\eqref{xi_sa} of $\xidr^z_\FS(\gr{\dist},\gr{r})$. Again the direction $\gr{z}$ is ambiguously defined, and is usually some average direction in the survey. In that case the dependence in $\gr{\dist}$ drops out, that is we use the approximation
\be\label{dzinReal}
\xidr^z_\FS(\gr{\dist},\gr{r}) \approx \xidr^z_\Kai(\gr{r}) =\sum_{\ell=0, 2, 4} \xidr_\ell^{(0)} (r) P_\ell(\mu_{\gr{\dist} \,\gr{z}})\,.
\ee
When going to the full Fourier space we get
\bea
\zetaKk^z_\Kai(\Delta\gr{k}, \gr{k}) & = & \int \frac{\dd^3 \gr{r}_1}{(2 \pi)^{3/2}} \int \frac{\dd^3 \gr{r}_2}{(2 \pi)^{3/2}} \xidr^z_\Kai(\gr{r}) \mathrm{e}^{ \ii \gr{k}_1 \cdot \gr{r}_1 - \ii \gr{k}_2 \cdot \gr{r}_2} \nonumber\\& = & \int \frac{\dd^3 \gr{\dist}}{(2 \pi)^{3/2}} \mathrm{e}^{ \ii \gr{\dist} \cdot \Delta \gr{k}} \int \frac{\dd^3 \gr{r}}{(2 \pi)^{3/2}} \mathrm{e}^{ - \ii \gr{k} \cdot \gr{r}} \xidr^z_\Kai(\gr{r})  \nonumber \\ & = & (2 \pi)^{3/2} \delta_D(\Delta\gr{k}) \int \frac{\dd^3 \gr{r}}{(2 \pi)^{3/2}} \xi^z_\Kai(\gr{r}) \mathrm{e}^{- \ii \gr{k} \cdot \gr{r}} \,,
\eea
and with the expansion~\eqref{dzinReal}, we also recover the Kaiser limit~\eqref{kaiser}.

%%%%%%%%%%%%%%%%%%%%%%%%%%%%%%%%%%%%%%%%%%%%%%
\subsection{Multipole decompositions and angular correlations}
\label{SecMultipoles}
%%%%%%%%%%%%%%%%%%%%%%%%%%%%%%%%%%%%%%%%%%%%%%

When calculating the effect of RSD on the correlation function in Fourier space, we derived that $C^z(\gr{k}_1, \gr{k}_2)$ is given in terms of the kernel $\kerK(\gr{k}_1, \gr{k}_2)$ by eq.~\eqref{corr_func_fourier}, and also that the kernel $\kerK$ admits a decomposition in a basis of spherical harmonics as presented in eq.~\eqref{kerk_l}. It is natural to ask if $C^z(\gr{k}_1, \gr{k}_2)$ inherits the decomposition property of the kernel $\kerK$. We shall show that this is precisely the case, and after defining the multipoles in the next section, we derive in section~\ref{SeclmFourier} the coefficients $C^z_{\ell}(k_1, k_2)$ of the multipoles correlations. Furthermore, we will show that the kernel $\kerK$ is also present on the expression of the distorted matter density field in configuration space, and we obtain the multipoles correlations $C^z_{\ell}(r_1, r_2)$ in section~\ref{SeclmReal}.

\subsubsection{Angular multipoles and Hankel transformations}

We decompose the distorted and underlying matter density fields on spherical harmonics basis as
\begin{align}\label{decomp_lm_delta}
\delta(\gr{k}) &= \sum_{\ell m}\delta_{\ell m}(k) \mathrm{Y}_{\ell m}(\hat{\gr{k}}) \,,
&\delta^z(\gr{k}) = \sum_{\ell m}\delta^z_{\ell m}(k) \mathrm{Y}_{\ell m}(\hat{\gr{k}})\,,\\
\delta(\gr{r}) &= \sum_{\ell m}\delta_{\ell m}(r) \mathrm{Y}_{\ell m}(\hat{\gr{r}}) \,,
&\delta^z(\gr{r}) = \sum_{\ell m}\delta^z_{\ell m}(r) \mathrm{Y}_{\ell m}(\hat{\gr{r}})\,.
\end{align}
From the statistic of the underlying density field~\eqref{DefPk}, we find that the associated multipoles statistics is simply
\be\label{EqStatMultipoles}
\langle \delta_{\ell m}(k) \delta_{\ell' m'}(k') \rangle  = \delta_{\ell \ell'} \delta_{m m'} \frac{\delta_D(k-k')}{k^2} \Pow(k)\,.
\ee
Furthermore, using the Rayleigh expansion~\eqref{rayleigh}, we find that the multipoles in configuration and Fourier spaces are related by a Hankel transformation
\be\label{DefHankel}
\delta_{\ell m}(k)=\sqrt{\frac{2}{\pi}} (-i)^\ell \int \dd r r^2 j_\ell(kr) \delta_{\ell m}(r) \,,\qquad
\delta_{\ell m}(r)=\sqrt{\frac{2}{\pi}} i^\ell \int \dd k k^2 j_\ell(kr) \delta_{\ell m}(k)
\ee
with similar relations between the distorted fields multipoles $\delta^z_{\ell m}(k)$ and $\delta^z_{\ell m}(r)$.
%%%%%%%%%%%%%%%%%%%%%%%%%%%%%%%%%%%%%%%%%%%
\subsubsection{Correlations of multipoles on Fourier space}
\label{SeclmFourier}
%%%%%%%%%%%%%%%%%%%%%%%%%%%%%%%%%%%%%%%%%%%
From eq.~\eqref{delta_z_k} and eq.~\eqref{gen_func_der} we find that the Fourier space multipoles are related by
\be
\label{delta_lm_k}
\delta^z_{\ell m}(k)  =  \left(1+\beta \right)\delta_{\ell m}(k) + \beta \int \dd k' \kerK_\ell(k,k') \delta_{\ell m}(k')  \, .
\ee
Because of the global rotational invariance, the correlation of multipoles is necessarily diagonal, even with the RSD effects. In particular, the correlations of Fourier space multipoles are of the form
\be
\langle \delta^z_{\ell m}(k)\delta^z_{\ell' m'}(k') \rangle= \delta_{\ell \ell'} \delta_{m m'} C^z_\ell(k,k')\,.
\ee
The $C^z_\ell(k,k')$ are related to the Fourier space correlation function thanks to
\be\label{C_l_k}
\Ckk^z(\gr{k}_1,\gr{k}_2)  =  \frac{1}{4\pi} \sum_\ell (2 \ell + 1) C_\ell^z(k_1, k_2) P_\ell (\hat{\gr{k}}_1 \cdot \hat{\gr{k}}_2) =\sum_{\ell m} C_\ell^z(k_1, k_2) \mathrm{Y}_{\ell m}(\hat{\gr{k}}_1) \mathrm{Y}^\star_{\ell m}(\hat{\gr{k}}_2) \, .
\ee
Inserting eq. \eqref{gen_func_der} into eq. \eqref{corr_func_fourier} we obtain, after integrating angular dependences,
\bea
\label{Clpp3}
C^z_\ell(k_1,k_2) &=& \left( 1+ \beta \right)^2\delta_D(k_1- k_2)\frac{\Pow(k_1)}{k_1^2}+\beta \left( 1+\beta \right) {\kerK}_\ell(k_1,k_2)\left[\frac{\Pow(k_1)}{k_1^2}+\frac{\Pow(k_2)}{k_2^2}\right] \nonumber\\ & & +\beta^2 \int \dd p \frac{\Pow(p)}{p^2} {\kerK}_\ell(k_1,p) \kerK_\ell(p, k_2) \nonumber\\ 
&=&  \frac{\Pow(k_1)}{k_1^2} \delta(k_1 - k_2) + \beta \kerH_\ell(k_1, k_2) \left( \frac{\Pow(k_1)}{k_1^2} + \frac{\Pow(k_2)}{k_2^2} \right) \nonumber\\ & & + \beta^2 \int \dd p \frac{\Pow(p)}{p^2} \kerH_\ell (k_1, p) \kerH_\ell(p, k_2) \, ,
\eea
where we defined a new kernel
\be
\label{comp_kers}
\kerH_\ell(k_1, k_2) \equiv \kerK_\ell(k_1, k_2) + \delta_D(k_1, k_2) \, .
\ee
Note also that eq.~\eqref{Clpp3} could be obtained directly from \eqref{delta_lm_k} using the statistics~\eqref{EqStatMultipoles}.
%%%%%%%%%%%%%%%%%%%%%%%%%%%%%%%%%%%%%%%%%%%%%
\subsubsection{Correlations of configuration space multipoles}
\label{SeclmReal}
%%%%%%%%%%%%%%%%%%%%%%%%%%%%%%%%%%%%%%%%%%%%%

We have obtained the relation~\eqref{delta_real_ang} from physical
arguments, and derived its Fourier counterpart~\eqref{delta_z_k}. We can, however, use eqs.~\eqref{delta_real_ang},~\eqref{TriangleOperators} and~\eqref{green_gen} to write
\be
\label{delta_z_r}
\delta^z(\gr{r}) = (1+\beta(r))\delta(\gr{r}) + \frac{\beta(r)}{r^2}\int \dd^3 \gr{r}'
\delta(\gr{r}') \kerK(\gr{r},\gr{r}') \, .
\ee
It is striking that the same kernel appears, both in the configuration and
Fourier spaces relations between the distorted and underlying matter
density fields. From eq.~\eqref{delta_z_r} we find that the configuration space multipoles are related thanks to
\be
\label{delta_lm_r}
\delta^z_{\ell m}(r) = \left(1+\beta(r)\right)\delta_{\ell m}(r) + \frac{\beta(r)}{r^2}
\int \dd r' r'^2 \kerK_\ell(r,r') \delta_{\ell m}(r')  \, .
\ee
Alternatively, in the case where $\beta(r)$ is approximated by a
constant function $\beta$ as we did when analyzing the Fourier space structure, this relation could have been obtained using a Hankel
transformation~\eqref{DefHankel} on eq.~\eqref{delta_lm_k}. This would only require to use eq.~\eqref{intjj} to express $\kerK_\ell(r, r')$.

Thanks to global rotational invariance, the correlation of configuration space multipoles is necessarily diagonal and thus takes the form
\be
\langle \delta^z_{\ell m}(r) \delta^z_{\ell' m'}(r') \rangle = \delta_{\ell \ell'} \delta_{m m'}C_\ell^z(r,r')\,.
\ee
The $C_\ell^z(r,r')$ are related to the configuration space correlation by
\be
\Crr^z(\gr{r}_1, \gr{r}_2) = \sum_{\ell m} C_\ell^z(r_1, r_2) \mathrm{Y}_{\ell m}(\hat{\gr{r}}_1) \mathrm{Y}^\star_{\ell m}(\hat{\gr{r}}_2) \, .
\ee
Using \eqref{j0_gegenbauer} in \eqref{Fullxireal1} the angular dependence can be easily integrated, and this leads to
\be
\label{Cl_real_v2}
C_{\ell}^z(r_1,r_2) = G_1 G_2\int k^2 \dd k\frac{2}{\pi} \Pow(k) \left[1 - \frac{\beta_1}{k^2}{\cal O}_{r_1}\right] \left[1 -\frac{\beta_2}{k^2}{\cal O}_{r_2}\right] j_{\ell}(kr_1) j_{\ell}(k r_2) \, .
\ee
This agrees with the $C_\ell$s given in \cite{Bonvin:2011bg} once restricted to RSD effects in wide angles. 

Using \eqref{sph_bessel_diff_eq}, it is easy to see that \eqref{Cl_real_v2} can also be written under the alternative form
\be
\label{Cl_real_v1}
C_{\ell}^z(r_1,r_2) = G_1 G_2\int k^2 \dd k\frac{2}{\pi} P(k) \left[1+\beta_1 - \beta_1\frac{\ell(\ell+1)}{(k r_1)^2} \right]  j_{\ell}(k r_1) \left[1+\beta_2 - \beta_2\frac{\ell(\ell+1)}{(k r_2)^2} \right] j_{\ell}(k r_2) \, .
\ee
\modif{Note that in the case where $\beta_1$ and $\beta_2$ can be
approximated by constants (and $G_1=G_2=1$),} then $C_\ell^z(r_1,
r_2)$ could also be obtained from $C_\ell^z(k_1, k_2)$ given by
eq.~\eqref{Clpp3} thanks to a double Hankel transformation, if we use eqs.~\eqref{intjj} and \eqref{int_kernel_j}. Alternatively, it is possible to obtain $C_\ell^z(r_1, r_2)$ directly from $\delta_{\ell m}^z(r)$. We need only to take the spherical harmonics decomposition of eq.~\eqref{Basisdeltaz} to obtain
\be
\label{delta_lm_r_bis}
\delta_{\ell m}^z(r) = G(r)\sqrt{\frac{2}{\pi}} \ii^\ell \int \dd k k^2 \delta_{\ell m}(k) \left[1- \frac{\beta(r)}{k^2} \mathcal{O}_r \right] j_\ell(kr) \, .
\ee
Equation~\eqref{Cl_real_v2} is then obtained easily if we use the statistics~\eqref{EqStatMultipoles}.

%%%%%%%%%%%%%%%%%%%%%%%%%%%%%%%%%%%%
\subsection{Representation of all derivation paths}
%%%%%%%%%%%%%%%%%%%%%%%%%%%%%%%%%%%%

\begin{figure}[!htb]
\begin{center}
\includegraphics[scale=0.4]{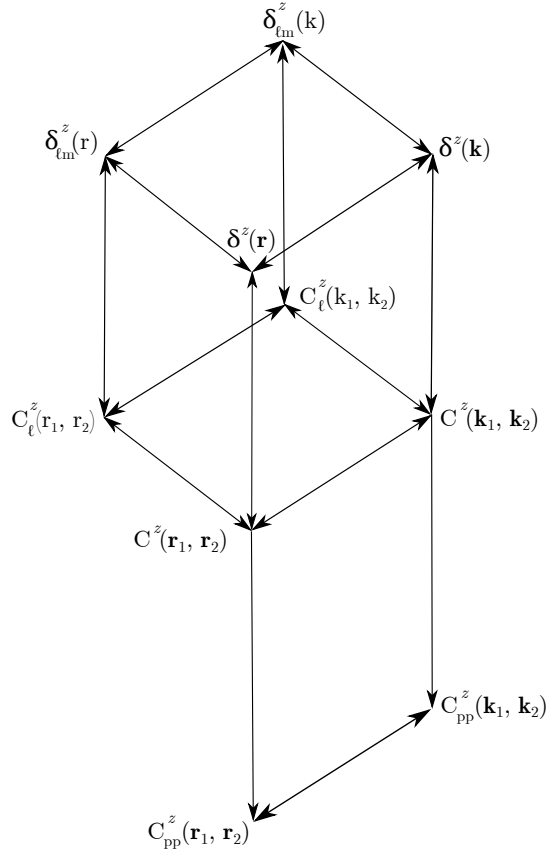}
\end{center}
\caption{Diagram representing the relations among quantities in configuration space, Fourier space, and their counterparts in harmonic space. We represent the various ways in which these objects are related, indicating invertible relations by double headed arrows. Plane-parallel limits are indicated with simple arrows.}
\label{box}
\end{figure}

We can perform a diagrammatic representation of all relations obtained so far as in figure~\ref{box}, so as to summarize the results obtained.

\begin{itemize}
\item In section~\ref{basics} we have obtained the general expression \eqref{delta_real_ang} of the distorted field $\delta^z(\gr{r})$ from the underlying field due to RSD effects.
\item This allowed to compute the correlation function in configuration space $\Crr^z(\gr{r}_1,\gr{r}_2)$ (one southbound step in the diagram) in eqs.~\eqref{EqBasicO1} and \eqref{Fullxireal1}.
\item In section~\ref{small_angle_real}, the plane-parallel limit $C_\FS^z(\gr{r}_1, \gr{r}_2) = \xidr_\FS^z(\gr{r})$ has then been obtained (another southbound step in the diagram), in eq.~\eqref{xi_sa}.
\item In section~\ref{SecFourier}, we obtained the general expression for the density field in Fourier space $\delta^z(\gr{k})$ in eq.~\eqref{delta_z_k} (one step in the north-east direction from $\delta^z(\gr{r})$).
\item The correlation function in Fourier space was then derived in eq.~\eqref{corr_func_fourier} (one southbound step).
\item The Fourier space plane-parallel limit was derived in section~\ref{small_angle_fourier} (bottom line of the diagram) but it was found in Secs.~\ref{small_angle_mixed} and \ref{SecKaiserOriginal} that we need to rely on a mixed configuration/Fourier space to be able to define correctly a Kaiser limit.
\item The angular multipoles were obtained in configuration and Fourier spaces in eqs.~\eqref{delta_lm_r} and \eqref{delta_lm_k} respectively (one step in the north-west direction from $\delta^z(\gr{r})$ and $\delta^z(\gr{k})$).
\item $\delta_{\ell m}^z(r)$ and  $\delta_{\ell m}^z(k)$ are related by Hankel transformations~\eqref{DefHankel}. 
\item The multipoles correlations $C^z_\ell$ were obtained from the correlation functions in configuration and Fourier spaces in eqs.~\eqref{Clpp3} and \eqref{Cl_real_v1} (closing the cube in the diagram from the bottom side), and they are related by Hankel transformations.
\item Furthermore, we explained that they can also be obtained by correlating $\delta_{\ell m}^z(r)$ and  $\delta_{\ell m}^z(k)$, thus closing the cube from the north-west side.
\end{itemize}

%%%%%%%%%%%%%%%%%%%%%%%%%%%%%%%%%%%%%%%%%
\section{Wide angle effects and plane-parallel corrections}
%%%%%%%%%%%%%%%%%%%%%%%%%%%%%%%%%%%%%%%%%

The remainder of this article is dedicated to the analysis of the corrections that should be added to the plane-parallel correlators to recover their general expressions, so as to take into account wide angle effects. We will thus explore in more details the bottom part of the diagram in figure \ref{box}. After expressing the general correlation function in a much more convenient form in the next section, we obtain the plane-parallel corrections in configuration space in section~\ref{SecWideAngleReal}. We then deduce the corrections in the mixed configuration/Fourier space in section~\ref{kaiser_fourier}, and then we explain that this is enough to obtain the plane-parallel corrections in the full Fourier space in section~\ref{SecDiscussion}.

%%%%%%%%%%%%%%%%%%%%%%%%%%%%%%%%%%%%%%%%%%%%%%%
\subsection{Improved correlation function in configuration space}
\label{SecNiceCorrelationReal}
%%%%%%%%%%%%%%%%%%%%%%%%%%%%%%%%%%%%%%%%%%%%%%%

The expression~\eqref{Fullxireal1} is general as it includes all the wide angle effects in configuration space. We now explicit the differential operators $\LLL{\nu}$ and $\LLL{\nu} \circ \LLL{\nu}$ (noting $\nu_{12}$ as $\nu$ in this section for simplicity) which appear in $I[j_0(kr)]$ so as to obtain an expression for which the corrections to the plane-parallel limit are more easily obtained. Defining
\be 
{\cal D}^n \equiv \left(-r_1 r_2\right)^n\left(\frac{1}{r}\frac{\dd}{\dd r}\right)^n \,,\qquad h_n \equiv \left(\frac{k r_1 r_2}{r}\right)^n j_n(kr)\,,
\ee
we can show that the operators take the form 
\bea
\LLL{\nu} &=& (1-\nu^2){\cal D}^2  -2 \nu{\cal D}\,,\\
\LLL{\nu} \circ \LLL{\nu} &=& (1-\nu^2)^2{\cal D}^4
-8\nu(1-\nu^2){\cal D}^3+(14\nu^2-6){\cal D}^2 + 4 \nu {\cal D} \,.
\eea
The differential operators ${\cal D}^n$ are closely related to those appearing in the expressions~\eqref{rodrigues} for the $j_\ell$. Therefore, 
\bea
\LLL{\nu}  j_0(kr)&=& (1-\nu^2)h_2  -2 \nu h_1\\
\LLL{\nu} \circ \LLL{\nu}  j_0(kr)&=& (1-\nu^2)^2 h_4 -8 \nu(1-\nu^2)h_3 +(14 \nu^2-6) h_2+4 \nu h_1 \, .
\eea
Using these results, we find that
\modif{\bea
\label{I_h}
I[j_0(kr)]&=&(1+\beta_1)(1+\beta_2) h_0 +
\left(\frac{\beta_1(1+\beta_2)}{(kr_1)^2}+\frac{\beta_2(1+\beta_1)}{(kr_2)^2}\right)\left[(1-\nu^2)h_2-2 \nu h_1 \right]\nonumber\\
&&+\frac{\beta_1 \beta_2}{(kr_1)^2 (kr_2)^2}\left[(1-\nu^2)^2 h_4-8 \nu (1-\nu^2)h_3 +(14 \nu^2-6)h_2+4 \nu h_1\right] \, .
\eea}
Using eqs.~\eqref{j2x2} it is then immediate to obtain an expression involving only $j_0$, $j_2$, and $j_4$
%\bea
%\label{GenCorrFunc}
%I[j_0(kr)] &=& j_0(kr) \left[ (1+\beta)^2 -\frac{2}{3} \beta(1+\beta) \left(\frac{r_1}{r_2}+\frac{r_2}{r_1}\right) \nu + \beta^2 \frac{14\nu^2-6}{15} \right] \nonumber\\ 
%& & + j_2(kr) \Bigg[  \beta(1+\beta) \left(-\frac{2}{3} \left(\frac{r_1}{r_2} +\frac{r_2}{r_1}\right) \nu +  (1-\nu^2)\frac{r_1^2+r_2^2}{r^2} \right)  \nonumber\\ & & + \beta^2 \left( -\frac{8}{%7}\nu(1-\nu^2)\frac{r_1 r_2}{r^2} +\frac{28\nu^2-12}{21} \right) \Bigg]  \nonumber\\ & & + j_4(kr) \, \beta^2 \left[(1-\nu^2)^2\frac{r_1^2 r_2^2}{r^4}-\frac{8}{7}\nu(1-\nu^2)\frac{r_1 r_2}{r^2}+\frac{14\nu^2-6}{35}\right] \nonumber\\ 
%& & + \beta^2 \frac{4 \nu r^2}{3 r_1 r_2} \left[ \frac{j_0(kr)}{(kr)^2} + \frac{j_2(kr)}{(kr)^2} \right]
%\eea
\modif{\bea
\label{GenCorrFunc}
&&I[j_0(kr)] = \nonumber\\
&&j_0(kr) \left[ (1+\beta_1)(1+\beta_2) -\frac{2 \nu}{3} \left(\frac{r_1}{r_2}\beta_2(1+\beta_1)+\frac{r_2}{r_1}\beta_1(1+\beta_2)\right) + \beta_1\beta_2 \frac{14\nu^2-6}{15} \right] \nonumber\\ 
& & + j_2(kr) \Bigg[ -\frac{2 \nu}{3}
\left(\frac{r_1}{r_2}\beta_2(1+\beta_1)+\frac{r_2}{r_1}\beta_1(1+\beta_2)\right)
+  (1-\nu^2)\left(\frac{r_1^2 \beta_2 (1+\beta_1)+r_2^2
    \beta_1(1+\beta_2)}{r^2} \right)  \nonumber\\ & & \qquad \qquad + \beta_1\beta_2
\left( -\frac{8}{7}\nu(1-\nu^2)\frac{r_1 r_2}{r^2}
  +\frac{28\nu^2-12}{21} \right) \Bigg]  \nonumber\\ 
& & + j_4(kr) \,
\beta_1 \beta_2 \left[(1-\nu^2)^2\frac{r_1^2 r_2^2}{r^4}-\frac{8}{7}\nu(1-\nu^2)\frac{r_1 r_2}{r^2}+\frac{14\nu^2-6}{35}\right] \nonumber\\ 
& & + \beta_1 \beta_2 \frac{4 \nu r^2}{3 r_1 r_2} \left[ \frac{j_0(kr)}{(kr)^2} + \frac{j_2(kr)}{(kr)^2} \right]\,.
\eea}
We prove in appendix~\ref{appendix_papai_szapudi} that
\eqref{GenCorrFunc} coincides with the two point correlation function
given in \cite{Papai:2008bd}. We have already obtained in section~\ref{small_angle_real} that in the plane-parallel limit the geometrical terms multiplying $j_0$, $j_2$ and $j_4$ reduce respectively to terms proportional  to the Legendre polynomials $P_0$, $P_2$, and $P_4$, leading to the Kaiser formula in configuration space. We will now examine the corrections to this plane-parallel limit and show that each of these geometrical terms is a linear combination of Legendre polynomials.

%%%%%%%%%%%%%%%%%%%%%%%%%%%%%%%%%%%%%%%%%%%%%%%
\subsection{Wide-angle effects corrections in configuration space}
\label{SecWideAngleReal}
%%%%%%%%%%%%%%%%%%%%%%%%%%%%%%%%%%%%%%%%%%%%%%%

\subsubsection{Geometry of correlation functions}\label{SecGeoDefv}

The geometry of the RSD effects depends on the shape of the triangle whose sides are $r_1$, $r_2$ and $r$. However there are alternative parametrizations for this triangle. A possible choice of parameterization which is examined in appendix~\ref{appendix_papai_szapudi} is to use two angular variables for the sources directions, and the distance $r$ between them. 

\begin{figure}[!htb]
\begin{center}
\includegraphics[scale=0.6]{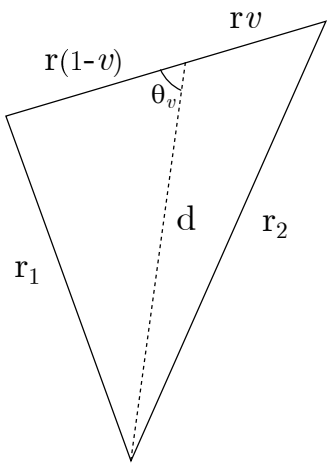}
\end{center}
\caption{Given the triangle with sides $r_1$, $r_2$, $r$, we define a family of parametrizations $r$, $\dist$, $\theta_v$, $0 \leq v \leq 1$. For each value of $v$, $\dist$ is the length of the segment joining the vertex associated to the observer to the opposite side of the triangle, and $\theta_v$ is the angle defined by the intersection of the lines $\dist$ and $r$.}
\label{triangled}
\end{figure}

However, ultimately we want to express our result as a power series in a small parameter which characterizes the departure from the plane-parallel limit. A first choice is to use the {\it median} parametrization
\be
\{r\,,\dist\,, \mu_{\gr{\dist} \, \gr{r}} \}\qquad {\rm with}\qquad\mu_{\gr{\dist} \, \gr{r}} \equiv \hat{\gr{\dist}}\cdot\hat{\gr{r}}\,,\qquad \gr{\dist}\equiv \frac{\gr{r}_1+\gr{r}_2}{2}\,.
\ee
More generally, a family of choices can be parametrized by a parameter $0 \leq v \leq 1$. We define $\dist$ as the distance from the observer's vertex to the point that divide the opposite side of the triangle, that is the line which connects the two sources, in the proportions $v/(1-v)$. This family of choices is illustrated in figure \ref{triangled}. For each possible choice $v$, an angle $\theta_v$ is defined as the angle between $\gr{r}$ and $\gr{\dist}$. In principle one should use the notation $\dist_v$ to specify that the average distance depends on the choice of the parametrization, but since no ambiguity can arise we always note it $\dist$.  If $v=1/2$ we recover the median parametrization.

Using that family of parametrizations, all the geometrical terms in eq. \eqref{GenCorrFunc} can be expressed as a power series in $r/\dist$. In the plane-parallel approximation, that is for distant sources, $\dist \gg r$ and this parameter is indeed small. We further expand the angular dependence of each order in this expansion using a multipole decomposition on Legendre polynomials. The correlation function is thus expanded in a double sum as
\begin{equation}
\label{two_point_power}
\xidr^z (\gr{\dist}, \gr{r}) \equiv  \Crr^z(\gr{\dist} - (1-v) \gr{r},\gr{\dist} + v \gr{r}) =  \sum_{n=0}^{\infty}\left(\frac{r}{\dist} \right)^n\xidr^{z(n)} (\gr{\dist}, \gr{r}) =\sum_{n=0}^{\infty} \left(\frac{r}{\dist} \right)^n \sum_{\ell=0}^{\infty}
\xidr_\ell^{(n)}(r) P_\ell(\mu_{\gr{\dist} \, \gr{r}} ) .
\end{equation}
In the following we consider three parametrizations. The simplest
choice is asymmetric and corresponds to $v = 0$. In that case the
distance to one of the sources is used as an average distance since
$\dist \equiv r_2$. We will show that this choice is not optimal. Then
we consider the median choice $v=1/2$ for which $\dist = |\gr{r}_1 +
\gr{r}_2|/2$. It is the choice described in section~\ref{SecOverview}
and it arises naturally given the structure of the Fourier
transform. Finally we also consider the choice in which $\dist$ is the
length of the line that bisects the angle $\phi$ as it might be the
easiest to implement observationally. \modif{For simplicity, we ignore
  in the remainder the variations of the bias $\beta(r_1)$ and $\beta(r_2)$ and we assume
a constant bias $\beta$. Similarly we neglect the variations of the
transfer fonction $G(r)$ and set $G=1$. This is will simplify
considerably the comparison with the results formulated in Fourier
space and help emphasizing the geometrical aspects of the RSD effects.}
For each case we expand the operator $I[j_0(kr)]$ in powers of $r/\dist$, restricting to second order corrections, as 
\be
I[j_0(kr)]= I^{(0)}[j_0(kr)] + \frac{r}{\dist}I^{(1)}[j_0(kr)]+\left(\frac{r}{\dist}\right)^2I^{(2)}[j_0(kr)] \, .
\ee
Once inserted in eq.~\eqref{GenCorrFunc} it allows to obtain the coefficients $\xidr_\ell^{(n)}(r)$ of our correlation function expansion~\eqref{two_point_power}.

%%%%%%%%%%%%%%%%%%%%%%%%%%%%%%%%%%%%%%
\subsubsection{Asymmetric parametrization ($v=0$)}
%%%%%%%%%%%%%%%%%%%%%%%%%%%%%%%%%%%%%%

From basic triangle geometry, we obtain
\be
r_1= r_2 \sqrt{1 - 2 \frac{r}{r_2} \mu + \left(\frac{r}{r_2} \right)^2} \, ,\qquad
\nu = \frac{1 - \frac{r}{r_2} \mu}{ \sqrt{1 - 2 \frac{r}{r_2} \mu + \left(\frac{r}{r_2} \right)^2}} \, ,
\qquad
1-\nu^2 = \frac{\left(\frac{r}{r_2} \right)^2 (1 - \mu^2)}{1 - 2 \frac{r}{r_2} \mu + \left(\frac{r}{r_2} \right)^2} \,\, ,
\ee
where we abbreviate $\mu_{\gr{\dist} \, \gr{r}}=\mu$. We first expand these relations in powers of $r/d=r/r_2$. We then need to insert them in the geometrical coefficients of eq.~\eqref{GenCorrFunc}. The relevant terms are collected in Table~\ref{triangle1}.

\begin{table}[!htb]
\begin{center}

\ifnotcompact

\begin{tabular}{c|c}
& \includegraphics[scale=0.4]{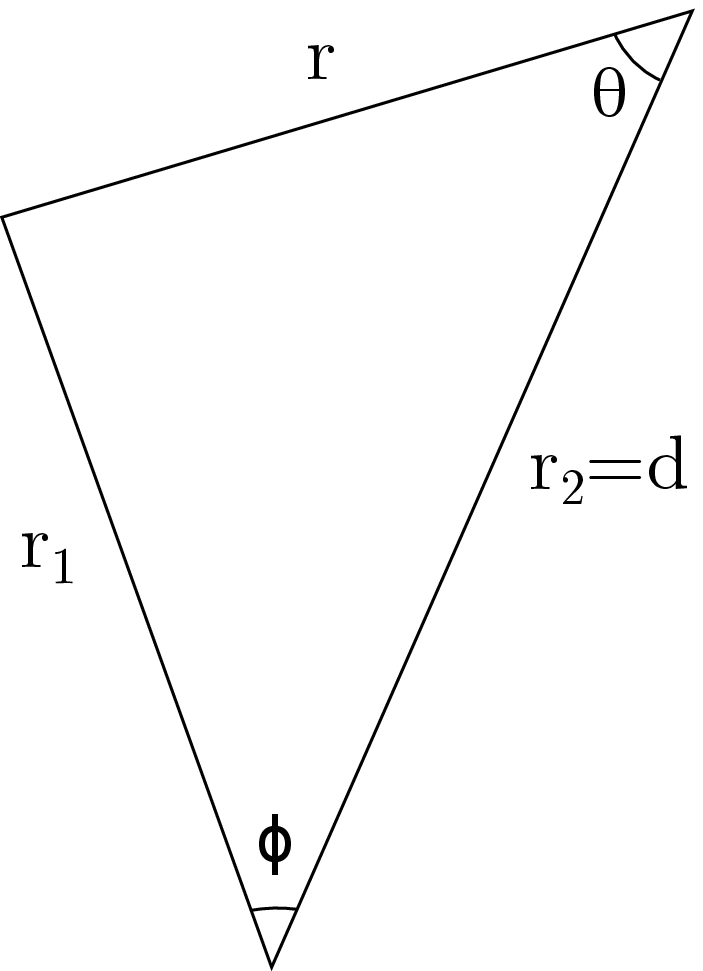}  \\
\hline
 & \\
$(1-\nu^2)\frac{r_1^2+r_2^2}{r^2}$ & $\frac{4}{3}\left[1-P_2(\mu)\right] + \frac{4}{5} 
\frac{r}{\dist} [P_1(\mu) - P_3(\mu)]  + \left(\frac{r}{\dist}\right)^2 \left[ -\frac{2}{5}  + \frac{22}{21} P_2(\mu) -\frac{32}{35} P_4(\mu) \right]$   \\
 & \\
$\nu \left( \frac{r_1}{r_2} + \frac{r_2}{r_1} \right)$ & $2 +\left(\frac{r}{\dist}\right)^2
\left[ -\frac{1}{3} + \frac{4}{3} P_2(\mu) \right]$  \\
 & \\
$(1-\nu^2)^2\frac{r_1^2 r_2^2}{r^4}$ & $\left[ \frac{8}{15} - \frac{16}{21} P_2(\mu) + \frac{8}{35} P_4(\mu) \right]
+\frac{r}{\dist} \left[ \frac{16}{35} P_1(\mu) -\frac{32}{45} P_3(\mu) + \frac{16}{63} P_5(\mu) \right] + \left(\frac{r}{\dist}\right)^2 \left[ - \frac{8}{35}  + \frac{16}{21} P_2(\mu) -\frac{312}{385} P_4(\mu) + \frac{64}{231} P_6(\mu) \right]$  \\
 & \\
$\nu(1-\nu^2)\frac{r_1 r_2}{r^2}$ & $(1 - \mu^2)+ \frac{r}{\dist} \left[ \frac{2}{5} P_1(\mu) - \frac{2}{5} P_3(\mu) \right] + \left(\frac{r}{\dist}\right)^2\left[ -\frac{2}{5}  + \frac{6}{7} P_2(\mu) -\frac{16}{35} P_4(\mu) \right]$  \\
 & \\
$14\nu^2-6$ & $8 +\frac{28}{3} \left(\frac{r}{\dist}\right)^2 \left[-1 + P_2(\mu) \right]$  \\
& \\
$\frac{\nu r^2}{ r_1 r_2}$ & $\left(\frac{r}{\dist}\right)^2 $ \\
\end{tabular}

\else
\begin{tabular}{c|c|c}
\hline
\multirow{20}{*}{\includegraphics[scale=0.5]{triangle1.eps}} 
&$\displaystyle(1-\nu^2)\frac{r_1^2+r_2^2}{r^2} $ & $\displaystyle\frac{4}{3}\left[1-P_2(\mu)\right] + \frac{4}{5} 
\frac{r}{\dist} [P_1(\mu) - P_3(\mu)]$ \\ &&$\displaystyle  + \left(\frac{r}{\dist}\right)^2 \left[ -\frac{2}{5}  + \frac{22}{21} P_2(\mu) -\frac{32}{35} P_4(\mu) \right]$  \\[0.4cm]
&$\displaystyle\nu \left( \frac{r_1}{r_2} + \frac{r_2}{r_1} \right) $ & $\displaystyle2 +\left(\frac{r}{\dist}\right)^2
\left[ -\frac{1}{3} + \frac{4}{3} P_2(\mu) \right]$  \\[0.4cm]
&$\displaystyle(1-\nu^2)^2\frac{r_1^2 r_2^2}{r^4}$ &  $\displaystyle\left[ \frac{8}{15} - \frac{16}{21} P_2(\mu) + \frac{8}{35} P_4(\mu) \right]$ \\ &&$\displaystyle
+\frac{r}{\dist} \left[ \frac{16}{35} P_1(\mu) -\frac{32}{45} P_3(\mu) + \frac{16}{63} P_5(\mu) \right]$ \\
&&$\displaystyle+ \left(\frac{r}{\dist}\right)^2 \left[ - \frac{8}{35}  + \frac{16}{21} P_2(\mu) -\frac{312}{385} P_4(\mu) + \frac{64}{231} P_6(\mu) \right]$  \\[0.4cm]
&$\displaystyle\nu(1-\nu^2)\frac{r_1 r_2}{r^2} $ & $\displaystyle(1 - \mu^2)+ \frac{r}{\dist} \left[ \frac{2}{5} P_1(\mu) - \frac{2}{5} P_3(\mu) \right]$\\ &&$\displaystyle + \left(\frac{r}{\dist}\right)^2\left[ -\frac{2}{5}  + \frac{6}{7} P_2(\mu) -\frac{16}{35} P_4(\mu) \right]$  \\[0.4cm]
&$\displaystyle14\nu^2-6 $ & $\displaystyle8 +\frac{28}{3} \left(\frac{r}{\dist}\right)^2 \left[-1 + P_2(\mu) \right]$  \\[0.4cm]
&$\displaystyle\frac{\nu r^2}{ r_1 r_2} $ & $\displaystyle\left(\frac{r}{\dist}\right)^2 $\\[0.4cm]
\hline
\end{tabular}
\fi

\caption{Expression of the geometrical coefficients in \eqref{GenCorrFunc} when $\dist$ and $\theta$ are defined in the asymmetric case $v=0$. }
\label{triangle1}
\end{center}
\end{table}

The zeroth order correction, corresponding to the plane-parallel limit is 
\be
I^{(0)}[j_0(kr)] = \left( 1 + \frac{2}{3} \beta + \frac{1}{5} \beta^2 \right) j_0(kr) P_0(\mu)  - \left(\frac{4}{3} \beta + \frac{4}{7} \beta^2 \right) j_2(kr) P_2(\mu)
+ \frac{8}{35} \beta^2 j_4(kr) P_4(\mu)\,.
\ee
Once inserted in eq.~\eqref{GenCorrFunc} we recover the plane-parallel result of eqs.~\eqref{xi_sa_all}. The non-vanishing coefficients $\xi_\ell^{(0)}(r)$ in this plane-parallel limit  are those of eqs.~\eqref{xi_sa_all}.

The correction linear $r/\dist$ is
\bea
I^{(1)}[j_0(kr)] & = & \left[\beta \frac{4}{5} \left( P_1(\mu) - P_3(\mu) \right) + \beta^2 \frac{12}{35} \left( P_1(\mu) - P_3(\mu) \right) \right] j_2(kr) \nonumber\\ & & + \beta^2 \frac{16}{63} \left[ -P_3(\mu) + P_5(\mu) \right] j_4(kr)\, .
\eea
Following the same method, the  non-vanishing coefficients $\xi_\ell^{(1)}(r)$ are 
\begin{displaymath}
\xi_1^{(1)}(r) = \left( \frac{4}{5} \beta + \frac{12}{35} \beta^2 \right) \MYXI_2^{0}(r)\,,\quad
\xi_3^{(1)}(r) = -\left ( \frac{4}{5} \beta + \frac{12}{35} \beta^2 \right) \MYXI_2^{0}(r) - \frac{16}{63} \beta^2 \MYXI_4^{0}(r)\,,
\end{displaymath}
\be
\label{coeff_corr_first_real}
\xi_5^{(1)}(r) = \frac{16}{63} \beta^2 \MYXI_4^{0}(r),
\ee
where the $\MYXI_\ell^{m}(r)$ are integrals on the power spectrum already defined in eq.~\eqref{DefXI}.

The quadratic corrections are
\bea
I^{(2)}[j_0(kr)] & = &  \left[ \beta \left( \frac{2}{9} P_0(\mu) - \frac{8}{9} P_2(\mu) \right) - \beta^2 \left( \frac{2}{5} P_0(\mu) + \frac{4}{15} P_2(\mu) \right) \right] j_0(kr)  \nonumber\\ & &  +\Bigg[ \beta \left( \frac{4}{45} P_0(\mu) + \frac{10}{63} P_2(\mu) - \frac{32}{35} P_4(\mu) \right) \nonumber\\ & & - \beta^2 \left(  \frac{12}{35} P_0(\mu) - \frac{10}{147} P_2(\mu) + \frac{96}{245} P_4(\mu) \right) \Bigg] j_2(kr)  \nonumber\\ & & + 
\beta^2 \left( -\frac{4}{105} P_0(\mu) + \frac{12}{245} P_2(\mu) -\frac{776}{2695}P_4(\mu) + \frac{64}{231} P_6(\mu) \right) j_4(\mu) \nonumber\\ & & +  \beta^2 \frac{4}{3} 
\left[ \frac{j_0(kr)}{(kr)^2} + \frac{j_2(kr)}{(kr)^2} \right] P_0(\mu) \, .
\eea
Following again the same method, one should be able to read the $\xidr_\ell^{(2)}$. At this order, only $\xidr_0^{(2)}$, $\xidr_2^{(2)}$, $\xidr_4^{(2)}$, $\xidr_6^{(2)}$ would be non-vanishing. We do not report them since in this parametrization the second order corrections are subdominant.

We remark that in this asymmetric parametrization the first order correction brings odd multipoles into play, as already observed by \cite{Raccanelli:2013dza, Bonvin:2013ogt}. For the second order correction there are only even multipoles, but beyond $\ell=0$, $\ell=2$, and $\ell=4$ corrections, there is a $\ell=6$ term. In general higher order corrections bring higher multipoles to the scene.

%%%%%%%%%%%%%%%%%%%%%%%%%%%%%%%%%%%%%%%%%%%%%%%%%
\subsubsection{Median parametrization ($v=1/2$)}\label{SecMedian}
%%%%%%%%%%%%%%%%%%%%%%%%%%%%%%%%%%%%%%%%%%%%%%%%%
\label{v1/2}

After some algebra, we find for this parametrization
\begin{subeqnarray}\label{geom_median_all}
r^2_1 &=& \dist^2+(r/2)^2 +\dist r \mu_{\gr{\dist} \, \gr{r}} \slabel{geom_median1} \\
r^2_2 &=& \dist^2+(r/2)^2 -\dist r \mu_{\gr{\dist} \, \gr{r}} \slabel{geom_median2} \\
r_1 r_2 \, \nu_{12} &=& \dist^2-(r/2)^2 \slabel{geom_median3}\\
r^2_1 r^2_2 (1-\nu_{12}^2) &=& \dist^2 r^2 (1-\mu^2_{\gr{\dist} \, \gr{r}}) \, . \slabel{geom_median4}
\end{subeqnarray}
From eqs.~\eqref{geom_median_all} expanded up to second order, we obtain the terms needed which are gathered in Table~\ref{triangle2}. The lowest order results are the same as it does not depend on the choice of parametrization. The first order corrections vanish in that median parametrization since $I^{(1)}[j_0(kr)]$ vanishes identically. We thus report only the second order which is

\begin{table}
\begin{center}

\ifnotcompact
\begin{tabular}{c | c}
& \includegraphics[scale=0.4]{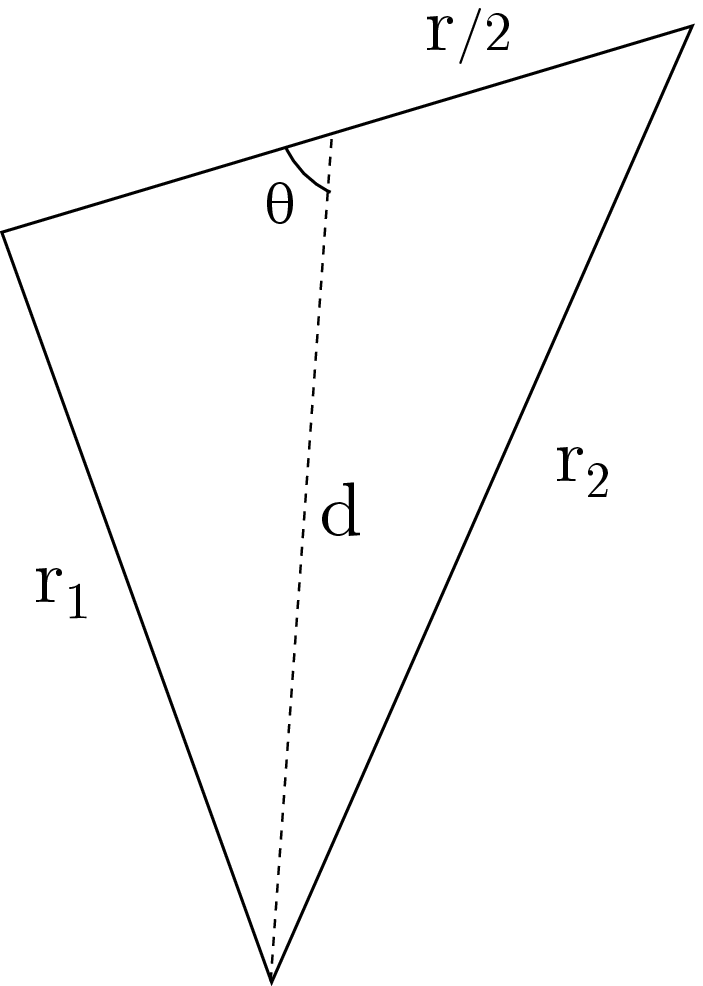} \\
\hline
$(1-\nu^2)\frac{r_1^2+r_2^2}{ r^2}$ & $\frac{4}{3}\left[1-P_2(\mu)\right] + \left(\frac{r}{\dist}\right)^2 \left[-\frac{1}{15} +\frac{11}{21}P_2(\mu) -\frac{16}{35}P_4(\mu) \right]$  \\
 & \\
$\nu \left( \frac{r_1}{r_2} + \frac{r_2}{r_1} \right)$ &  $2 +\left(\frac{r}{\dist}\right)^2
\left[  -\frac{1}{3} + \frac{4}{3} P_2(\mu)\right]$  \\
 & \\
$(1-\nu^2)^2\frac{r_1^2 r_2^2}{r^4}$ &  $\left[ \frac{8}{15} -\frac{16}{21}P_2(\mu) + \frac{8}{35}P_4(\mu) \right]+\left(\frac{r}{\dist}\right)^2\left[-\frac{4}{21} +\frac{8}{21}P_2(\mu) -\frac{20}{77}P_4(\mu) + \frac{16}{231}P_6(\mu)\right]$  \\
 & \\
$\nu(1-\nu^2)\frac{r_1 r_2}{r^2}$ & $(1 - \mu^2)  + \left(\frac{r}{\dist}\right)^2 \left[ -\frac{11}{30} +\frac{25}{42}P_2(\mu) -\frac{8}{35}P_4(\mu)\right]$  \\
 & \\
$14\nu^2-6$ &  $8 +\frac{28}{3}\left(\frac{r}{\dist}\right)^2 \left[-1 + P_2(\mu)\right]$  \\
 & \\
$\frac{\nu r^2}{ r_1 r_2}$ & $\left(\frac{r}{\dist}\right)^2$ \\
\end{tabular}

\else
\begin{tabular}{c|c|c}
\hline
\multirow{15}{*}{\includegraphics[scale=0.5]{triangle2.eps}}
& $\displaystyle(1-\nu^2)\frac{r_1^2+r_2^2}{ r^2}$ & $\displaystyle\frac{4}{3}\left[1-P_2(\mu)\right] + \left(\frac{r}{\dist}\right)^2 \left[-\frac{1}{15} +\frac{11}{21}P_2(\mu) -\frac{16}{35}P_4(\mu) \right]$  \\[0.4cm]
&$\displaystyle\nu \left( \frac{r_1}{r_2} + \frac{r_2}{r_1} \right)$ &  $\displaystyle2 +\left(\frac{r}{\dist}\right)^2
\left[  -\frac{1}{3} + \frac{4}{3} P_2(\mu)\right]$  \\[0.4cm]
&$\displaystyle(1-\nu^2)^2\frac{r_1^2 r_2^2}{r^4}$ &  $\displaystyle\left[ \frac{8}{15} -\frac{16}{21}P_2(\mu) + \frac{8}{35}P_4(\mu) \right]$\\ &&$\displaystyle +\left(\frac{r}{\dist}\right)^2\left[-\frac{4}{21} +\frac{8}{21}P_2(\mu) -\frac{20}{77}P_4(\mu) + \frac{16}{231}P_6(\mu)\right]$  \\[0.4cm]
&$\displaystyle\nu(1-\nu^2)\frac{r_1 r_2}{r^2}$ & $\displaystyle(1 - \mu^2)  + \left(\frac{r}{\dist}\right)^2 \left[ -\frac{11}{30} +\frac{25}{42}P_2(\mu) -\frac{8}{35}P_4(\mu)\right]$  \\[0.4cm]
&$\displaystyle14\nu^2-6$ &  $\displaystyle8 +\frac{28}{3}\left(\frac{r}{\dist}\right)^2 \left[-1 + P_2(\mu)\right]$  \\[0.4cm]
&$\displaystyle\frac{\nu r^2}{ r_1 r_2}$ & $\displaystyle\left(\frac{r}{\dist}\right)^2$\\[0.4cm]
\hline
\end{tabular}
\fi

\caption{Expression of the geometrical coefficients in \eqref{GenCorrFunc} when $\dist$ and $\theta$ are defined according to the median parametrization $v=1/2$. }
\label{triangle2}
\end{center}
\end{table}

\bea
\label{I2vhalf}
I^{(2)}[j_0(kr)] & = & \left[  
\beta \left( \frac{2}{9} P_0(\mu) - \frac{8}{9} P_2(\mu) \right) - \beta^2 \left( \frac{2}{5} P_0(\mu) + \frac{4}{15} P_2(\mu) \right) \right] j_0(kr)  \nonumber\\ & & 
+\Bigg[ \beta \left(  \frac{7}{45} P_0(\mu) -\frac{23}{63} P_2(\mu) - \frac{16}{35} P_4(\mu) \right) 
\nonumber\\ & &- \beta^2 \left( \frac{11}{35} P_0(\mu) + \frac{23}{147} P_2(\mu) + \frac{48}{245} P_4(\mu) \right)  \Bigg] j_2(kr)  \nonumber\\ & & + 
\beta^2 \left( -\frac{4}{105} P_0(\mu) - \frac{8}{245} P_2(\mu) +\frac{4}{2695}P_4(\mu) + \frac{16}{231} P_6(\mu) \right) j_4(\mu) \nonumber\\ & & +  \beta^2 \frac{4}{3} 
\left[ \frac{j_0(kr)}{(kr)^2} + \frac{j_2(kr)}{(kr)^2} \right] P_0(\mu) \,.
\eea
The coefficients $\xi_{\ell}^{(2)}$ can be read from eq. \eqref{I2vhalf}, using eq.~\eqref{j2x2} to express $j_2(kr)/(kr)^2$ in terms of $j_0$, $j_2$ and $j_4$, and the non-vanishing ones read
\bea
\xi_0^{(2)}(r) & = & \left(\frac{2}{9}\beta - \frac{14}{15}\beta^2 \right) \MYXI_0^{0}(r) + \left(\frac{7}{45} \beta - \frac{59}{315}\beta^2 \right) \MYXI_2^{0}(r) + \frac{4}{3} \beta^2 \MYXI_0^{2}(r) \, , \\
\xi_2^{(2)}(r) & = & - \left(\frac{8}{9} \beta + \frac{4}{15} \beta^2 \right) \MYXI_0^{0}(r) - \left(\frac{23}{63} \beta + \frac{23}{147} \beta^2 \right) \MYXI_2^{0}(r) - \frac{8}{245} \beta^2 \MYXI_4^{0}(r) \, , \\
\xi_4^{(2)}(r) & = & - \left( \frac{16}{35} \beta + \frac{48}{248} \beta^2 \right) \MYXI_2^{0}(r) + \frac{4}{2695} \beta^2 \MYXI_4^{0}(r) \, , \\
\xi_6^{(2)}(r) & = & \frac{16}{231} \beta^2 \MYXI_4^{0}(r) \, .
\eea

Because of the simplicity of the relations~\eqref{geom_median_all}, we can explicitly verify that all the terms on the left column of Table~\ref{triangle2} are expanded only with even powers of $r/d$ and $\mu$, to all orders. It implies obviously that $I^{(n)} = 0$ when $n$ is odd, and in particular that the first order corrections vanish. Furthermore, since only even powers of $\mu$ appear in the series expansions at all orders, the parity of Legendre multipoles implies that only even $\ell$s appear in the decomposition~\eqref{two_point_power}. Finally, we verify from the structure of the terms on Table~\ref{triangle2} with the rules~\eqref{geom_median_all} that, for a given even order $n$ in the series expansion, the highest possible power of the angular dependence is $\mu^{n+4}$. It follows that for a given $n$, the sum in $\ell$ in eq.~\eqref{two_point_power} does not extend to infinity, but is limited by $\ell_{max} = n+4$.

%%%%%%%%%%%%%%%%%%%%%%
\subsubsection{Angle bisector}
%%%%%%%%%%%%%%%%%%%%%%
If we take $d$ to be the length of the line that bisects the angle $\phi$, $v$ will depend on the sides of the triangle. With the help of two basic triangle geometry results known as Stewart's and angle bisector theorems, we relate the variables 
\begin{equation}
r_2 = \sqrt{\frac{1}{1-v} d^2 + v r^2 + r_1^2 \left(1 - \frac{1}{1-v} \right)} \, ,
\qquad
\frac{1-v}{r_1} = \frac{v}{r_2} \, ,
\end{equation}
and then determine the geometric factors of eq.~\eqref{GenCorrFunc} in Table \ref{triangle4}.
\begin{table}
\begin{center}

\ifnotcompact
\begin{tabular}{c | c}
& \includegraphics[scale=0.4]{triangle4.eps} \\
\hline
$(1-\nu^2)\frac{r_1^2+r_2^2}{ r^2}$ & $\frac{4}{3}\left[1-P_2(\mu)\right] + \left(\frac{r}{\dist}\right)^2 \left[-\frac{1}{5} +\frac{3}{7}P_2(\mu) -\frac{8}{35}P_4(\mu) \right]$  \\
 & \\
$\nu \left( \frac{r_1}{r_2} + \frac{r_2}{r_1} \right)$ &  $2 +\left(\frac{r}{\dist}\right)^2
\left[  -\frac{1}{3} + \frac{4}{3} P_2(\mu)\right]$  \\
 & \\
$(1-\nu^2)^2\frac{r_1^2 r_2^2}{r^4}$ &  $\left[ \frac{8}{15} -\frac{16}{21}P_2(\mu) + \frac{8}{35}P_4(\mu) \right]+\left(\frac{r}{\dist}\right)^2\left[-\frac{4}{15} +\frac{8}{21}P_2(\mu) -\frac{4}{35}P_4(\mu) \right]$  \\
 & \\
$\nu(1-\nu^2)\frac{r_1 r_2}{r^2}$ & $(1 - \mu^2)  + \left(\frac{r}{\dist}\right)^2 \left[ -\frac{13}{30} +\frac{23}{42}P_2(\mu) -\frac{4}{35}P_4(\mu)\right]$  \\
 & \\
$14\nu^2-6$ &  $8 +\frac{28}{3}\left(\frac{r}{\dist}\right)^2 \left[-1 + P_2(\mu)\right]$  \\
 & \\
$\frac{\nu r^2}{ r_1 r_2}$ & $\left(\frac{r}{\dist}\right)^2$ \\
\end{tabular}

\else
\begin{tabular}{c|c|c}
\hline
\multirow{15}{*}{\includegraphics[scale=0.5]{triangle4.eps}} &$\displaystyle(1-\nu^2)\frac{r_1^2+r_2^2}{ r^2}$ & $\displaystyle\frac{4}{3}\left[1-P_2(\mu)\right] + \left(\frac{r}{\dist}\right)^2 \left[-\frac{1}{5} +\frac{3}{7}P_2(\mu) -\frac{8}{35}P_4(\mu) \right]$  \\[0.4cm]
&$\displaystyle\nu \left( \frac{r_1}{r_2} + \frac{r_2}{r_1} \right)$ &  $\displaystyle2 +\left(\frac{r}{\dist}\right)^2
\left[  -\frac{1}{3} + \frac{4}{3} P_2(\mu)\right]$  \\[0.4cm]
&$\displaystyle(1-\nu^2)^2\frac{r_1^2 r_2^2}{r^4}$ &  $\displaystyle\left[ \frac{8}{15} -\frac{16}{21}P_2(\mu) + \frac{8}{35}P_4(\mu) \right]$\\ &&$\displaystyle+\left(\frac{r}{\dist}\right)^2\left[-\frac{4}{15} +\frac{8}{21}P_2(\mu) -\frac{4}{35}P_4(\mu) \right]$  \\[0.4cm]
&$\displaystyle\nu(1-\nu^2)\frac{r_1 r_2}{r^2}$ & $\displaystyle(1 - \mu^2)  + \left(\frac{r}{\dist}\right)^2 \left[ -\frac{13}{30} +\frac{23}{42}P_2(\mu) -\frac{4}{35}P_4(\mu)\right]$  \\[0.4cm]
&$\displaystyle14\nu^2-6$ &  $\displaystyle8 +\frac{28}{3}\left(\frac{r}{\dist}\right)^2 \left[-1 + P_2(\mu)\right]$  \\[0.4cm]
&$\displaystyle\frac{\nu r^2}{ r_1 r_2}$ & $\displaystyle\left(\frac{r}{\dist}\right)^2$ \\[0.4cm]
\hline
\end{tabular}
\fi

\caption{Expression of the geometrical coefficients in \eqref{GenCorrFunc} when $\dist$ and $\theta$ are defined according to the bisector parametrization.}
\label{triangle4}
\end{center}
\end{table}
As for the median parametrization, the first order corrections vanish identically. The second order corrections are then
\bea
\label{I2bisector}
I^{(2)}[j_0(kr)] & = & \left[  
\beta \left( \frac{2}{9} P_0(\mu) - \frac{8}{9} P_2(\mu) \right) - \beta^2 \left( \frac{2}{5} P_0(\mu) + \frac{4}{15} P_2(\mu) \right) \right] j_0(kr)  \nonumber\\ & & 
+\Bigg[ \beta \left(  \frac{1}{45} P_0(\mu) -\frac{29}{63} P_2(\mu) - \frac{8}{35} P_4(\mu) \right) \nonumber\\ & & 
- \beta^2 \left( \frac{13}{35} P_0(\mu) + \frac{29}{147} P_2(\mu) + \frac{24}{245} P_4(\mu) \right)  \Bigg] j_2(kr)  \nonumber\\ & & + 
\beta^2 \left( -\frac{4}{105} P_0(\mu) + \frac{16}{735} P_2(\mu) +\frac{4}{245}P_4(\mu)  \right) j_4(\mu)  \nonumber\\ & & +  \beta^2 \frac{4}{3} 
\left[ \frac{j_0(kr)}{(kr)^2} + \frac{j_2(kr)}{(kr)^2} \right] P_0(\mu) \, .
\eea

It follows from \eqref{I2bisector} that the non-vanishing coefficients $\xi_{\ell}^{(2)}(r)$ are expressed as
\bea
\xi_0^{(2)}(r) & = & \left(\frac{2}{9}\beta - \frac{14}{15}\beta^2 \right) \MYXI_0^{0}(r)+ \left(\frac{1}{45} \beta - \frac{11}{45}\beta^2 \right) \MYXI_2^{0}(r)+ \frac{4}{3} \beta^2 \MYXI_0^{2}(r) \, , \\
\xi_2^{(2)}(r) & = & - \left(\frac{8}{9} \beta + \frac{4}{15} \beta^2 \right) \MYXI_0^{0}(r) - \left(\frac{29}{63} \beta + \frac{29}{147} \beta^2 \right) \MYXI_2^{0}(r) - \frac{4}{245} \beta^2 \MYXI_4^{0}(r)\, , \\
\xi_4^{(2)}(r) & = & - \left( \frac{8}{35} \beta + \frac{24}{248} \beta^2 \right) \MYXI_2^{0}(r) + \frac{4}{245} \beta^2 \MYXI_4^{0}(r) \,  .
\eea

We observe that in this parametrization only corrections to the monopole, the quadrupole, and the hexadecapole are introduced. The fact that different multipoles appear in different parametrizations is expected as it is related to the natural sensitivity to the choice of the origin in multipole expansions. In fact, we can verify also in this case that only corrections for $n$ even appear, and that only even multipoles contribute but, for given $n$, $\ell_{max}=n+2$ in the expansion~\eqref{two_point_power}.

To summarize, in configuration space wide angle effects are encoded in corrections linear in $r/\dist$ if the parametrization used is not symmetric enough, e.g. for the case $v=0$. The largest effect in that case is the dipole correction ($\ell=1$) which could be used to constrain $\beta$.
However for a balanced parametrization such as the median or the bisector parametrizations, the corrections are quadratic in $r/\dist$.  In all cases, the only integrals on the power spectrum that need to be performed up to second order are
\be
\MYXI_\ell^{0}(r)=\int \frac{\dd k k^2}{2\pi^2} \Pow(k) j_\ell(kr) \quad{\rm with}\quad\ell=0,2,4, \quad{\rm and}\quad \MYXI_0^{2}(r)=\int \frac{dk}{2\pi^2 r^2} \Pow(k) j_0(kr) \, .
\ee

%%%%%%%%%%%%%%%%%%%%%%%%%%%%%%%%%%%%%%%%%%%%%%%%%%%%%
\subsection{Wide-angle corrections in mixed configuration/Fourier space}
\label{kaiser_fourier}
%%%%%%%%%%%%%%%%%%%%%%%%%%%%%%%%%%%%%%%%%%%%%%%%%%%%%

In the plane-parallel limit, we have already emphasized in section~\ref{small_angle_mixed} that the standard results appear in the mixed configuration/Fourier space and not in the Fourier space. In that case we need to Fourier transform the dependence in $\gr{r}$ while keeping the dependence in the average distance $\gr{\dist}$. This leads to define a \emph{power spectrum at a distance $\dist$}.
The dependence on $\gr{d}$ in the configuration space correlation function~\eqref{two_point_power} appears through its norm $\dist$ and through the angle $\mu_{\gr{\dist \, r}} = \hat{\gr{d}} \cdot \hat{\gr{r}}$. We then find that in the mixed space, the natural perturbative expansion is in powers of $1/(k\dist)$ as
\be
\label{P_dist_def}
\xidk^z(\gr{\dist},\gr{k})  \equiv  \int \frac{\dd^3 \gr{r}}{(2 \pi)^{3/2}} {\rm e}^{-\ii \gr{k}\cdot \gr{r} }  \xidr^z(\gr{\dist} ,\gr{r})\equiv \sum_{n=0}^\infty \frac{1}{(k\dist)^n} \xidk^{z(n)}(k,\mu_{\gr{k \, \dist}})
\ee
with an angular multipoles expansion given by
\be\label{P_dist_def2}
\xidk^{z(n)}(k,\mu_{\gr{k \, \dist}}) \equiv \sum_{\ell=0}^\infty \Pow_\ell^{(n)}(k)P_\ell(\mu_{\gr{k \, \dist}})\,,\quad \Pow_\ell^{(n)}(k)= \left[ \sqrt{\frac{2}{\pi}} (-\ii)^\ell \int r^2 \dd r (kr)^n j_\ell (k r) \xidr_\ell^{(n)}(r) \right] \, ,
\ee
with $\mu_{\gr{k \, \dist}}=\hat{\gr{k}}\cdot \hat{\gr{\dist}}$. Equation~\eqref{P_dist_def2} is obtained by expanding the exponential in spherical waves and using the addition theorem for Legendre polynomials. Given the global rotational invariance, we find that from the apparent six degrees of freedom of $\xidk^z(\gr{\dist},\gr{k})$, only three remain as it depends only on $\dist$, $k$ and $\mu_{\gr{k \, \dist}}$. 

At lowest order, that is considering the term $n=0$ in \eqref{P_dist_def},  we get
\bea
\label{P_dist_zeroth}
\xidk^{z(0)}(k,\mu_k) &=& \frac{\Pow(k)}{(2 \pi)^{3/2}} \left\{\left[1+ \frac{2}{3} \beta + \frac{1}{5}\beta^2 \right] P_0(\mu_{\gr{k \, \dist}})  +\left[\frac{4}{3}\beta + \frac{4}{7} \beta^2 \right]P_2(\mu_{\gr{k \, \dist}}) + \beta^2 \frac{8}{35}P_4(\mu_{\gr{k \, \dist}})\right\}   \nonumber\\ 
& = & \frac{\Pow(k)}{(2 \pi)^{3/2}} (1 + \beta \mu^2_{\gr{k \, \dist}})^2\,,
\eea
thus recovering the plane-parallel result~\eqref{zpk_sum}.

The wide angle corrections can also be computed for each type of parametrization.
\begin{itemize}
\item {\it Asymmetric parametrization ($v=0$)}

The first order correction is
\bea
\xidk^{z(1)}(k,\mu_{\gr{k \, \dist}})  &=& -\frac{\ii k}{(2 \pi)^{3/2}}  \int p^2 \dd p \Pow(p) \left\{  \left(\frac{4}{5} \beta + \frac{12}{35}\beta^2 \right)\IntI_{2\,1\,-1}(p,k) P_1(\mu_{\gr{k \, \dist}}) \right. \nonumber\\
&&\left.+ \left[\left(\frac{4}{5}\beta + \frac{12}{35}\beta^2 \right)\IntI_{2\,3\,-2}(p,k)  +\frac{16}{63}\beta^2  \IntI_{4\,3\,-3}(p,k)\right] P_3(\mu_{\gr{k \, \dist}})\right. \nonumber\\ & &\left. + \frac{16}{63} \beta^2 \IntI_{5\, 4\,-4}(p,k) P_5(\mu_{\gr{k \, \dist}}) \right\} \, ,
\eea
where the integrals $\IntI_{pqn}$ are defined in appendix~\ref{cal_Int} and listed in \ref{integrals_used}. After integration by parts we obtain
\bea\label{Power1}
\xidk^{z(1)}(k,\mu_{\gr{k \, \dist}})&=& \frac{\ii}{(2\pi)^{3/2}} \Bigg[ \Pow(k)\Bigg(\frac{64}{63} \beta^2  P_5(\mu_{\gr{k \, \dist}})+\left(\frac{8}{5}\beta+\frac{88}{45}\beta^2\right) P_3(\mu_{\gr{k \, \dist}}) \nonumber\\& &  +\left(\frac{12}{5}\beta+\frac{36}{35}\beta^2\right) P_1(\mu_{\gr{k \, \dist}})\Bigg) + k \Pow'(k) \Bigg(-\frac{16}{63} \beta^2 P_5(\mu_{\gr{k \, \dist}}) \nonumber\\ & & +\left(-\frac{4}{5}\beta-\frac{4}{45}\beta^2\right) P_3(\mu_{\gr{k \, \dist}}) + \left(\frac{4}{5}\beta+\frac{12}{35}\beta^2\right)P_1(\mu_{\gr{k \, \dist}})\Bigg) \Bigg]\,.
\eea
We observe that this contribution to the power spectrum is purely imaginary, as observed by \cite{McDonald:2009ud}, and is expressed in terms of $\Pow(k)$ and $\Pow'(k)$. Furthermore, the coefficients $\Pow_\ell^{(1)}(k)$ can then be read directly from it. As a check, we perform the same analysis for $v=0$ using cylindrical coordinates in appendix~\ref{SecCylinder}.

At second order, we obtain
\bea
\label{P2v0}
\xidk^{z(2)}(k,\mu_{\gr{k \, \dist}}) & = & \frac{k^2}{(2\pi)^{3/2}}  \int p^2 \dd p \Pow(p) \Bigg\{  \left[\left(\frac{2}{9} \beta -\frac{2}{5}\beta^2 \right)\IntI_{00\,1}(p,k) \right. \nonumber\\
&& \left. + \left(\frac{4}{45} \beta -\frac{12}{35}\beta^2 \right)\IntI_{200}(p,k)-\frac{4}{105} \beta^2 \IntI_{40\,-1}(p,k) \right. \nonumber\\
&& \left. + \frac{2}{3} \frac{\beta^2}{p^2} \left( \IntI_{00\,0}(p,k) + \IntI_{02\,-1}(p,k)\right)\right] P_0(\mu_{\gr{k \, \dist}})  + \left[\left(\frac{8}{9}\beta + \frac{4}{15} \beta^2 \right)\IntI_{020}(p,k) \right. \nonumber\\
&& \left. - \left(\frac{10}{63}\beta +\frac{10}{147} \beta^2 \right)\IntI_{22\,-1}(p,k) - \frac{12}{245} \beta^2 \IntI_{42\,-2}(p,k)\right] P_2(\mu_{\gr{k \, \dist}}) \nonumber\\
&& - \left[  \left(\frac{32}{35} \beta + \frac{96}{245} \beta^2 \right)\IntI_{24\, -2}(p,k) + \frac{776}{2695} \beta^2 \IntI_{44 \, -3}(p,k)\right] P_4(\mu_{\gr{k \, \dist}}) \nonumber\\ & & - \frac{64}{231} \beta^2 \IntI_{46\,-4}(p,k) P_6(\mu_{\gr{k \, \dist}}) \Bigg\}\,.
\eea
Similarly the integrals $\IntI_{pqn}$ are defined in appendix~\ref{cal_Int} and listed in \ref{integrals_used}. We do not report the final result at second order since it is subdominant in that case.

\item {\it Median parametrization ($v = 1/2$)}

The first order correction vanishes, and at second order we get
\bea
\label{P2vhalf}
\xidk^{z(2)}(k,\mu_{\gr{k \, \dist}}) & = & \frac{k^2}{(2\pi)^{3/2}} \int p^2 \dd p \Pow(p) \Bigg\{  \left[\left(\frac{2}{9} \beta -\frac{2}{5}\beta^2 \right)\IntI_{00\,1}(p,k) \right. \nonumber\\
&&  \left. + \left(\frac{7}{45} \beta -\frac{11}{35}\beta^2 \right)\IntI_{200}(p,k) -\frac{4}{105} \beta^2 \IntI_{40\,-1}(p,k) \right. \nonumber\\ &&  \left.
+ \frac{4}{3} \frac{\beta^2}{p^2} \left( \IntI_{00\,0}(p,k) + \IntI_{02\,-1}(p,k)\right)\right] P_0(\mu_{\gr{k \, \dist}}) \nonumber\\
&&  + \left[\left(\frac{8}{9}\beta + \frac{4}{15} \beta^2 \right)\IntI_{020}(p,k) + \left(\frac{23}{63}\beta +\frac{23}{147} \beta^2 \right)\IntI_{22\,-1}(p,k) \right. \nonumber\\
&&  \left. + \frac{8}{245} \beta^2 \IntI_{42\,-2}(p,k)\right] P_2(\mu_{\gr{k \, \dist}}) + \left[ - \left(\frac{16}{35} \beta + \frac{48}{245} \beta^2 \right)\IntI_{24\, -2}(p,k) \right. \nonumber\\
&&  \left. +\frac{4}{2695} \beta^2 \IntI_{44 \, -3}(p,k)\right] P_4(\mu_{\gr{k \, \dist}})  - \frac{16}{231} \beta^2 \IntI_{46\,-4}(p,k) P_6(\mu_{\gr{k \, \dist}}) \Bigg\}\,.
\eea
Following the same method, after integrations by parts we obtain the second order correction
\bea\label{Power2_spherical}
\xidk^{z(2)}(k,\mu_{\gr{k \, \dist}})&=&\frac{\Pow(k)}{(2\pi)^{3/2}}\left[-\frac{128}{77}\beta^2 P_6(\mu_{\gr{k \, \dist}}) -\left(\frac{128}{35}\beta + \frac{592}{385}\beta^2 \right)P_4(\mu_{\gr{k \, \dist}}) \right. \nonumber\\
&&  \left. +\left(\frac{46}{21}\beta+\frac{10}{7}\beta^2\right)P_2(\mu_{\gr{k \, \dist}})+ \left(\frac{7}{15}\beta+\frac{27}{35}\beta^2 \right) P_0(\mu_{\gr{k \, \dist}}) \right]\nonumber\\ & & + \frac{k \Pow'(k)}{(2\pi)^{3/2}} \left[\frac{48}{77}\beta^2 P_6(\mu_{\gr{k \, \dist}}) +\left(\frac{16}{7}\beta+ \frac{376}{385} \beta^2\right) P_4(\mu_{\gr{k \, \dist}}) \right. \nonumber\\ &&  \left. -\left(\frac{34}{7}\beta+\frac{2}{7}\beta^2\right)P_2(\mu_{\gr{k \, \dist}})+ \left( \frac{1}{3}\beta-\frac{11}{35}\beta^2 \right) P_0(\mu_{\gr{k \, \dist}}) \right] \nonumber\\ & & + \frac{k^2 \Pow''(k)}{(2\pi)^{3/2}} \left[-\frac{16}{231}\beta^2 P_6(\mu_{\gr{k \, \dist}}) -\left(\frac{16}{35}\beta+\frac{76}{385}\beta^2\right)P_4(\mu_{\gr{k \, \dist}}) \right. \nonumber\\ &&  \left. +\left(\frac{11}{21}\beta+\frac{1}{7}\beta^2\right)P_2(\mu_{\gr{k \, \dist}}) + \left( -\frac{1}{15}\beta+\frac{13}{105}\beta^2 \right) P_0(\mu_{\gr{k \, \dist}})\right] \,.
\eea
The coefficients $\Pow_\ell^{(2)}(k)$ are read directly from this result. We remark that it is expressed in terms of $\Pow(k)$, $\Pow'(k)$, and $\Pow''(k)$. In general at $n$-th order it will involve the $n$-th derivative of $\Pow(k)$. As a check, we perform the same analysis for $v=1/2$ using cylindrical coordinates in appendix~\ref{SecCylinder}. 

\item {\it Bisector parametrization}

The first order correction also vanishes, and at second order we get
\bea
\label{P2bisect}
\xidk^{z(2)}(k,\mu_{\gr{k \, \dist}}) & = & \frac{k^2}{(2\pi)^{3/2}} \int p^2 \dd p \Pow(p) \Bigg\{   \left[\left(\frac{2}{9} \beta -\frac{2}{5}\beta^2 \right)\IntI_{00\,1}(p,k) \right. \nonumber\\
&& \left. + \left(\frac{1}{45} \beta -\frac{13}{35}\beta^2 \right)\IntI_{200}(p,k)  -\frac{4}{105} \beta^2 \IntI_{40\,-1}(p,k) \right. \nonumber\\ && \left.
+ \frac{4}{3} \frac{\beta^2}{p^2} \left( \IntI_{00\,0}(p,k) + \IntI_{02\,-1}(p,k)\right)\right] P_0(\mu_{\gr{k \, \dist}}) \nonumber\\
&&  + \left[\left(\frac{8}{9}\beta + \frac{4}{15} \beta^2 \right)\IntI_{020}(p,k) + \left(\frac{29}{63}\beta +\frac{29}{147} \beta^2 \right)\IntI_{22\,-1}(p,k) \right. \nonumber\\
&& \left. + \frac{8}{245} \beta^2 \IntI_{42\,-2}(p,k)\right] P_2(\mu_{\gr{k \, \dist}}) + \left[ - \left(\frac{8}{35} \beta + \frac{24}{245} \beta^2 \right)\IntI_{24\, -2}(p,k) \right. \nonumber\\
&& \left. +\frac{4}{245} \beta^2 \IntI_{44 \, -3}(p,k)\right] P_4(\mu_{\gr{k \, \dist}}) \Bigg\}\,.
\eea
Following the same method we are able to find that the second order corrections are
\bea
\label{Power2_spherical_bisect}
\xidk^{z(2)}(k,\mu_{\gr{k \, \dist}}) &=& \frac{1}{(2\pi)^{3/2}} \Bigg\{ \Pow(k) \left[ \left(-\frac{64}{35}\beta- \frac{16}{35}\beta^2 \right) P_4(\mu_{\gr{k \, \dist}}) + \left(\frac{58}{21} \beta + \frac{82}{49}\beta^2\right) P_2(\mu_{\gr{k \, \dist}}) \right. \nonumber\\
& & \left. + \left( \frac{1}{15}\beta+\frac{3}{5}\beta^2\right) P_0(\mu_{\gr{k \, \dist}}) \right]  + k \Pow'(k) \left[ \left(\frac{8}{7}\beta + \frac{16}{35} \beta^2\right) P_4(\mu_{\gr{k \, \dist}}) \right. \nonumber\\ & & \left. +\left(-\frac{38}{21}\beta-\frac{18}{49}\beta^2\right) P_2(\mu_{\gr{k \, \dist}}) - \left( \frac{1}{3}\beta + \frac{3}{5}\beta^2\right) P_0(\mu_{\gr{k \, \dist}}) \right] \nonumber\\ & & + k^2 \Pow''(k) \left[\left(-\frac{8}{35}\beta- \frac{4}{35}\beta^2\right)P_4(\mu_{\gr{k \, \dist}})  +\left(\frac{3}{7}\beta+\frac{5}{49}\beta^2\right)P_2(\mu_{\gr{k \, \dist}}) \right. \nonumber\\
& & \left.  - \left( \frac{1}{5}\beta - \frac{1}{15}\beta^2 \right) P_0(\mu_{\gr{k \, \dist}}) \right]\Bigg\} \,,
\eea
from which the corresponding $\Pow_\ell^{(2)}(k)$ are easily read.
\end{itemize}

As we can observe from section~\ref{integrals_used}, the integrals $\IntI_{pqn}$ appearing in 
\eqref{P2v0}, \eqref{P2vhalf}, and \eqref{P2bisect} are all expressed in terms of Dirac delta function and its first and second derivatives, except for $\IntI_{40\,-1}(p,k)$ and $\IntI_{02\,-1}(p, k)$, which also involve Heaviside step functions. The numerical coefficients, however, are such that the contributions from the Heaviside step functions always cancel out.

%%%%%%%%%%%%%%%%%%%%%%%%%%%%%%%%%%%%%%%%%%%%%%%%%%%%%%
\subsection{Discussion and Fourier space correlations}\label{SecDiscussion}
%%%%%%%%%%%%%%%%%%%%%%%%%%%%%%%%%%%%%%%%%%%%%%%%%%%%%%

We notice that among the coefficients $\xidr_\ell^{(n)}(r)$ of the plane-parallel expansion, some are always vanishing. Indeed they only take non-vanishing values if $\ell$ and $n$ are either both odd or both even. For instance at lowest order the non-vanishing coefficients are $\xidr_0^{(0)}(r)$, $\xidr_2^{(0)}(r)$ and $\xidr_4^{(0)}(r)$. At first order in the asymmetric case, the non-vanishing coefficients are $\xidr_1^{(1)}(r)$, $\xidr_3^{(1)}(r)$ and $\xidr_5^{(1)}(r)$ whereas at second order we have only $\xidr_0^{(2)}(r)$, $\xidr_2^{(2)}(r)$, $\xidr_4^{(2)}(r)$ and $\xidr_6^{(2)}(r)$. 

In fact, we have shown at the end of section~\ref{SecMedian} that if we use the median parametrization the only non-vanishing coefficients are those for which $\ell$ and $n$ are both even at all orders. Including second-order corrections would lead to an expression which is correct up to fourth-order corrections. The number of terms to be included is limited since 
$\ell\le n+4$. The coefficients in the mixed space follow obviously the same structure since they are related by the Hankel-type transformation~\eqref{P_dist_def2}. The inverse transformation takes the form
\be
\xidr_\ell^{(n)}(r) = \left[\sqrt{\frac{2}{\pi}}\ii^\ell \int k^2 \dd k (kr)^{-n} j_\ell(kr) \Pow_\ell^{(n)}(k)\right]\,.
\ee
Given that the coefficients in the mixed space are expressed only in terms of the matter power spectrum and its derivatives, then provided it has nice convergence properties at $k \dist \to 0$ and $k \dist \to \infty$, the transformations between configuration space and mixed space coefficients are in general not worrisome.

However, if we want to build an expansion of the correlation function in the full Fourier space, it is not obvious at first sight that it is well defined. Indeed, following the logic that the Fourier mode associated with the difference of points is the average Fourier mode, and the Fourier modes associated with the average of points is the difference of Fourier modes, we are lead to consider that the small parameter in the plane-parallel expansion in Fourier space is $|\Delta \gr{k}|/k$. We are thus tempted to look for an expansion of the type
\be\label{AttemptExpansion}
\zetaKk^z(\Delta\gr{k},\gr{k}) \overset{?}{=}\frac{1}{|\Delta\gr{k}|^3} \sum_{n=0}^\infty \left(\frac{|\Delta \gr{k}|}{k}\right)^n \sum_{\ell=0}^\infty \zeta_\ell^{(n)}(k) P_\ell(\mu_{\gr{k \, \Delta} })\,\qquad {\rm with} \qquad \mu_{\gr{k \, \Delta}} \equiv \hat{\gr{k}}\cdot\hat{\Delta\gr{k}}\,,
\ee
the prefactor $1/|\Delta\gr{k}|^3$ being introduced from dimensional arguments.
Performing a Fourier transform on $\xidk(\gr{d},\gr{k})$, and using the Rayleigh formula~\eqref{rayleigh} to expand the exponential, would lead to Hankel-type relations between the coefficients in mixed spaced and Fourier space in the form
\bea
\zetaKk_\ell^{(n)}(k)&\overset{?}{=}&\Pow_\ell^{(n)}(k)\sqrt{\frac{2}{\pi}}(-\ii)^\ell \int  x^{2-n} j_\ell(x)\dd x  \,,\\
\Pow_\ell^{(n)}(k)&\overset{?}{=}&\zetaKk_\ell^{(n)}(k)\sqrt{\frac{2}{\pi}}\ii^\ell \int x^{n-1} j_\ell(x)  \dd x\,.
\eea
From the Weber integrals of appendix~\ref{AppWeber}, we already see that this would not be defined if $\ell$ and $n$ were not both odd or both even, that is if they would not have the same parity, given that one would possibly encounter the poles of the $\Gamma$ functions, e.g. for $n\ge\ell+3$. Since we have shown that $\ell$ and $n$ always have the same parity, such problem never arises. A second problem remains for $\ell=n=0$ though, since it involves $\Gamma(0)$. If we inspect the plane-parallel limit obtained in eqs.~\eqref{FSFourierShouldBe} and \eqref{zetaFS}, we notice however that there is a monopole multiplying a Dirac delta function, and it differs from the attempted expansion~\eqref{AttemptExpansion}. Hence, we should consider instead an expansion of the form~\eqref{SuperFourierExpansion}
where the factor $4\pi$ has only been introduced for convenience. Using eqs.~\eqref{WeberUseful}, we then recover that the relation between the Fourier space and mixed space multipoles is given by eqs.~\eqref{PlnOFzetaln}. It can be checked that at lowest order, that is in the plane-parallel limit, the Fourier space coefficients~\eqref{zetaFS} and the mixed space coefficients~\eqref{zpk_coeff} are indeed related by eqs.~\eqref{PlnOFzetaln}. Since we have already shown in section~\ref{SecKaiserOriginal} that the Kaiser formula naturally arises in the mixed space, we conclude that one should always rely on a mixed space expansion rather than on a full Fourier space expansion when apprehending the RSD effects in Fourier space.

%%%%%%%%%%%%%%%%%%%%
\section{Conclusion}
%%%%%%%%%%%%%%%%%%%%

The wide angle effects on RSD have a rich geometrical structure, both in configuration space and Fourier space. This richness is highly simplified in the plane-parallel limit where we get an expression for the two-point function that does not depend on the distance from the observer to the pair of galaxies and, on the other side of the coin, independence of Fourier modes in the Kaiser formula.  

Our initial goal was to detail how such tremendous simplifications occur, and to 
provide intermediate formulas between the full wide angle treatment and the plane-parallel limits. Instead of using the formulation introduced by \cite{Papai:2008bd} where a functional basis built from the monopole of the tensor product of three sets of angular functions is used, we derived a very simple expression for the general correlation function, given in eq.~\eqref{Fullxireal1}, in terms of an integral on the product of the power spectrum $\Pow(k)$ with the action of differential operators on the spherical Bessel function $j_0(kr)$, where $r$ encodes all the information about the triangle formed by the observer and the pair of galaxies. Beyond its simplicity, this formulation of the problem proved to be very powerful to demonstrate relations between configuration and Fourier spaces.

The Fourier conjugate of eq.~\eqref{Fullxireal1}, which is eq.~\eqref{corr_func_fourier}, paves the way for unveiling the geometrical structure of the approximations leading the Kaiser formula. Indeed, the presence of the kernel $\kerK$ both in configuration and Fourier spaces umbilically connects both formulations, and its (singular) structure leads to the expected plane-parallel limit. However, we argued that a mixed configuration/Fourier space has to be employed, since it is only in this space that the plane-parallel limit is meaningful. Indeed, if one looks at the Kaiser formula, the dependence on the angle between a Fourier mode and the line of sight cannot be well defined neither in Fourier space nor in configuration space. It is the structure of the plane-parallel limit in the mixed configuration/Fourier space that gives a precise meaning to this phantasmal dependence. 

Going back to the general expression for the two-point function given in eq.~\eqref{Fullxireal1}, we worked out the action of the differential operators on the spherical Bessel function to obtain an explicit formula in terms of lengths and angles given by eq.~\eqref{GenCorrFunc}. We first observed that this expression coincides with the general result presented by \cite{Papai:2008bd} (as demonstrated on appendix~\ref{appendix_papai_szapudi}), and then proposed a perturbative representation to this general result, as presented in eq.~\eqref{two_point_power}. On this perturbative solution we looked at contributions classified by powers of $r/\dist$, where $r$ is the distance between two galaxies on a pair and $\dist$ is a mean distance from the observer to the pair. In the plane-parallel limit this ratio goes to zero, and only the term $n=0$ contributes on the sum. The coefficients $\xi_{\ell}^{(0)}$ are the plane-parallel limit ones, and are only non-vanishing for $\ell=0, 2, 4$.
Departing from the plane-parallel limit, we found the $r/\dist$ corrections, and the corresponding corrections in $1/(k\dist)$ in the mixed configuration/Fourier space. However, these corrections depend on the parametrization used to express the geometry of the triangle in eq.~\eqref{GenCorrFunc}. If we choose an asymmetric parametrization we can show that dipolar terms appear at first order, as shown in eq.~\eqref{coeff_corr_first_real}. Such a parametrization could be used if one is interested in probing the values of $\beta$.

As for symmetric parametrizations, we have two natural options for the distance $\dist$, namely the median and the length of the angle bisector. We show that if these parametrizations are used, the first correction to the plane-parallel limit is quadratic in $r/\dist$, and therefore the full wide angle corrections are reduced  even for large coverage surveys, as already found numerically by \cite{Raccanelli2010,Samushia2012,Yoo2015}. If one includes second order corrections, the parametrization of the angle bisector proves to be the most immediate extension to the plane-parallel limit, since only $\ell=0, 2, 4$ contribute in eq.~\eqref{two_point_power}.

Finally, we have shown that it is possible to define a plane-parallel expansion in the full Fourier space. However, its coefficients are related to the mixed space coefficients by simple numerical factors and one should instead consider the mixed space expansion as the spectrum at a given distance is the natural quantity to extend the notion of power spectrum.  
%%%%%%%%%%%%%%%%%%%%%
\section*{Acknowledgments}
%%%%%%%%%%%%%%%%%%%%%
The authors thank Lado Samushia, Marc Manera and Raul Abramo for discussions about redshift-space distortions. We are indebted towards G. Faye for his help about the structure of the RSD kernels. This work has been done within the Labex ILP (reference ANR-10-LABX-63) part of the Idex SUPER, and received financial state aid managed by the Agence Nationale de la Recherche, as part of the programme Investissements d'avenir under the reference ANR-11-IDEX-0004-02, and also under the reference ANR-12-BS05-0002. P. R. also thanks FAPESP for financial support.

\appendix

%%%%%%%%%%%%%%%%%%%
\section{Special functions}
%%%%%%%%%%%%%%%%%%%
\label{Bessel_app}
We collect the some properties of Bessel functions and Legendre polynomials in this appendix, since they are frequently employed throughout the main text. 
%%%%%%%%%%%%%%%%%%%%%%%%%%%%%%%%%
\subsection{Bessel functions of the first kind}
%%%%%%%%%%%%%%%%%%%%%%%%%%%%%%%%%
Bessel functions of the first kind are solutions of Bessel's differential equation
\be
\frac{1}{x} \frac{\dd}{\dd x} \left[x \frac{\dd}{\dd x} J_\ell(x)\right] = \left[\frac{\ell(\ell-1)}{x^2}-1\right]J_\ell(x)
\ee
regular at $r=0$. For a negative index, the definition is extended
through $J_{-\ell} = (-1)^n J_{\ell}$. They can be obtained from the 
integral representation 
\be\label{JasIntegral}
J_\ell (r) = \frac{1}{2\pi} \int_{2\pi} \dd \, \theta \,  {\rm e}^{\ii r \sin \theta -\ii \ell \theta} =(-\ii)^\ell\frac{1}{2\pi} \int_{2\pi} \dd \, \theta \, {\rm e}^{\ii r \cos \theta -\ii \ell \theta}\,.
\ee
This is particularly useful to write the Fourier series of ${\rm e}^{\ii r \cos \theta}$. Indeed from this integral representation, we obtain
\be\label{RayleighCylinder}
{\rm e}^{\ii r \cos \theta} = \sum_{\ell=-\infty}^\infty \ii^\ell J_\ell(r) {\rm
e }^{\ii \ell\theta} \,.
\ee
Two recurrence relations satisfied by $J_\ell$ are of particular interest
\be
J'_\ell(x) =\frac{ J_{\ell-1}(x)- J_{\ell+1}(x)}{2} \,\qquad\frac{ J_{\ell}(x)}{x} = \frac{J_{\ell+1}(x)+J_{\ell-1}(x)}{2\ell} \, .
\ee
Finally, for all $\ell \in \mathbb{N}$, the Bessel functions can be
generated from $J_0$ from
\be
J_\ell(x) =x^\ell \left(-\frac{\dd}{x \dd x}\right)^\ell J_0(x) \, .
\ee

%%%%%%%%%%%%%%%%%%%%%%%%%%%%
\subsection{Spherical Bessel functions}
%%%%%%%%%%%%%%%%%%%%%%%%%%%%
The spherical Bessel functions are related to Bessel functions of the first kind by
\be
\label{relation_j_J}
j_\ell(x) \equiv \sqrt{\frac{\pi}{2 x}} J_{\ell+1/2}(x)
\ee
and they are the solutions regular at the origin of the differential equation
\be
\label{sph_bessel_diff_eq}
\frac{1}{r^2}\frac{\dd}{\dd r}\left[r^2 \frac{\dd}{\dd r} j_\ell(r)\right] = \left(\frac{\ell(\ell+1)}{r^2}-1\right) j_\ell(r) \, .
\ee
The spherical Bessel functions appear naturally in a spherical geometry
when a plane wave is expanded into spherical harmonics or Legendre polynomials. Indeed, this
is given by the Rayleigh expansion  which is
\be
\label{rayleigh}
\mathrm{e}^{\ii \gr{r} \cdot \gr{k}} = 4 \pi \sum_{\ell m} \ii^\ell
j_\ell(kr) \mathrm{Y}_{\ell m} (\hat{\gr{r}}) \mathrm{Y}_{\ell
  m}^\star (\hat{\gr{k}})=\sum_{\ell} (2\ell+1) \ii^\ell j_\ell(kr)
P_{\ell} (\hat{\gr{k}}\cdot \hat{\gr{r}})
\ee
where the $\mathrm{Y}_{\ell m}$ are the spherical harmonics and the
$P_\ell$ are the Legendre polynomials.
Two recurrence relations satisfied by $j_{\ell}(x)$ are particularly useful
\be
\label{rec_rel}
\frac{j_{\ell}(x)}{x} = \frac{j_{\ell+1}(x) + j_{\ell-1}(x)}{2 \ell+1} 
\ee
\be
\label{rec_rel_derivative}
(2\ell+1)j_{\ell}'(x) = \ell j_{\ell-1}(x) -(\ell+1) j_{\ell+1}(x) \, .
\ee
Spherical Bessel functions for any $n \in \mathbb{N}$ can be obtained
from $j_0(x)$ thanks to
\be
\label{rodrigues}
j_\ell(x) = (-1)^\ell x^\ell \left(\frac{1}{x} \frac{\dd}{\dd x} \right)^\ell j_0(x) \, , \qquad j_0(x) = \frac{\sin x}{x} \, .
\ee
If $\gr{r} = \gr{r}_2 - \gr{r}_1$, then $\mathrm{e}^{\ii \gr{k} \cdot \gr{r}} = \mathrm{e}^{\ii \gr{k} \cdot (\gr{r}_2-\gr{r}_1)}$. The Rayleigh expansion~\eqref{rayleigh} and the addition theorem for
Legendre polynomials \eqref{addition_thm} imply that
\be
\label{j0_gegenbauer}
j_0(kr) = \sum_\ell (2\ell + 1) j_\ell(kr_1) j_\ell(kr_2) P_\ell(\hat{\gr{r}}_1 \cdot \hat{\gr{r}}_2)
\ee
where $r$, $r_1$, and $r_2$ must form a triangle ($r^2 = r_1^2 + r_2^2
- 2r_1 r_2 \hat{\gr{r}}_1 \cdot \hat{\gr{r}}_2$). Note that this is a
special case of Gegenbauer addition theorem (eq. (3) on section 11.4 of~\cite{Watson}). 

The spherical Bessel functions satisfy the orthogonality relation
\be
\label{orth_j}
\int \dd r r^2 j_\ell(a r) j_\ell(b r) = \frac{\pi}{2ab} \delta_D(a-b) \, .
\ee

%%%%%%%%%%%%%%%%%%%%%%%%%%%%%%%%%%%%%%%%%%%%%
\subsection{Useful relations among spherical Bessel functions}
%%%%%%%%%%%%%%%%%%%%%%%%%%%%%%%%%%%%%%%%%%%%%
The relations involving spherical Bessel functions of low
order and their derivatives which we have used in this article are
\be
\label{j2x2}
\frac{j_1(x)}{x} = \frac{j_2(x) + j_0(x)}{3} \quad \frac{j_2(x)}{x^2} =\frac{1}{15}j_0(x) + \frac{2}{21}j_2(x) + \frac{1}{35} j_4(x) \quad \frac{j_3(x)}{x} = \frac{j_4(x) + j_2(x)}{7}
\ee

\be
\label{j0_prime_x}
\frac{j_0'(x)}{x}  = - \frac{1}{3} (j_0(x) + j_2(x))  \, , \qquad
j_0''(x)  = - \frac{1}{3} j_0(x) + \frac{2}{3} j_2(x) 
\ee
\be
\label{j3_x}
\frac{j_0^{(3)}(x)}{x} = \frac{1}{5} j_0(x) + \frac{1}{7} j_2(x) - \frac{2}{35} j_4(x) \, , \qquad
j_0^{(4)}(x)  = \frac{1}{5} j_0(x) - \frac{4}{7} j_2(x) + \frac{8}{35}j_4(x)  \, .
\ee

%%%%%%%%%%%%%%%%%%%%%
\subsection{Weber integrals}
%%%%%%%%%%%%%%%%%%%%%
\label{AppWeber}
From eq. (1) on section 13.24 of \cite{Watson} we obtain the following Weber integrals
\be
\int_0^\infty x^{p} j_\ell(x) \dd x = \sqrt{\pi}2^{p-1}\frac{\Gamma[(p+\ell+1)/2]}{\Gamma(\ell/2-p/2+1)}\,.
\ee
In particular, if $\ell$ and $n$ are both odd or both even, such that we can define $s$ by $\ell=2+n+2s$, we obtain
\be\label{WeberUseful}
\int j_\ell(x) x^{n-1 }\dd x = \left[\frac{2}{\pi}\int j_\ell(x)x^{2-n}\dd x\right]^{-1} = \frac{2^{n+s}(n+s)!}{(2s+3)!!}\,.
\ee
The main Weber integrals used in this article are
\begin{align}
\label{W_n_minus_one}
&\int j_2(x) x^{-1} \dd x = \frac{1}{3}\,,\quad &\int j_4(x) x^{-1} \dd x = \frac{2}{15}\,, \\
&\frac{2}{\pi}\int j_2(x)x^2\dd x=3\,,\quad &\frac{2}{\pi}\int j_2(x)x^2\dd x=\frac{15}{2}\,.
\end{align}

%%%%%%%%%%%%%%%%%%%%%%%%%
\subsection{Legendre polynomials}
%%%%%%%%%%%%%%%%%%%%%%%%%

Legendre polynomials are solutions regular at the origin of
\be
\label{legendre_eq}
\frac{\dd}{\dd \mu} \left[ (1-\mu^2) \frac{d}{d \mu}  P_\ell(\mu) \right] = -\ell(\ell+1) P_\ell(\mu) \, ,
\ee
 where $-1 \leq \mu \leq 1$. They obey the parity relation
\be
\label{parity}
P_\ell(-\mu) = (-1)^\ell P_\ell(\mu) \, .
\ee
The more relevant Legendre polynomials in this work are
\be
\label{p024}
P_0(\mu) =1 \, , \quad P_2(\mu) = \frac{3\mu^2 -1}{2} \, , \quad P_4(\mu) = \frac{35 \mu^4 - 30 \mu^2 +3}{8} \, .
\ee
Legendre polynomials satisfy an addition theorem, expressed in terms of spherical harmonics $\mathrm{Y}_{\ell m}$ as:
\be
\label{addition_thm}
\frac{2\ell+1}{4 \pi} P_\ell(\hat{\gr{r}}_1 \cdot \hat{\gr{r}}_2) = \sum_{m} \mathrm{Y}_{\ell m} ( \hat{\gr{r}}_1 ) \mathrm{Y}^\star_{\ell m} ( \hat{\gr{r}}_2 )\,.
\ee
They are also used to define a closure relation for the angular part of a Dirac delta function with
\be
\frac{\delta_D(r_1 - r_2)}{r_1^2} \sum_{\ell} \frac{2\ell+1}{4 \pi} P_\ell(\hat{\gr{r}}_1 \cdot \hat{\gr{r}}_2) = \delta_D(\gr{r}_1 - \gr{r}_2) \, .
\ee
For a function $f$ of a variable $\mu \in [-1, 1]$, using the orthogonality relation
\be\label{OrthoPl}
\int_{-1}^1 P_\ell(\mu) P_{\ell'}(\mu)\dd \mu=
\frac{2}{2\ell+1}\delta_{\ell \ell'}
\ee
we can write a Legendre series decomposition as
\be
\label{legendre_series}
f(\mu) = \sum_{\ell=0}^{\infty} c_\ell P_\ell(\mu) \, , \qquad c_\ell = \frac{2\ell+1}{2} \int_{-1}^1 \dd \mu \, f(\mu)  P_\ell(\mu) \, .
\ee

%%%%%%%%%%%%%%%%%%%%%%%%
\section{The kernel $\kerK_\ell$}
%%%%%%%%%%%%%%%%%%%%%%%%
\label{prop_kernel}
In order to obtain important identities satisfied by the kernel
coefficients $\kerK_\ell$, we first consider the integral
\be
\label{int_prime_def}
\frac{2}{\pi} \int \dd r r^2 j_\ell'(kr) j_\ell'(k'r) k k' \, ,
\ee
where the prime indicates differentiation with respect to the argument
of the functions. We will compute this expression using two different
methods. First we will use integration by parts, and then we will use recurrence relations. By means of integration by parts, we can show with the help of \eqref{sph_bessel_diff_eq} that
\be
\label{M_1}
\frac{2}{\pi} \int \dd r r^2 j_\ell'(kr) j_\ell'(k'r) k k' = \delta_D(k-k') - \frac{2}{\pi} \ell(\ell + 1) \int \dd r j_\ell(kr) j_\ell(k'r) \, ,
\ee
where we also have used \eqref{orth_j}.
For the second method, we use eqs.~\eqref{rec_rel}, \eqref{rec_rel_derivative}, and \eqref{orth_j} to obtain
\bea
\label{M_2}
\frac{2}{\pi} \int \dd r r^2 j_\ell'(kr) j_\ell'(k'r) k k' & = & \delta_D(k-k') - \frac{2}{\pi} \frac{\ell(\ell+1)}{2 \ell + 1} \left(k' \int \dd r r j_\ell(kr) j_{\ell-1}(k'r) \right. \nonumber\\ & & \left. + k \int \dd r r j_\ell(k' r) j_{\ell-1}(kr) \right)\,.
\eea
Each of the integrals on the right hand side of \eqref{M_2} belongs to
a class of discontinuous integrals of Bessel functions. 
The equation (8) on section 13.42 of \cite{Watson} can be used to show that
\be
\label{heaviside_watson}
k' \int \dd r r j_\ell(kr) j_{\ell-1}(k'r) + k \int \dd r r j_\ell(k' r) j_{\ell-1}(kr) = \begin{cases} 
\frac{1}{k} \left( \frac{k'}{k} \right)^\ell & \mbox{ if } k' < k \\
\frac{1}{k} & \mbox{ if } k'=k \\
\frac{1}{k'} \left( \frac{k}{k'} \right)^\ell & \mbox{ if } k' > k
\end{cases}
\ee
Hence, we find that
\be
\frac{2}{\pi} \int \dd r r^2 j_\ell'(kr) j_\ell'(k'r) k k' = \delta_D(k - k') - \frac{l(l+1)}{(2l+1)}  \frac{k_<^\ell}{k_>^{\ell+1}} \, ,
\ee
where $k_< \equiv \mathrm{min} \{ k, k' \}$ and $k_> \equiv \mathrm{max} \{k, k' \}$, and using the definition~\eqref{kerk_l} it is simply expressed as
\be
\label{M_final}
\frac{2}{\pi} \int \dd r r^2 j_\ell'(kr) j_\ell'(k'r) k k' = \delta_D(k - k') + \kerK_\ell(k, k') \, .
\ee
Comparing \eqref{M_1} and \eqref{M_final}, we conclude that
\be
\label{intjj}
\kerK_\ell(k, k') = - \frac{2}{\pi} \ell(\ell + 1) \int \dd r j_\ell(kr) j_\ell(k'r) \, .
\ee
We finally obtain a useful relation derived from the orthogonality relation \eqref{intjj}
\be
\label{int_kernel_j}
\int \dd k k^2 \kerK_\ell(k, k') j_\ell(kr) = -\frac{\ell(\ell+1)}{r^2} j_\ell(k'r) \, .
\ee

%%%%%%%%%%%%%%%%%%%%%%%%%%%%%%%%
\section{Comparison with \cite{Papai:2008bd}}
%%%%%%%%%%%%%%%%%%%%%%%%%%%%%%%%
\label{appendix_papai_szapudi}

We shall now demonstrate that the two point correlation function given
in \eqref{GenCorrFunc} coincides with the expression given in
\cite{Papai:2008bd} \modif{when evaluated in the case of constant bias ($\beta_1=\beta_2=\beta$)}.  The two-point correlation function in configuration space is written in that reference as
\begin{eqnarray}
\label{xi_papai}
\xidr^z(r, \phi_1, \phi_2) & = & \frac{1}{2 \pi^2} \int dk k^2 P(k) \Bigg[
\sum_{n_1, n_2 = 0, 1, 2} a_{n_1 \, n_2} \cos( n_1 \, \phi_1) \cos( n_2 \, \phi_2) 
\nonumber\\ & &
 + b_{n_1 \, n_2} \sin(n_1 \, \phi_1) \sin(n_2 \, \phi_2) \Bigg] \,, 
\end{eqnarray}
where the non-vanishing coefficients are given by
\bea
a_{00} &=& \left( 1 + \frac{2}{3} \beta + \frac{2}{15} \beta^2 \right) j_0(kr)  - \left(\frac{1}{3} \beta+ \frac{2}{21} \beta^2 \right) j_2(kr) + \frac{3}{140}  \beta^2 \, j_4(kr)\\
a_{02} &=& a_{20} = \left( -\frac{1}{2} \beta - \frac{3}{14} \beta^2 \right) j_2(kr) + \frac{1}{28} \beta^2 \, j_4(kr) \\
a_{22} & = & \left(\frac{1}{15}  j_0(kr) - \frac{1}{21}  j_2(kr) +  \frac{19}{140}  j_4(kr) \right) \beta^2 \\
b_{22}  & = & \left( \frac{1}{15}  j_0(kr) - \frac{1}{21}  j_2(kr) -  \frac{4}{35}  j_4(kr) \right) \beta^2\\
a_{10} &= & \frac{1}{g_1} \left[\left( 2 \beta+ \frac{4}{5} \beta^2 \right)  \frac{j_1(kr)}{(kr)} - \frac{1}{5}  \beta^2 \frac{j_3(kr)}{(kr)} \right] \\
a_{01} & = & - \frac{1}{g_2} \left[ \left( 2 \beta + \frac{4}{5} \beta^2 \right) \frac{j_1(kr)}{(kr)} - \frac{1}{5} \beta^2 \frac{j_3(kr)}{(kr)} \right]
\eea
\begin{align}
a_{11} &= \frac{1}{g_1 g_2} \left( \frac{4}{3}   \frac{j_0(kr)}{(kr)^2}- \frac{8}{3}  \frac{j_2(kr)}{(kr)^2} \right) \beta^2
\qquad  &b_{11} =& \frac{1}{g_1 g_2} \left( \frac{4}{3}  \frac{j_0(kr)}{(kr)^2} + \frac{4}{3} 
\frac{j_2(kr)}{(kr)^2} \right) \beta^2\\
a_{12} &= \frac{1}{g_1} \left( \frac{2}{5}   \frac{j_1(kr)}{(kr)} - \frac{3}{5} \frac{j_3(kr)}{(kr)} \right) \beta^2
\qquad &a_{21} =& - \frac{1}{g_2} \left( \frac{2}{5}   \frac{j_1(kr)}{(kr)} - \frac{3}{5} \frac{j_3(kr)}{(kr)} \right) \beta^2\\
b_{12} &= \frac{1}{g_1} \left( \frac{2}{5}  \frac{j_1(kr)}{(kr)} + \frac{2}{5}  \frac{j_3(kr)}{(kr)} \right) \beta^2
\qquad &b_{21} =& - \frac{1}{g_2} \left( \frac{2}{5}  \frac{j_1(kr)}{(kr)} + \frac{2}{5} \frac{j_3(kr)}{(kr)} \right) \beta^2 \, ,
\end{align}
with
\begin{equation}
g_1 \equiv \frac{\sin(\phi_2)}{\sin(\phi_2 - \phi_1)} \qquad \qquad 
g_2 \equiv \frac{\sin(\phi_1)}{\sin(\phi_2 - \phi_1)}  \, .
\end{equation}

\begin{table}[!htb]
\begin{center}

\ifnotcompact
\begin{tabular}{c | c}
& \includegraphics[scale=0.4]{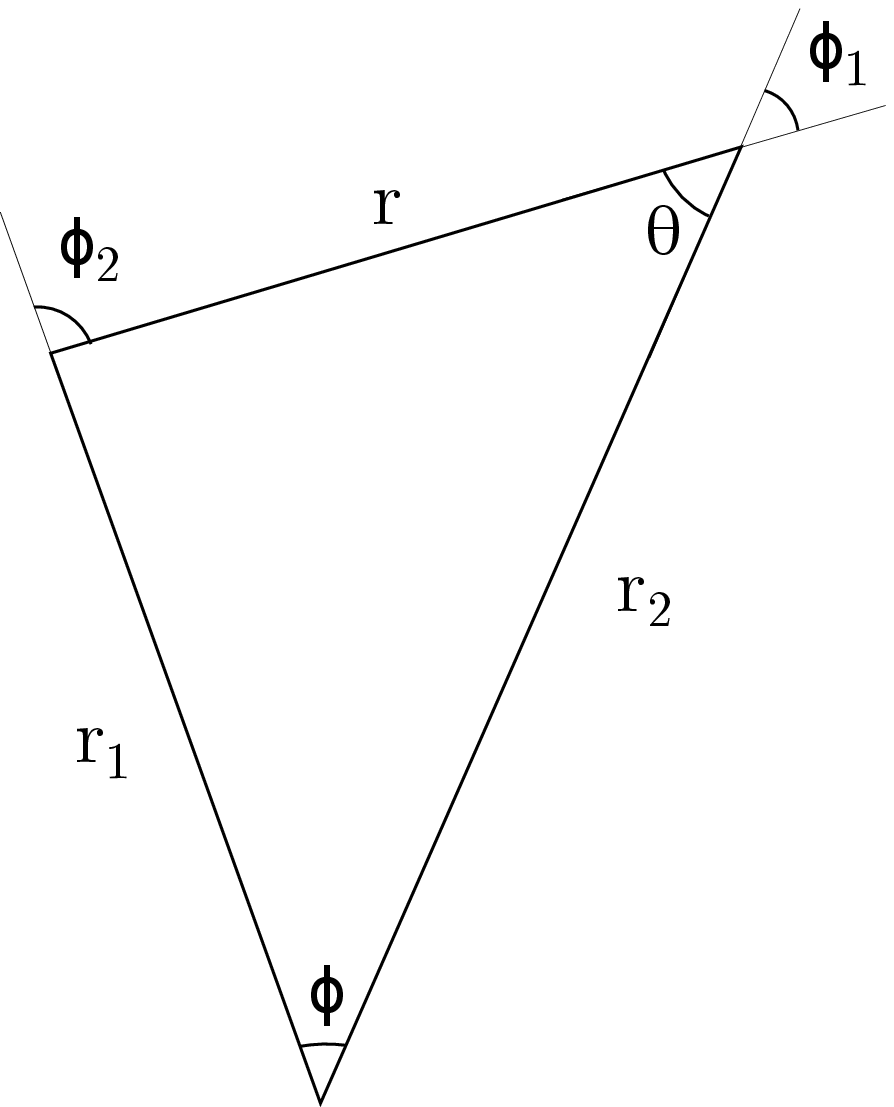} \\
\hline
$(1-\nu^2)\frac{r_1^2+r_2^2}{ r^2}$ & $1 - \frac{1}{2}[ \cos(2\phi_1) + \cos(2 \phi_2)]$  \\
 & \\
$\nu \left( \frac{r_1}{r_2} + \frac{r_2}{r_1} \right)$ &  $\left( \frac{\sin \phi_1}{\sin \phi_2} + \frac{\sin \phi_2}{\sin \phi_1} \right) \cos(\phi_2 - \phi_1)$\\
 & \\
$(1-\nu^2)^2\frac{r_1^2 r_2^2}{r^4}$  &  $\sin^2 \phi_1 \sin^2 \phi_2$\\
 & \\
$\nu(1-\nu^2)\frac{r_1 r_2}{r^2}$  & $2 -4 \, [\cos (2 \phi_1) + \cos (2 \phi_2)] -2 \, \sin^2 (\phi_2 - \phi_1)$\\
 & \\
$14\nu^2-6$  & $8 -14 \sin^2 (\phi_2 - \phi_1)$\\
 & \\
$\frac{\nu r^2}{ r_1 r_2}$ &  $\frac{ \sin^2(\phi_2 - \phi_1) \cos(\phi_2 - \phi_1)}{\sin \phi_1 \sin \phi_2}$ \\ 
\end{tabular}

\else
\begin{tabular}{c | c | c}
\hline
\multirow{6}{*}{\includegraphics[scale=0.43]{triangle3.eps}} &
$\displaystyle(1-\nu^2)\frac{r_1^2+r_2^2}{ r^2}$ & $\displaystyle1 - \frac{1}{2}[ \cos(2\phi_1) + \cos(2 \phi_2)]$  \\[0.4cm]
&$\displaystyle\nu \left( \frac{r_1}{r_2} + \frac{r_2}{r_1} \right)$ &  $\displaystyle\left( \frac{\sin \phi_1}{\sin \phi_2} + \frac{\sin \phi_2}{\sin \phi_1} \right) \cos(\phi_2 - \phi_1)$\\[0.4cm]
&$\displaystyle(1-\nu^2)^2\frac{r_1^2 r_2^2}{r^4}$  &  $\displaystyle\sin^2 \phi_1 \sin^2 \phi_2$\\[0.4cm]
&$\displaystyle\nu(1-\nu^2)\frac{r_1 r_2}{r^2}$  & $\displaystyle2 -4 \, [\cos (2 \phi_1) + \cos (2 \phi_2)] -2 \, \sin^2 (\phi_2 - \phi_1)$\\[0.4cm]
&$\displaystyle14\nu^2-6$  & $\displaystyle8 -14 \sin^2 (\phi_2 - \phi_1)$\\[0.4cm]
&$\displaystyle\frac{\nu r^2}{ r_1 r_2}$ &  $\displaystyle\frac{ \sin^2(\phi_2 - \phi_1) \cos(\phi_2 - \phi_1)}{\sin \phi_1 \sin \phi_2}$ \\[0.4cm]
\hline
\end{tabular}
\fi

\caption{Expression of the geometrical coefficients in \eqref{GenCorrFunc} when $\phi_1$, $\phi_2$ and $r$ are used as parameters.}
\label{geom_papai_szapudi}
\end{center}
\end{table}

Our strategy is to sum explicitly all the terms in eq.~\eqref{xi_papai} using some properties of spherical Bessel functions and trigonometric identities to obtain eq.~\eqref{GenCorrFunc}. For this sake, we consider the partial sums
\be
S_i \equiv \sum_{(n_1, n_2) \in \Lambda_i} a_{n_1 \, n_2} \cos(n_1 \phi_1) \cos(n_2 \phi_2) + b_{n_1 \, n_2} \sin(n_1 \phi_1) \sin(n_2 \phi_2) \, , \quad i=0, 1, 2
\ee
with  $\Lambda_0 = \{(0, 0)\}$, 
$\Lambda_1 = \{ (0, 1), (1, 0), (1, 2), (2, 1), (1, 1) \}$,
and
$\Lambda_2 = \{ (0, 2), (2, 0), (2, 2) \}$.
With the help of eqs.~\eqref{rec_rel}, \eqref{j2x2} and the trigonometric relations
\begin{align}
&2 \left[ \frac{\cos \phi_1 + \cos \phi_1 \cos (2 \phi_2) }{g_1}
- \frac{ \cos \phi_2 + \cos (2 \phi_1) \cos \phi_2}{g_2} \right] = - 4 \, \frac{\sin^2 (\phi_2 - \phi_1)}{\sin \phi_1 \sin \phi_2} \cos \phi_1 \cos \phi_2 
\, ,\label{rel_ratio_sin}\\
&\frac{\cos \phi_1}{g_1} - \frac{ \cos \phi_2}{g_2} = 2 - \left( \frac{\sin \phi_1}{\sin \phi_2}
+ \frac{\sin \phi_2}{\sin \phi_1} \right) \cos (\phi_2 - \phi_1) \, ,\label{rel_cos1}\\
&\frac{ \cos (2 \phi_1) \cos \phi_2}{g_2} - \frac{\cos \phi_1 \cos (2 \phi_2)}{g_1}  =
-2 + 2 \sin^2 (\phi_2 - \phi_1) + \left( \frac{ \sin \phi_1}{\sin \phi_2} +
\frac{\sin \phi_2}{\sin \phi_1} \right) \cos(\phi_2 - \phi_1) \, ,\label{rel_cos2}\\
&
\frac{\sin \phi_1 \sin (2 \phi_2)}{g_1} - \frac{ \sin (2 \phi_1) \sin \phi_2}{g_2} =
- 2 \sin^2(\phi_2 - \phi_1) \, ,\label{rel_sin}
\end{align}
we can express $S_1$ as
\bea
S_1 & = &
\frac{4}{3} \, \frac{\sin^2(\phi_2 - \phi_1) \cos(\phi_2 - \phi_1)}{\sin \phi_1 \sin \phi_2}
\left( \frac{j_0(kr)}{(kr)^2} + \frac{j_2(kr)}{(kr)^2} \right) \beta^2 
\nonumber\\ & & 
- \frac{2}{3} \left( \frac{\sin \phi_1}{\sin \phi_2} + \frac{ \sin \phi_2}{\sin \phi_1} \right)
\cos (\phi_2 - \phi_1) (j_0 (kr) + j_2(kr) ) \beta(1+\beta) \nonumber\\ & &  
+ \frac{4}{3} (j_0(kr) + j_2(kr)) \beta(1+\beta) 
+ \sin^2 (\phi_2 - \phi_1) \left[ -\frac{4}{5} \, j_0(kr) - \frac{6}{7} \, j_2(kr) - \frac{2}{35} \, j_4(kr) \right] \beta^2 \, . \nonumber
\end{eqnarray}
The trigonometric identities
\bea
\cos( 2 \phi_1) \cos (2 \phi_2) &=& -1 + (\cos (2 \phi_1) + \cos (2 \phi_2)) + 4 \sin^2 \phi_1 \sin^2 \phi_2 \, ,\\
\sin(2\phi_1) \sin(2\phi_2) &=& 2 - 2\sin^2(\phi_2-\phi_1) - (\cos(2\phi_1) + \cos(2\phi_2) - 4\sin^2\phi_1 \sin^2\phi_2 \, ,
\eea
allow us to write $S_2$ as
\bea
S_2 &=& \left(\frac{1}{5} - \frac{2}{15} \sin^2(\phi_2 - \phi_1) \right) \beta^2 j_0(kr)
+ \left( 1 -\frac{1}{2} ( \cos(2\phi_1) + \cos(2\phi_2)) \right)\beta j_2(kr) \nonumber\\
& & +\left(-\frac{3}{21} -\frac{3}{14} (\cos(2\phi_1) + \cos(2 \phi_2)) + \frac{2}{21} \sin^2(\phi_2 - \phi_1) \right) \beta^2 j_2(kr) \nonumber\\
& & +\left(-\frac{12}{35} + \frac{8}{35} \sin^2(\phi_2-\phi_1) + \frac{2}{7} ( \cos(2 \phi_1) + \cos(2 \phi_2) ) + \sin^2 \phi_1 \sin^2 \phi_2 \right)\beta^2 j_4(kr) \, . \nonumber
\eea
Finally, using the identities in Table~\ref{geom_papai_szapudi}, we can check that $\xidr^z(r, \phi_1, \phi_2)  = \frac{1}{2 \pi^2} \int dk k^2 P(k) [S_0+S_1+S_2]$ is precisely eq.~\eqref{GenCorrFunc}.

%%%%%%%%%%%%%%%%%%%%%%
\section{Cylindric method}
%%%%%%%%%%%%%%%%%%%%%%
\label{SecCylinder}

We first write the configuration space correlation function as
in~\cite{Papai:2008bd}. \modif{For simplicity we neglect the variations of
$\beta(r_1)$ and $\beta(r_2)$ and consider a constant bias $\beta$. We
also neglect the variations of the transfer function and set
$G=1$).} We get
\bea
\xidr^z(\gr{\dist},\gr{r}) & \equiv & \Crr^z(\gr{r}_1,\gr{r}_2) = \int \frac{\dd^3 \gr{k}}{(2 \pi)^{3}} \left[1 -\frac{\beta}{k^2} {\cal O}_{r_1} \right] \left[1 -\frac{\beta}{k^2}{\cal O}_{r_2} \right] {\rm e}^{-\ii \gr{k} \cdot \gr{r}} \nonumber\\
&=& \int \frac{\dd^3 \gr{k}}{(2 \pi)^{3}}\left[1 - \frac{2 \ii \gr{k}\cdot \hat{\gr{r}}_1}{k^2 r_1} + \frac{(\gr{k}\cdot\hat{\gr{r}}_1)^2}{k^2}\right] \left[1 + \frac{2 \ii \gr{k} \cdot \hat{\gr{r}}_2}{k^2 r_2} + \frac{(\gr{k}\cdot\hat{\gr{r}}_2)^2}{k^2}\right]{\rm e}^{- \ii \gr{k}\cdot\gr{r}}\,.
\eea
The operator which applies on the exponential is thus~\cite{Montanari:2012}
\be
\label{DefOr1r2}
{\cal  O}_{r_1 r_2}\equiv \left[1 - \frac{2 \ii \gr{k}\cdot \hat{\gr{r}}_1}{k^2 r_1}+\frac{(\gr{k}\cdot\hat{\gr{r}}_1)^2}{k^2}\right] \left[1 + \frac{2 \ii \gr{k}\cdot \hat{\gr{r}}_2}{k^2 r_2}+\frac{(\gr{k}\cdot\hat{\gr{r}}_2)^2}{k^2}\right]\,.
\ee
Like in the spherical method, we expand this operator in orders of plane-parallel perturbations as
\be
{\cal  O}_{r_1 r_2} = {\cal  O}^{(0)}_{r_1 r_2}  + \frac{r}{\dist}{\cal
  O}^{(1)}_{r_1 r_2} + \left(\frac{r}{\dist}\right)^2{\cal  O}^{(2)}_{r_1 r_2}  \,.
\ee
However, the difference is that we perform the integral on $\gr{k}$ using cylindrical coordinates instead of spherical coordinates. This implies that we need to decompose everything using these cylindrical coordinates. The natural choice for the azimuthal axis is the direction defined by $\hat{\gr{\dist}}$. The separation vector $\gr{r}\equiv \gr{r}_2-\gr{r}_1$ is then decomposed into an axial part which is $r \mu_r$ with $\mu_r \equiv \hat{\gr{\dist}}\cdot \hat{\gr{r}}$, and a polar part $\gr{r}_\perp$ whose length is just $r \omega_r$ where $\omega_r^2 \equiv 1-\mu_r^2$, that is
\be
\gr{r} = \mu_r \hat{\gr{\dist}}+\gr{r}_\perp\,,\qquad r_\perp = r\omega_r\,,\quad r_\parallel = r \mu_r\,.
\ee

A general Fourier mode $\gr{k}$ is decomposed similarly in cylindric coordinates with an axial part which is along $\hat{\gr{\dist}}$, and a polar part which is orthogonal to this axial direction. The polar part has a length $k\omega_k$ with $\omega_k^2 = 1-\mu_k^2$, but it is not necessarily aligned with $\gr{r}_\perp$. That is we use
\bea
\gr{k} &=& k \mu_k \hat{\gr{\dist}}+\gr{k}_\perp\,,\qquad k_\perp = k \omega_k \,,\qquad k_\parallel = k \mu_k \,,\qquad \cos\theta_k\equiv \hat{\gr{r}}_\perp\cdot \hat{\gr{k}}_\perp\\
\gr{k}\cdot\gr{r} &=& k_\parallel r_\parallel + k_\perp r_\perp \cos \theta_k = kr(\mu_r \mu_k + \omega_r \omega_k \cos \theta_k )\,.
\eea
In order to perform the integration in cylindrical coordinates, we need to express all products of the type $\hat{\gr{k}}\cdot\hat{\gr{r}}_1$ or $\hat{\gr{k}}\cdot\hat{\gr{r}}_2$ in the operator~\eqref{DefOr1r2} in terms of the radial and polar parts of $\gr{k}$, $\hat{\gr{r}}_1$ and $\hat{\gr{r}}_2$. This depends on the choice of parametrization, that is on the choice of $\gr{\dist}$ which is given by the parameter $v$ (see section~\ref{SecGeoDefv}). 

If we define $\Phi_1$ as the angle between $\hat{\gr{r}}_1$ and $\hat{\gr{\dist}}$, and similarly for $\Phi_2$, then from basic trigonometry we get
\begin{align}
&\cos\Phi_1\equiv\gr{\dist}\cdot \hat{\gr{r}}_1=\sqrt{1-\left[\frac{r(1-v)}{r_1}\right]^2\omega_r }\\
&\sin \Phi_1 = \frac{(1-v )r }{r_1}\omega_r = \frac{(1-v)(r/\dist) \omega_r}{\sqrt{1+(r/\dist)^2 (1-v)^2 -2 (r/\dist) (1-v)\mu}}\\
&\cos \Phi_2 \equiv\gr{\dist}\cdot \hat{\gr{r}}_2=\sqrt{1-\left[\frac{rv}{r_2}\right]^2\omega_r} \\
&\sin \Phi_2 =\frac{r v}{r_2}\omega_r = \frac{(r/\dist) v \omega_r}{\sqrt{1+(r/\dist)^2 v^2 + 2 (r/\dist) v \mu }}\,.
\end{align}
Note that in the limit where the separation is much larger than the average distance $\dist$, the directions $\gr{n}_1$ and $\gr{n}_2$ are nearly $\hat{\gr{\dist}}$ since $\Phi_1\approx\Phi_2\approx0$, meaning that at lowest order $ \hat{\gr{k}}\cdot \hat{\gr{r}}_1 = \mu_k$ and $ \hat{\gr{k}}\cdot \hat{\gr{r}}_2 = \mu_k$. Expanding in powers of $r/\dist$ we find up to first order
\be
\hat{\gr{k}}\cdot \hat{\gr{r}}_1 = \mu_k -(1-v) \frac{r}{\dist}\, \omega_k\,\omega_r\, \cos
\theta_k \,,\qquad \hat{\gr{k}}\cdot \hat{\gr{r}}_2 = \mu_k +v  \frac{r}{\dist}\, \omega_k\,\omega_r\, \cos \theta_k \,.
\ee
If we consider the particular case of the median parametrization $v=1/2$, we get up to second order
\bea
\hat{\gr{k}}\cdot \hat{\gr{r}}_1 &=& \mu_k -\frac{r}{2\dist} \, \omega_k\,\omega_r\, \cos
\theta_k +\frac{r^2}{8 \dist^2}\left( -\mu_k \omega_r^2 -2 \mu_r\,
  \omega_k\, \omega_r \cos \theta_k\right)\\
\hat{\gr{k}}\cdot \hat{\gr{r}}_2 &=& \mu_k +\frac{r}{2 \dist} \, \omega_k\,\omega_r\, \cos
\theta_k +\frac{r^2}{8 \dist^2}\left( -\mu_k \omega_r^2 -2 \mu_r\, \omega_k\, \omega_r\, \cos \theta_k\right)\,.
\eea
Using these expansions, we can find the operator~\eqref{DefOr1r2} order by order, in terms of the axial and polar parts. The lowest order is just the Kaiser operator
\be
{\cal  O}^{(0)}_{r_1 r_2} = (1+\beta \mu_k^2)^2\,.
\ee
At first order, if we choose the asymmetric parametrization ($v=0$) we get
\be
{\cal  O}^{(1)}_{r_1 r_2} = -2 \,\omega_k \mu_k (1+\beta \mu_k^2)\, \omega_r \cos \theta_k\,.
\ee
In the median case ($v=1/2$), there are no first order corrections, but the second order ones read
\bea
{\cal  O}^{(2)}_{r_1 r_2} &=&\frac{4 \beta^2 \mu_k^2}{(k r)^2}+\frac{2 \ii \beta}{kr}\left[\mu_r \mu_k (1+\beta
  \mu_k^2) -\omega_k(1-\beta \mu_k^2) \omega_r \cos
  \theta_k\right]\\
&&+\frac{\beta}{2}\left[-\omega_r^2
  \mu_k^2(1+\beta \mu_k^2)-2 \omega_k \mu_k (1+\beta \mu_k^2) \mu_r \omega_r
  \cos \theta_k+\omega_k^2 (1-\beta\mu_k^2)(\omega_r \cos \theta_k)^2\right]\nonumber\,.
\eea

The correlation function is found from these operators by integration over $\theta_k$ using the property~\eqref{JasIntegral} of Bessel functions. At lowest order, that is in the plane-parallel limit, we obtain
\be\label{BeautifulCylinderZero}
\xidr^{z(0)}(\gr{\dist},\gr{r}) =
\int \frac{\dd k_\parallel  k_\perp\dd
  k_\perp}{(2\pi)^2}\Pow(k)\left(1+\beta\mu_k^2\right)^2 J_0(k_\perp r_\perp)
{\rm  e}^{\ii k_\parallel r_\parallel}\,.
\ee
The corrections are defined using the expansion~\eqref{two_point_power}. At first order in the asymmetric case $v=0$, we get
\be\label{BeautifulCylinderOne}
\xidr^{z(1)}(\gr{\dist},\gr{r})=
\int \frac{\dd k_\parallel  k_\perp\dd k_\perp}{(2\pi)^2}\Pow(k) \frac{2 \ii \beta}{r}\omega_k \mu_k
  (1+\beta \mu_k^2) \partial_{k_\perp}J_0(k_\perp
r_\perp) {\rm  e}^{\ii k_\parallel r_\parallel}\,.
\ee
At second order in the median parametrization ($v=1/2$) there are no first order corrections and the second order corrections read
\bea\label{BeautifulCylinder}
\xidr^{z(2)}(\gr{\dist},\gr{r})&=&
\int \frac{\dd k_\parallel  k_\perp\dd k_\perp}{(2\pi)^2}
\Pow(k)\frac{\beta}{r^2}\left\{\frac{4}{k^2} \mu_k^2 +\frac{2}{k}\mu_k(1+\beta
\mu_k^2) \partial_{k_r}-\frac{2}{k}\omega_k(1-\beta
\mu_k^2)\partial_{k_\perp} \right.\nonumber\\
&&\left. -\frac{1}{2}\omega_k^2(1-\beta \mu_k^2) \partial_{k_\perp}^2  +\omega_k \mu_k (1+\beta \mu_k^2)\partial_{k_r}\partial_{k_\perp} \right.\nonumber\\
&&\left. -\frac{1}{2}\mu_k^2(1+\beta
\mu_k^2) r_\perp^2\right\} J_0(k_\perp r_\perp) {\rm  e}^{\ii k_\parallel r_\parallel}\,.
\eea
We can then check that the correlation functions found in eqs.~\eqref{BeautifulCylinderZero}~\eqref{BeautifulCylinderOne} and~\eqref{BeautifulCylinder} match the ones obtained using the spherical method if we use the relation
\be
\int_{-1}^1 \frac{\dd \mu_k}{2} \,{\rm e}^{\ii k_\parallel r_\parallel}\, \ii^m \,J_m\left(k
r_\perp \sqrt{1-\mu_k^2}\right)  P_\ell^m(\mu_k) = i^\ell j_\ell(k r) P_\ell^m (\mu_r)\,.
\ee
This general relation is simply found by expanding ${\rm e}^{\ii {\bf k}\cdot {\bf r}}$ either in spherical coordinates with eq.~\eqref{rayleigh} or in cylindrical coordinates with eq.~\eqref{RayleighCylinder} and by integrating both expressions. Indeed, the integrand $k_\perp \dd k_\perp$ in eq.~\eqref{BeautifulCylinder} can be changed to $k \dd k$, and then the integration on $k_\parallel$ can be performed using this formula, eventually leading to the results obtained with the spherical method in section~\ref{SecWideAngleReal}.

The spectrum at a given distance $\dist$ is defined as in the spherical case~\eqref{P_dist_def}, but the integral of the Fourier transform is done using cylindrical coordinates, leading to
\be
\xidk^z(\gr{\dist},\gr{k})=\frac{1}{(2\pi)^{3/2}} \int 2 \pi \dd r_\parallel r_\perp \dd r_\perp
J_0(p_\perp r_\perp ) \xidr^z(\gr{\dist},\gr{r}) {\rm e}^{-\ii k_\parallel r_\parallel}\,,
\ee
where the integration over the polar angle in spherical coordinates has already been performed. 
We use the integrals of appendix~\ref{AppIJ} in order to perform the integral on $r_\perp$, and for the integral on
$r_\parallel$ we use the property
\be
\int \dd r_\parallel \,\partial_{k_\parallel}^n {\rm e}^{\ii
  (k_\parallel-p_\parallel) r_\parallel} = 2\pi \delta_D^{\{n\}}(k_\parallel-p_\parallel)\,.
\ee
The final step consists in taking integrations by parts.

With this method, we recover that the spectrum at lowest order is the Kaiser limit in Fourier space. For this we need only $\IpowJ_{00}(p_\perp,k_\perp)$ as we get after integrating over $r_\parallel$ and $r_\perp$ 
\bea
\xidk^{z(0)}(k,\mu_k) &=& \int \frac{\dd p_\parallel p_\perp \dd p_\perp}{(2\pi)^{3/2}} (1+\beta \mu_p^2)^2\IpowJ_{00}(k_\perp,p_\perp) \delta_D(p_\parallel-k_\parallel)\,,\\
&=& \frac{1}{{(2\pi)^{3/2}}}(1+\beta \mu_k^2)^2 \Pow(k)\,.
\eea
We can compute the spectrum at first order for the asymmetric case $v=0$. For this we need only $\widetilde \IpowJ_{10}(p_\perp,k_\perp)$ defined in appendix~\ref{SecIntProdJ}, and applying the same method as for the background order, and after integration by parts, we get exactly the same expression as in eq.~\eqref{Power1}.
Finally, in the median parametrization ($v=1/2$), there are no first order corrections and the second order corrections are exactly the same as those of  eq.~\eqref{Power2_spherical}.

When deriving this result, the intermediate steps involve $\IpowJ_{0}(p_\perp,k_\perp)$, $\IpowJtilde_{10}(p_\perp,k_\perp)$, $\IpowJtilde_{20}(p_\perp,k_\perp)$ and $\IpowJ_{01}(p_\perp,k_\perp)$ of appendix~\ref{SecIntProdJ} which are handled with integration by parts. 

%%%%%%%%%%%%%%%%%%%%%%%%%%%%%%%%%%%%%
\section{Integrals over products of Bessel functions}\label{SecIntProdJ}
%%%%%%%%%%%%%%%%%%%%%%%%%%%%%%%%%%%%%

%%%%%%%%%%%%%%%%%%%%%
\subsection{General Method}
%%%%%%%%%%%%%%%%%%%%%

We follow the method developed in the appendix of \cite{Bernardeau:2010ac} to compute general integrals
involving a product of Bessel functions. We start from the simplest orthogonality relation
\be
\int x \dd x J_0(a x) J_0(b x) = \frac{\delta_D(a-b)}{b}\equiv \calJ_0(a-b)\,.
\ee
Then we define the following integrals
\bea
\IpowJ_{np} &\equiv& \int x \dd x J_n(ax) J_0(bx) x^{n+2p}\\
\IpowJtilde_{np} &\equiv& \int x \dd x \left[\partial_a^n J_0(ax)\right] J_0(bx) x^{n+2p}\,.
\eea
From the recurrence relations of the Bessel functions, we find that
these integrals are also related by recurrence relations, which are
\bea
\IpowJ_{n+1\,\,p}(a,b) &=&
-\frac{1}{a^{-n}} \partial_a\left[a^{-n}\IpowJ_{n\,p}(a,b)\right]\\
\IpowJ_{p-1\,\,n+1}(a,b) &=& \frac{1}{a^{p+1}} \partial_a\left[a^{p+1}\IpowJ_{p\,n}(a,b)\right]\\
\IpowJtilde_{n+1\,\,p}(a,b) &=& \partial_a\IpowJtilde_{n\,p}(a,b)\,.
\eea

\subsection{Integrals used in this article}\label{AppIJ}

In this paper we have used only a few integrals of this type which are
\bea
\IpowJtilde_{10}(a,b)&=& \partial_a \calJ_0(a,b)=\frac{\delta_D'(a-b)}{b}\\
\IpowJtilde_{20}(a,b)&=& \partial_a \partial_a \calJ_0(a,b)=\frac{\delta_D''(a-b)}{b}\\
\IpowJ_{01}(a,b)&=& - \partial_a \partial_a
\calJ_0(a,b)-\frac{1}{a}\partial_a
\calJ_0(a,b)+\frac{1}{a^2}\calJ_0(a,b)\nonumber\\ & = & -\frac{\delta_D''(a-b)}{b}-\frac{\delta_D'(a-b)}{ab}
+ \frac{\delta_D'(a-b)}{b a^2}\,.
\eea

%%%%%%%%%%%%%%%%%%%%%%%%%%%%%%%%%%%
\section{Integrals over products of spherical Bessel functions}\label{SecIntProdj}
%%%%%%%%%%%%%%%%%%%%%%%%%%%%%%%%%%%

\subsection{General method}
\label{cal_Int}
We follow a similar method for integrals of products of
spherical Bessel functions. However, the class of functions that are
required for this article is slightly more general. Let us define the integrals
\be
\IntI_{p\,q\,n}(a,b)=\frac{2}{\pi}\int_0^\infty \dd r r^{2+2 n + p + q}j_p(ar)
j_q(br)  \, ,
\ee
\be
{}^{n}_{p} \IntI_{\ell}(a,b)=\frac{2}{\pi}\int_0^\infty \dd r r^{n
}j_{\ell+p}(ar) j_\ell(br) \,\quad \Rightarrow \quad\, {}^{n}_{p} \IntI_{\ell}(a,b) = \IntI_{(\ell+p)\,\,\,\ell\,\,\,(n/2-1-\ell-p/2)}(a,b)\,.
\ee
Obviously, we have $\IntI_{pqn}(a,b) = \IntI_{qpn}(b,a)$. We will
derive recursive relations using the recurrence relations for the
spherical Bessel functions. From
\be
j_\ell'(x)= j_{\ell-1}(x) -\frac{\ell+1}{x} j_\ell(x) \,,\qquad j_\ell'(x)=-j_{\ell+1}(x)+\frac{\ell}{x}j_\ell(x)
\ee
we get the following recursions
\bea
\IntI_{p-1\,\, q\,\,n+1}(a,b) &=&
\frac{1}{a^{p+1}} \partial_a\left[a^{p+1} \IntI_{pqn}(a,b) \right]\\
\IntI_{p\,\, q-1\,\,n+1}(a,b) &=& \frac{1}{b^{q+1}} \partial_b\left[b^{q+1} \IntI_{pqn}(a,b) \right]\\
\IntI_{p+1\,\, q\,\,n}(a,b) &=& -\frac{1}{a^{-p}} \partial_a\left[a^{-p}
  \IntI_{pqn}(a,b) \right]\\
\IntI_{p\,\, q+1\,\,n}(a,b) &=& -\frac{1}{b^{-q}} \partial_b\left[b^{-q} \IntI_{pqn}(a,b) \right]
\eea
which can be generalized to give
\be
a^{-(p+n)}\IntI_{p+n\,\,q\,n}(a,b) =\left(-\frac{1}{a}\frac{\partial}{\partial
    a}\right)^n [a^{-p} \IntI_{p\,q\,n}(a,b)]\, ,
\ee
or in particular
\be
a^{-(p+n)}\IntI_{p+n\,\,p\,-p}(a,b)
=\left(-\frac{1}{a}\frac{\partial}{\partial a}\right)^n [a^{-p} \IntI(a,b)]\,.
\ee
In order to use these recursions, we need to start from a known
integral. The simplest starting point is just eq.~\eqref{orth_j},
which in the notation of this section is written as
\be\label{Start1}
\IntI(a,b) \equiv {}^{2}_0\IntI_p(a,b) = \IntI_{p\,p\,-p}(a,b) = \frac{\delta_D(a-b)}{a^2} \, .
\ee
Other starting points are necessary in this paper as we cannot reach
all integrals from the previous starting point and the recursion
relations. We find them by using eq.~(8) of section 13.42 of \cite{Watson}. It leads to
\be
{}^{1}_{1} \IntI_p(a,b)=I_{p+1 \,p\,-(p+1)}(a,b) = \frac{b^{p-1}}{a^{p+1}}H(a-b) \, ,
\ee
where $H$ is the Heaviside step functions which satisfies $H(0)=1/2$,
$H(x)=1$ if $x>0$ and $H(x)=0$ if $x<0$. Then, the recurrence relation
\eqref{rec_rel} implies that
\be\label{Start2}
{}^{2}_{2} \IntI_{p-1}(a,b)=\IntI_{p+1 \,p-1\,-p}(a,b) =
(2p+1)\frac{b^{p-1}}{a^{p+2}}H(a-b)-\frac{\delta_D(a-b)}{b^2} \, .
\ee 
Comparing \eqref{heaviside_watson}, \eqref{M_final}, and
\eqref{intjj}, we also conclude that
\be
{}^{0}_0 \IntI_\ell(a,b)=\IntI_{\ell\,\ell\,-(\ell+1)}(a,b)=\frac{1}{2\ell+1}\left[\frac{b^\ell}{a^{\ell+1}}H(a-b)+\frac{a^\ell}{b^{\ell+1}}H(b-a)\right] \, .
\ee

%%%%%%%%%%%%%%%%%%%%%%%%%%%%%
\subsection{Integrals used in this article}
%%%%%%%%%%%%%%%%%%%%%%%%%%%%%
\label{integrals_used}
With this method and starting from eq.~\eqref{Start1}, we obtain for instance
\bea
\IntI_{2\,2\,-1}(a,b) &=&\left(a\frac{\partial}{\partial a} \frac{1}{a}\right)
\left(b\frac{\partial}{\partial b} \frac{1}{b}\right)\IntI(a,b)\\
\IntI_{4\,4\,-3}(a,b) &=&\left(a^3\frac{\partial}{\partial a} \frac{1}{a^3}\right)
\left(b^3\frac{\partial}{\partial b} \frac{1}{b^3}\right)\IntI(a,b)\\
\IntI_{0\,0\,1}(a,b)&=&-\frac{1}{a^2}\partial_a a^2 \partial_a \IntI(a,b)\,.
\eea
If we start instead from eq.~\eqref{Start2} we also get
\bea
\IntI_{2\,0\,-1}(a,b) &=&\frac{3}{a^3} H(a-b) -\frac{\delta_D(a-b)}{b^2}
\equiv {}^2_2 \IntI(a,b)\\
a^{-4}\IntI_{4\,0\,-1}(a,b)&=&\left(-\frac{1}{a}\frac{\partial}{\partial a}\right)^2
a^{-2} \,{}^2_2 \IntI(a,b)\,.
\eea
Using recursions similar to these two examples, we are able to compute the integrals needed in this article which read
\bea
k^2 \IntI_{0\,0\,0}(k,p) &=& \delta_D(k-p)\\
k^2 \IntI_{0\,0\,1}(k,p) &=& -2\frac{\delta_D(k-p)}{k^2}+ 2 \frac{\delta_D'(k-p)}{k} -\delta_D''(k-p)\\
k^2 \IntI_{2\,0\,0}(k,p) &=& 8 \frac{\delta_D(k-p)}{k^2} - 5 \frac{\delta_D'(k-p)}{k} +\delta_D''(k-p)\\
k^2 \IntI_{2\,0\,-1}(k,p) &=& 3 \frac{H(k-p)}{k}-\delta_D(k-p)\\
k^2 \IntI_{4\,0\,-1}(k,p) &=& 105 \frac{H(k-p)}{k^3} -57 \frac{\delta_D(k-p)}{k^2} +12 \frac{\delta_D'(k-p)}{k} -\delta_D''(k-p)
\eea
\bea
k^2 \IntI_{2\,2\,-1}(k,p) &=& 3 \frac{\delta_D(k-p)}{kp} +3 \frac{\delta_D'(k-p)}{k} -\frac{\delta_D'(k-p)}{p} - \delta_D''(k-p)\\
k^2 \IntI_{0\,2\,0}(k,p) &=& \frac{\delta_D'(k-p)}{p}+\delta_D''(k-p)\\
k^2 \IntI_{4\,2\,-2}(k,p) &=& 24 \frac{\delta_D(k-p)}{k^2} -9 \frac{\delta_D'(k-p)}{k} +\delta_D''(k-p)\\
k^2 \IntI_{2\,4\,-2}(k,p) &=& 8\frac{\delta_D(k-p)}{p^2} + 5 \frac{\delta_D'(k-p)}{p} +\delta_D''(k-p)\\
k^2 \IntI_{4\,4\,-3}(k,p) &=& 15 \frac{\delta_D(k-p)}{(kp)} +5 \frac{\delta_D'(k-p)}{k} -3 \frac{\delta_D'(k-p)}{p} -\delta_D''(k-p)\\
k^2 \IntI_{4\,6\,-4}(k,p) &=& 24 \frac{\delta_D(k-p)}{p^2} + 9 \frac{\delta_D'(k-p)}{p} +\delta_D''(k-p)
\eea
\bea
k^2 \IntI_{2\,1\,-1}(k,p) &=& 3 \frac{\delta_D(k-p)}{k} -\delta_D'(k-p)\\
k^2 \IntI_{4\,3\,-3}(k,p) &=& 5 \frac{\delta_D(k-p)}{k} -\delta_D'(k-p)\\
k^2 \IntI_{2\,3\,-2}(k,p) &=&2 \frac{\delta_D(k-p)}{k} +\delta_D'(k-p)\\
k^2 \IntI_{4\,5\,-4}(k,p) &=&4 \frac{\delta_D(k-p)}{k}+\delta_D'(k-p)\,.
\eea

%%%%%%%%%%%%%%%%%%%%%%%%%%%%%%%%%%%%%%%%%%%%%%%%%

%\ifrmp
%\else
%\bibliographystyle{mn2e}
%\fi

\bibliographystyle{JHEP}
\bibliography{BiblioRSD}

\end{document}